\documentclass{ws-ijmpa}
\usepackage[super,compress]{cite}
\usepackage{graphics}
\usepackage{color}
\usepackage{bbm}
\begin{document}

\markboth{Fran\c cois Gelis}{Color Glass Condensate and
  Glasma}
\def\draftnote{}

\catchline{}{}{}{}{}
%
%%%%%%%%%%%%%%%%%%%%%%%%%%%%%%%%%%%%%%%%%%%%%%%%%%%%%%%%%%%%%%%%%%%%

\def\colora{}
\def\colorb{}
\def\colorc{}
\def\colord{}
\def\colore{}
\def\colorf{}

\def\empile#1\over#2{\mathrel{\mathop{\kern 0pt#1}\limits_{#2}}}
\newcommand{\slvarepsilon}{\raise.15ex\hbox{$/$}\kern-.53em\hbox{$\varepsilon$}}
\newcommand{\slL}{\raise.15ex\hbox{$/$}\kern-.53em\hbox{$L$}}
\newcommand{\slP}{\raise.15ex\hbox{$/$}\kern-.53em\hbox{$P$}}
\newcommand{\slD}{\raise.15ex\hbox{$/$}\kern-.53em\hbox{$D$}}
\newcommand{\slp}{\raise.1ex\hbox{$/$}\kern-.63em\hbox{$p$}}
\newcommand{\slq}{\raise.1ex\hbox{$/$}\kern-.53em\hbox{$q$}}
\newcommand{\slv}{\raise.1ex\hbox{$/$}\kern-.63em\hbox{$v$}}
\newcommand{\slR}{\raise.15ex\hbox{$/$}\kern-.53em\hbox{$R$}}
\newcommand{\slQ}{\raise.15ex\hbox{$/$}\kern-.53em\hbox{$Q$}}
\newcommand{\slK}{\raise.15ex\hbox{$/$}\kern-.53em\hbox{$K$}}
\newcommand{\slk}{\raise.15ex\hbox{$/$}\kern-.53em\hbox{$k$}}
\newcommand{\slSigma}{\raise.15ex\hbox{$/$}\kern-.53em\hbox{$\Sigma$}}
\newcommand{\slcalP}{\raise.15ex\hbox{$/$}\kern-.63em\hbox{$\cal P$}}
\newcommand{\slcalA}{\raise.15ex\hbox{$/$}\kern-.63em\hbox{$\cal A$}}
\newcommand{\slA}{\raise.15ex\hbox{$/$}\kern-.73em\hbox{$A$}}
\newcommand{\slbfA}{\raise.15ex\hbox{$/$}\kern-.73em\hbox{${\imb A}$}}
\newcommand{\slpartial}{\raise.15ex\hbox{$/$}\kern-.53em\hbox{$\partial$}}
\newcommand{\sla}{\raise.15ex\hbox{$/$}\kern-.53em\hbox{$a$}}
\newcommand{\slb}{\raise.15ex\hbox{$/$}\kern-.53em\hbox{$b$}}
\newcommand{\slc}{\raise.15ex\hbox{$/$}\kern-.53em\hbox{$c$}}
\newcommand{\slC}{\raise.15ex\hbox{$/$}\kern-.63em\hbox{$C$}}

\def\p{{\boldsymbol p}}
\def\q{{\boldsymbol q}}
\def\P{{\boldsymbol P}}
\def\l{{\boldsymbol l}}
\def\k{{\boldsymbol k}}
\def\m{{\boldsymbol m}}
\def\n{{\boldsymbol n}}
\def\x{{\boldsymbol x}}
\def\y{{\boldsymbol y}}
\def\X{{\boldsymbol X}}
\def\r{{\boldsymbol r}}
\def\s{{\boldsymbol s}}
\def\u{{\boldsymbol u}}
\def\v{{\boldsymbol v}}
\def\w{{\boldsymbol w}}
\def\z{{\boldsymbol z}}
\def\b{{\boldsymbol b}}
\def\a{{\boldsymbol a}}
\def\A{{\boldsymbol A}}
\def\E{{\boldsymbol E}}
\def\B{{\boldsymbol B}}
\def\cc{{\boldsymbol c}}

\def\bs{\boldsymbol}

\def\wt#1{\widetilde{#1}}

\def\vec{}

%%%%%%%%%%%%%%%%%%%%%%%%%%%%%%%%%%%%%%%%%%%%%%%%%%%%%%%%%%%%%%%%%%%%

\title{\bf COLOR GLASS CONDENSATE AND GLASMA\footnote{Based on lectures given at the 22nd Jyv\"askyl\"a
    Summer School, August 2012, Jyv\"askyl\"a, Finland.}}

\author{FRAN\c COIS  GELIS}

\address{Institut de Physique Th\'eorique, CEA Saclay\\
91191 Gif sur Yvette cedex, FRANCE\\
francois.gelis@cea.fr}

\maketitle

\begin{history}
%\received{Day Month Year}
%\revised{Day Month Year}
\end{history}

\begin{abstract}
  We review the Color Glass Condensate effective theory, that
  describes the gluon content of a high energy hadron or nucleus, in
  the saturation regime. The emphasis is put on applications to high
  energy heavy ion collisions. After describing initial state
  factorization, we discuss the Glasma phase, that precedes the
  formation of an equilibrated quark-gluon plasma. We end this review
  with a presentation of recent developments in the study of the
  isotropization and thermalization of the quark-gluon plasma.

\end{abstract}

\ccode{PACS numbers: 12.20.-m, 11.10.Hi, 11.15.Kc, 11.80.Fv, 11.80.La, 13.85.Hd}

\section{Introduction}
\label{sec:intro}
Heavy Ion Collisions at high energy are performed at the Relativistic
Heavy Ion Collider (RHIC) and the Large Hadron Collider (LHC), in
order to study the properties of nuclear matter under extreme
conditions of density and temperature. From lattice
simulations\cite{KarscC1}, it is known that above a certain
critical temperature, the protons and neutrons should melt into a
plasma made of their constituents, the Quark-Gluon Plasma (QGP), and
the critical temperature of this deconfinement transition is expected
to be reached in the collisions performed at RHIC and LHC.

\begin{figure}[htbp]
\begin{center}
\resizebox*{!}{1.7cm}{\includegraphics{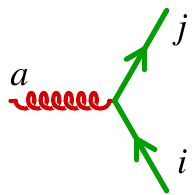}}\hfil
\resizebox*{!}{1.7cm}{\includegraphics{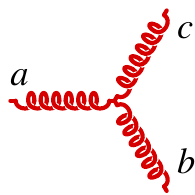}}\hfil
\resizebox*{!}{1.7cm}{\includegraphics{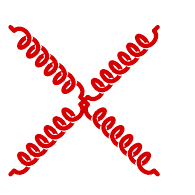}}
\end{center}
\caption{\label{fig:qcd}Elementary couplings between quarks and
  gluons in QCD.}
\end{figure}
\begin{figure}[htbp]
\begin{center}
\resizebox*{11cm}{!}{\includegraphics{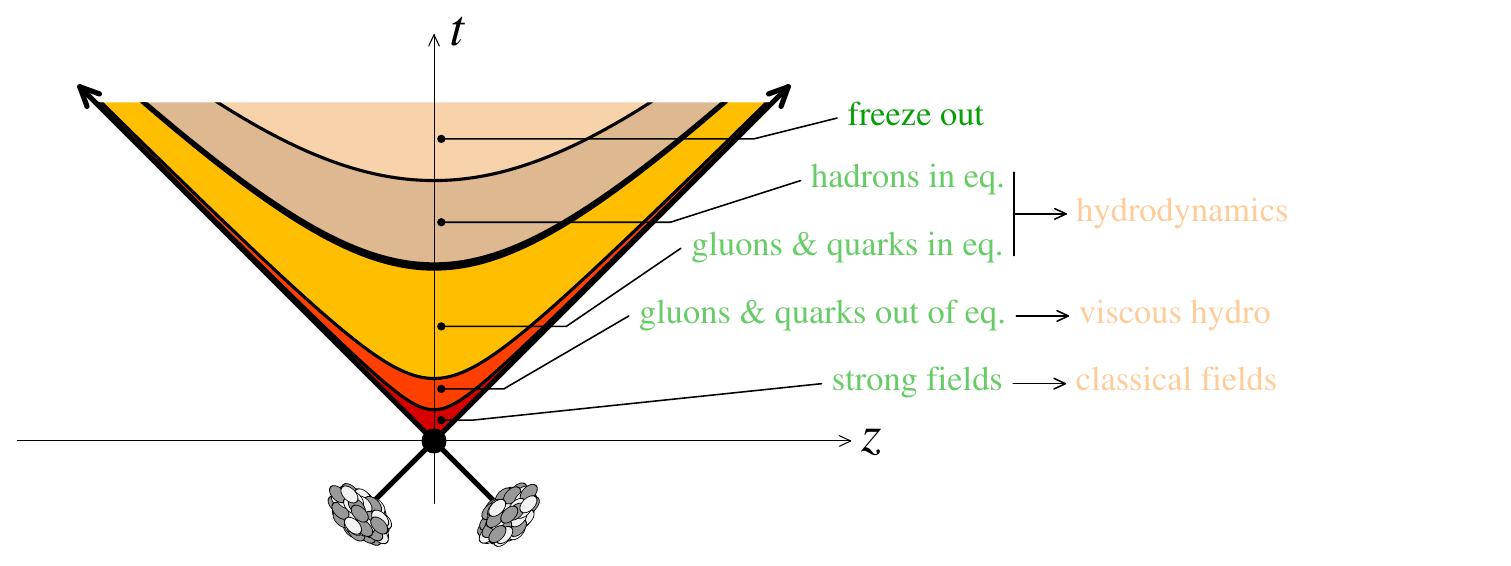}}
\end{center}
\caption{\label{fig:collision1}Successive stages of a high energy
  heavy ion collision, and the most widely used approaches to describe
  each stage.}
\end{figure}

Quantum Chromodynamics (QCD), the theory of the microscopic
interactions between the quarks and the gluons (see the figure
\ref{fig:qcd}), should in principle also be applicable to heavy ion
collisions. However, since the strong interaction coupling constant
becomes large at low momentum, it is not obvious a priori that heavy
ion collisions can be studied by weak coupling techniques. This is
certainly possible for the rare large-momentum processes that take
place in these collisions, but quite questionable for the bulk of the
particle production processes. Moreover, since the system produced in
such a collision expands rapidly along the collision axis (see the
figure \ref{fig:collision1}), its characteristic momentum scales
(e.g. its temperature if it reaches thermal equilibrium) decrease with
time.
Therefore, there will always be a time beyond which the coupling is
strong and weak coupling approaches are useless. This is obviously the
case near the phase transition that sees the quarks and gluons
recombine in order to form hadrons. In the best of cases, we can only
hope for a weak coupling treatment of the early stages of these
collisions (say up to a couple of fm/c after the collision).

\begin{figure}[htbp]
\begin{center}
\resizebox*{2.5cm}{!}{\includegraphics{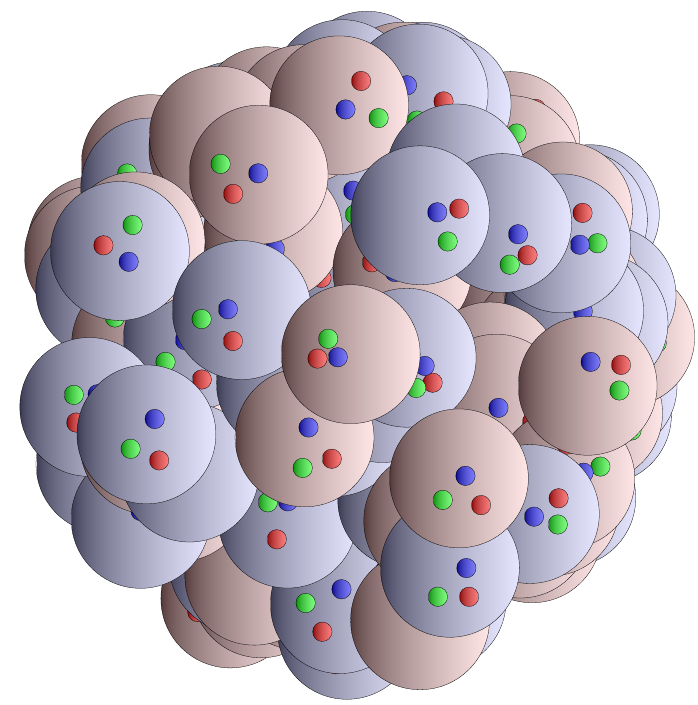}}\hfil
\resizebox*{8cm}{!}{\includegraphics{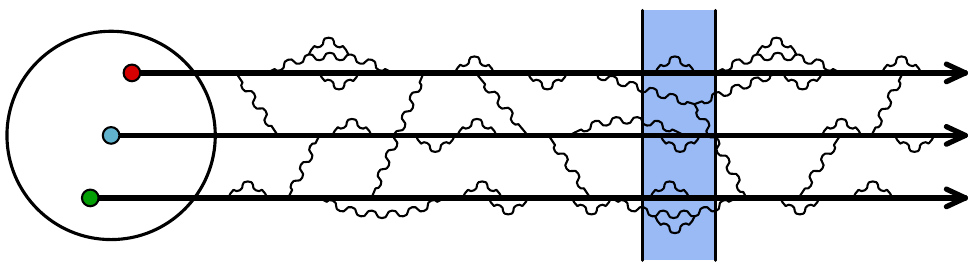}}
\end{center}
\caption{\label{fig:boost1}Parton content of a nucleon or nucleus at
  low energy. Left: cartoon of a nucleus at low energy and its valence
  quarks. Right: the thick lines represent the valence quarks, and the
  wavy lines are gluons. Virtual quark-antiquark pairs are not
  represented.}
\end{figure}
When applying QCD to the study of hadronic collisions, an essential
ingredient is the quark and gluon content of the hadrons that are
being collided, since the elementary degrees of freedom in QCD are
partons rather than hadrons. On the surface, a nucleon is made of three
valence quarks, bound by gluons. However, these quarks can
also temporarily fluctuate into states that have additional gluons and
quark-antiquark pairs (see the figure \ref{fig:boost1}). These
fluctuations are short lived, with a lifetime that is inversely
proportional to their energy. The largest possible lifetime of these
fluctuations is comparable to the nucleon size, and they can be
arbitrarily short lived. However, in a given reaction that probes the
nucleon, there is always a characteristic time scale set by the
resolution power of the probe (for instance by the frequency of the
virtual photon that probes the nucleon in Deep Inelastic Scattering).
Only the fluctuations that are longer lived than the resolution in
time of the probe can actually be seen in the process. The shorter
lived fluctuations are present, but do not influence the reaction.  
In collisions involving a low energy nucleon, only its valence quarks
and a few of these fluctuations are visible. Moreover, in a low energy
nucleon, there will typically be interactions between its constituents
during the collision with the probe, thus making low energy reactions
very complicated.

\begin{figure}[htbp]
\begin{center}
\resizebox*{2.5cm}{!}{\includegraphics{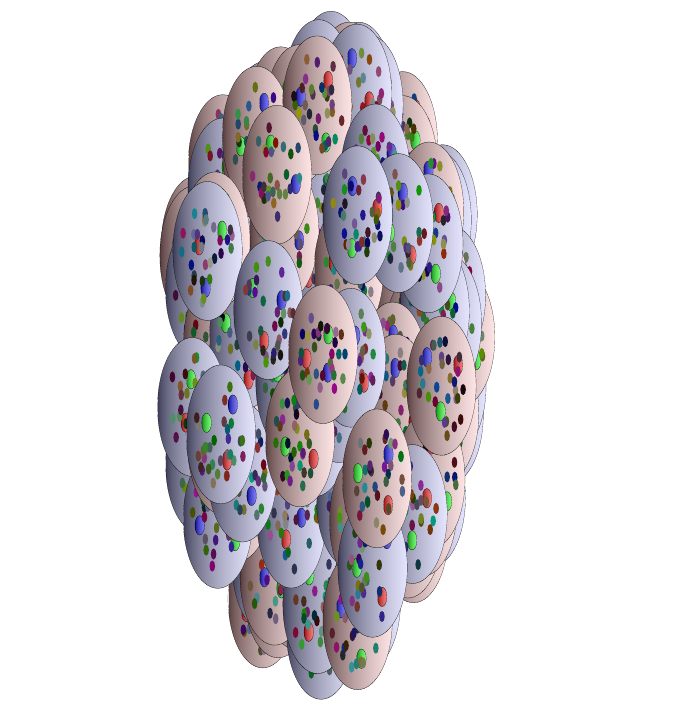}}\hfil
\resizebox*{8cm}{!}{\includegraphics{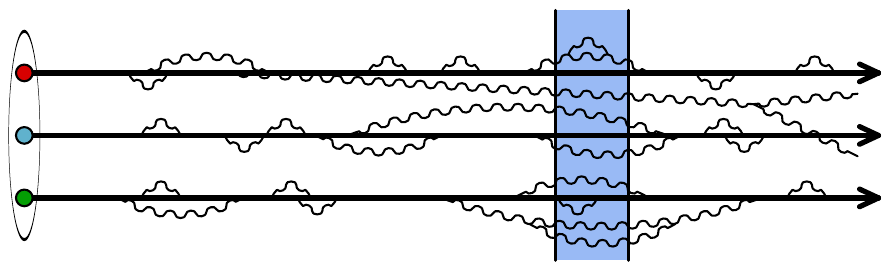}}
\end{center}
\caption{\label{fig:boost2}Parton content of a nucleon or nucleus at
  high energy, as seen in the laboratory frame. Left: boosted nucleus
  and its gluonic content. Right: fluctuations inside a boosted nucleon.}
\end{figure}
However, this picture is dramatically modified when the reaction
involves a high energy nucleon, due to relativistic kinematics (see
the figure \ref{fig:boost2}). Firstly, the geometry of the nucleon
changes due to Lorentz contraction: at very high energy, the nucleon
appears essentially two-dimensional in the laboratory
frame\footnote{The Lorentz gamma factor is $\gamma\sim 100$ at RHIC
  and $\gamma\sim 1000$ in heavy ion collisions at the
  LHC.}. Simultaneously, all the internal timescales of the nucleon
--in particular the lifetimes of the fluctuations and the duration of
the interactions among the constituents-- are multiplied by the same
Lorentz factor. The first consequence of this time dilation is that
the partons are now unlikely to interact precisely during the time
interval probed in the reaction: the constituents of a high energy
nucleon appear to be free during the collision. Secondly, since the
lifetimes of the fluctuations are also dilated, more fluctuations are
now visible by the probe: the number of gluons seen in a reaction
increases with the energy of the collision.
\begin{figure}[htbp]
\begin{center}
\resizebox*{7cm}{!}{\includegraphics{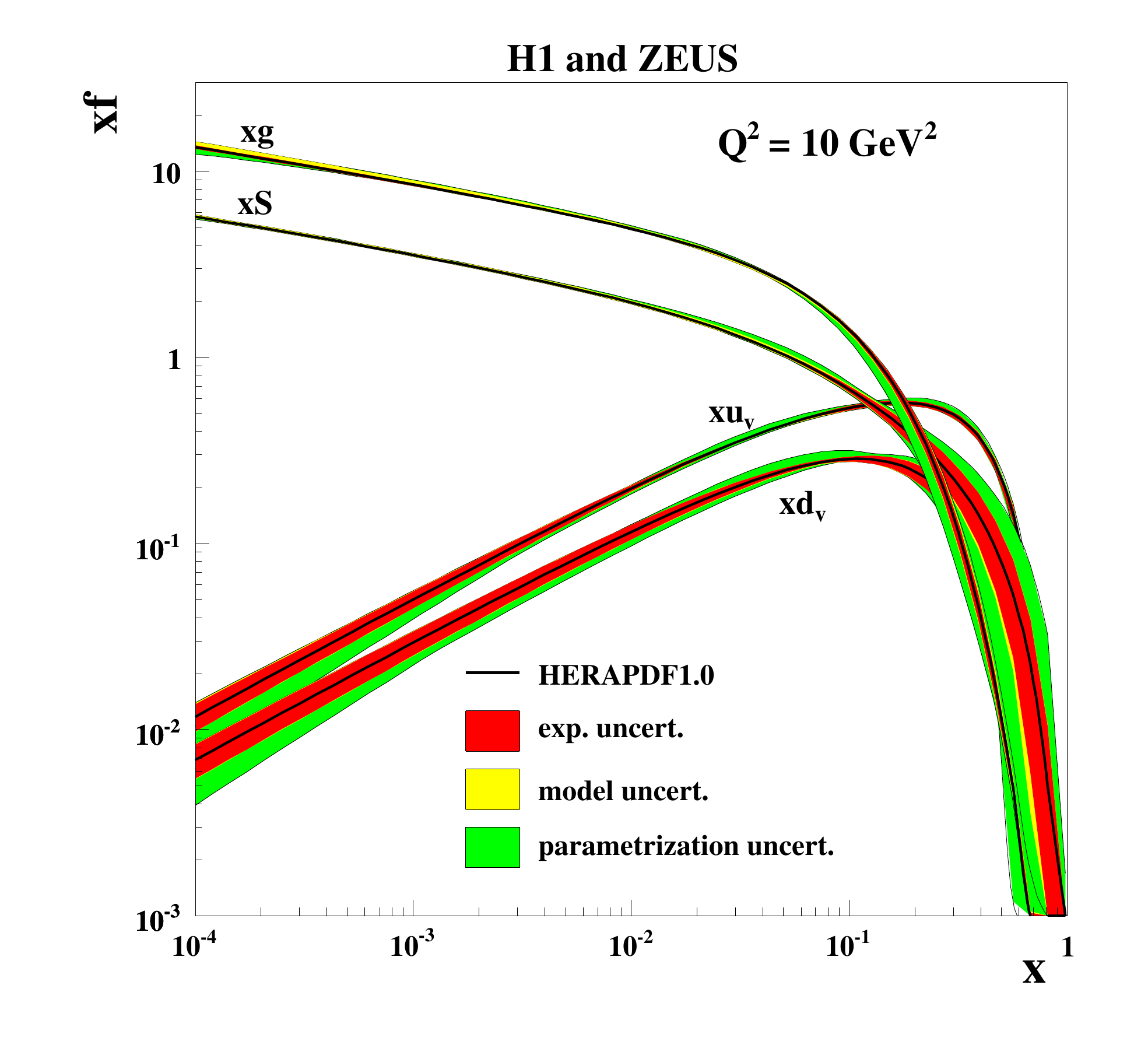}}
\end{center}
\caption{\label{fig:hera}Parton distributions in a proton, measured in
  Deep Inelastic Scattering at HERA\cite{Aarona2}.}
\end{figure}
This increase with energy of the number of gluons in a nucleon has
been observed experimentally in Deep Inelastic Scattering (DIS), for
instance at HERA. This is shown in the figure \ref{fig:hera} for a
proton. Note that in this plot, high energy corresponds to small
values of the longitudinal momentum fraction $x$ carried by the
parton, $x\equiv p_z/\sqrt{s}$. The other important feature of the
parton distributions at high energy/small $x$ is that the gluons are
outnumbering all the other parton species -- the valence quarks are
completely negligible in this kinematical region, and the sea quarks
are suppressed by one power of the coupling $\alpha_s$, since they are
produced from the gluons by the splitting $g\to q\overline{q}$.

\begin{figure}[htbp]
\begin{center}
\resizebox*{5cm}{!}{\includegraphics{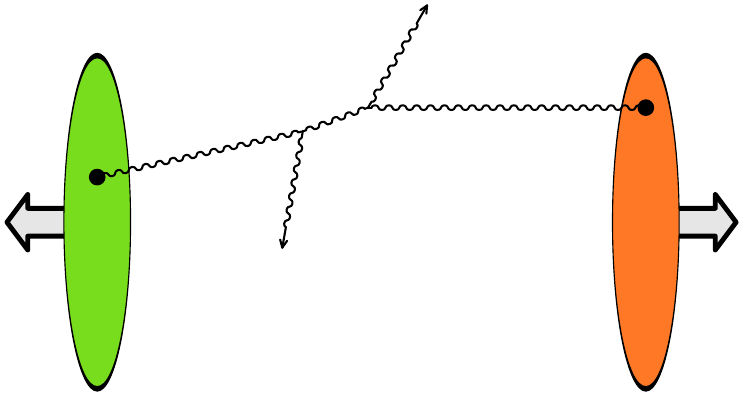}}\hfil
\resizebox*{5cm}{!}{\includegraphics{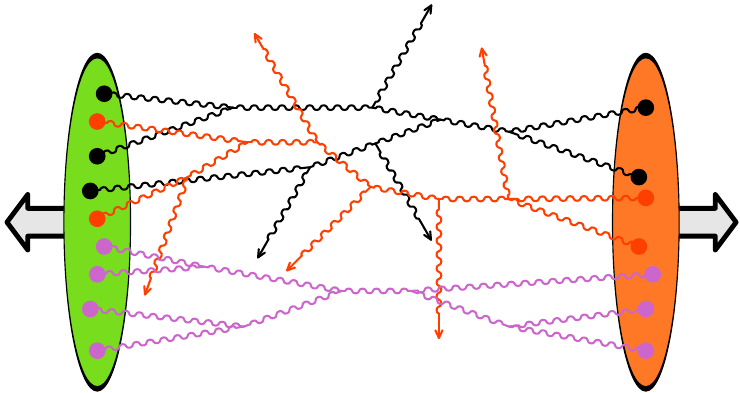}}
\end{center}
\caption{\label{fig:coll}Left: typical process in a hadronic collision in the dilute regime. Right: typical process in the dense regime.}
\end{figure}
This increase of the gluon distribution at small $x$ leads to a major
complication when applying QCD to compute processes in this
regime. Indeed, the usual tools of perturbation theory are well
adapted to the situation where the parton distributions are small (see
the left figure \ref{fig:coll}) and where a fairly small number of
graphs contribute at each order. On the contrary, when the parton
distributions increase, processes involving many partons become more
and more important, as illustrated in the right panel of the figure
\ref{fig:coll}. The extreme situation arises when the gluon occupation
number is of order $1/\alpha_s$: in this case, an infinite number of
graphs contribute at each order. This regime of high parton densities
is non-perturbative, even if the coupling constant is weak -- the
non-perturbative features arise from the fact that the large parton
density compensates the smallness of the coupling constant.  In
addition to requiring the summation of an infinite set of Feynman
diagrams, this situation requires some knowledge about the probability
of occurrence of multigluon states in the wavefunctions of the two
colliding projectiles, something which is not provided by the usual
parton distributions (they give only the single parton density).

A major progress in dealing with this situation has been the
realization that weak coupling methods can be used in this problem,
thanks to the dynamical generation of a scale that is much larger than
the non-perturbative scale $\Lambda_{_{QCD}}$ (the scale at which the
strong interactions become really strong, and where confinement
effects are crucial). This scale, known as the {\sl saturation
  momentum} and denoted $Q_s$, is due to the non-linear interactions
among the gluons. Roughly speaking, it is defined as the coupling
constant $\alpha_s$ times the gluon density per unit area (because the
Lorentz contraction makes the nucleus look like a flat sheet of gluons
in the laboratory frame), and it is a measure of the strength of the
gluon recombination processes that may occur when the gluon density
becomes large\cite{GriboLR1,MuellQ1,BlaizM2}. Any process involving
momenta smaller than $Q_s$ may be affected by {\sl gluon saturation}.

Moreover, an effective theory --known as the Color Glass Condensate
(CGC)-- has been developed in order to organize the calculations of
processes in the saturation regime. This effective theory approximates
the description of the fast partons in the wavefunction of a hadron by
exploiting the fact that their dynamics is slowed down by Lorentz time
dilation, and provides a way to track the evolution with energy of the
multigluon states that are relevant in the dense regime.  This
framework has been applied to a range of reactions at high energy:
DIS, proton-nucleus collisions and nucleus-nucleus collisions.  At
leading order, these calculations correspond to a classical field
description of the system. In nucleus-nucleus collisions, this
classical field remains coherent for a brief amount of time after the
collision, forming a state know as the {\sl Glasma}. A central question in
heavy ion collisions is to understand how these classical fields lose
their coherence in order to form a plasma of quarks and gluons in
local thermal equilibrium.

The purpose of these lectures is to expose the physics of the Color
Glass Condensate and of the Glasma. The main focus will be heavy ion
collisions\footnote{For this reason, the topic of pomeron loops will
  not be covered here, because these effects are important only in the
  dilute
  regime\cite{IancuT1,IancuT2,MuellSW3,KovneL2,KovneL3,HattaIMST1,Balit4}.},
except for the first sections where the simpler example of DIS is used
in order to introduce the concept of gluon saturation and the CGC
framework.  Our emphasis is to present a consistent framework to
describe heavy ion collisions from the pre-collision initial state to
the time at which the system may be described by
hydrodynamics\footnote{Theoretical and phenomenological aspects of the
  Color Glass Condensate, with slightly different points of view, have
  also been addressed in other
  reviews\cite{IancuV1,Lappi6,Weige2,GelisIJV1}.}.  The outline is the
following:
\begin{itemize}
\item In the section \ref{sec:small-x}, we discuss the evolution of the
  wavefunction of a color singlet dipole, and derive the BFKL and BK
  equations. This will provide a first glimpse at gluon saturation,
  and allow us to explain some interesting scaling properties for the
  inclusive DIS cross-section.
\item In the section \ref{sec:CGC}, we introduce the Color Glass
  Condensate effective theory, as a more general way to describe the
  saturation regime.
\item The section \ref{sec:factorization} is devoted to a discussion
  of the factorization of the logarithms that appear in loop
  calculations in the CGC framework, for heavy ion collisions. We show
  how these logarithms can be absorbed into universal distributions
  that describe the multigluon content of the two colliding nuclei.
\item In the section \ref{sec:pheno}, we present the main ideas of the
  Glasma picture, and discuss some phenomenological consequences
  that follow from it.
\item The section \ref{sec:glasma} discusses the final state
  evolution, that leads from the coherent Glasma fields to the
  thermalized quark gluon plasma. The main focus is on the
  instabilities of the classical solutions of the Yang-Mills
  equations, their resummation, and their possible role in the
  isotropization and thermalization processes.
\end{itemize}

\section{Hadron wavefunction at high energy}
\label{sec:small-x}
In order to introduce the physics of gluon saturation, we consider in
this section the case of Deep Inelastic Scattering. This situation is
one of the simplest, since only one of the two projectiles is a hadron,
the other being a lepton that interacts only electromagnetically via a
photon exchange. DIS can viewed as an interaction between a photon
with a negative virtuality ($q_\mu q^\mu = -Q^2<0$) and the hadronic
target.

\begin{figure}[htbp]
\begin{center}
\resizebox*{8cm}{!}{\includegraphics{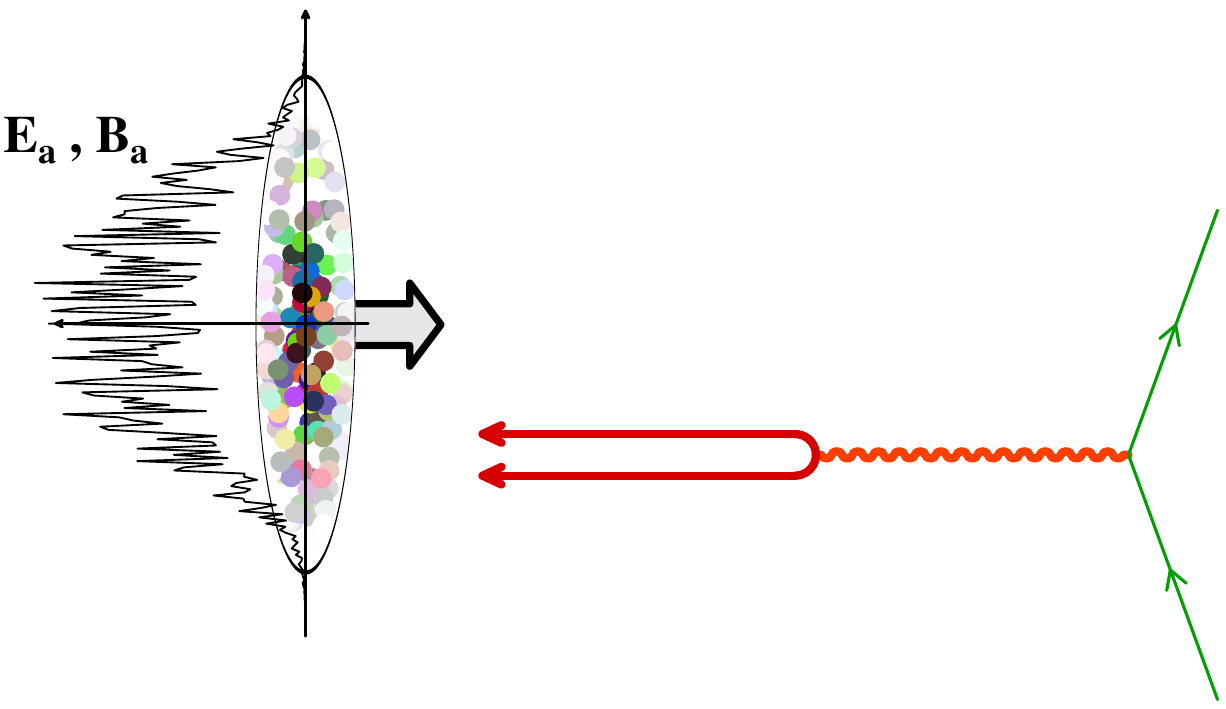}}
\end{center}
\caption{\label{fig:dis}Cartoon of Deep Inelastic Scattering in a
  frame where the photon fluctuates into a dipole.}
\end{figure}
The cross-section for this process is frame independent, but in this
section we will analyze it in a frame where the photon has a large
longitudinal momentum, while the hadron has only a moderate momentum.
In such a frame, the photon can fluctuate into a quark-antiquark
pair\footnote{Of course, the photon may also fluctuate into more
  complicated Fock states, such as a $q\overline{q}g$ state, but the
  probability of occurrence of these states is suppressed by at least
  one extra power of the coupling $\alpha_s$ and therefore they do not
  contribute at leading order.} in a color singlet state (called a
{\sl color dipole}), and it is this pair that interacts with the
colored constituents of the target. Moreover, at high energy, this
dipole sees the constituents (quarks and gluons) as frozen due to time
dilation. Therefore, its interactions with these constituents can be
approximated as interactions with the (static) color field that they
create. Moreover, thanks to confinement, it is legitimate to assume
that this target color field occupies a bounded region of space.

\subsection{Eikonal approximation}
We will compute the total cross-section between the dipole and the
target field by using the optical theorem, that relates the total
cross-section to the forward scattering amplitude. In the limit where
the longitudinal momentum of the dipole is very large, this amplitude
becomes very simple and is given by the {\sl eikonal
  approximation}. Since this is an extremely important result in the
study of high energy scattering, let us derive it in detail. Consider
an $S$-matrix element,
\begin{equation}
S_{\beta\alpha}\equiv\big<\beta{}_{\rm out}\big|\alpha{}_{\rm in}\big>
=
{\colorc\big<\beta_{\rm in}}\big|U(+\infty,-\infty)\big|{\colorc\alpha_{\rm in}\big>}\; ,
\end{equation}
for the transition between two arbitrary states made of quarks,
antiquarks and gluons, $\alpha$ and $\beta$. In the second equality,
$U(+\infty,-\infty)$ is the evolution operator from the initial to the
final state. It can be expressed as the time ordered exponential
of the interaction part of the Lagrangian,
\begin{equation}
U(+\infty,-\infty)=T\,\exp\Big[i\int d^4x\;{\cal L}_{\rm int}(\phi_{\rm in}(x))\Big]\; ,
\end{equation}
where $\phi_{\rm in}$ denotes generically the fields in the
interaction picture. In our problem, ${\cal L}_{\rm int}$ contains both
the self-interactions of the fields, and their interaction with the
target color field.
We want to compute the high energy limit of this scattering amplitude,
\begin{equation}
    S_{\beta\alpha}^{(\infty)}\equiv\lim_{\omega\to +\infty}
    {\colorc\big<\beta_{\rm in}}\big|
    {\colorb e^{-i\omega K^3}}U(+\infty,-\infty)
    \underbrace{{\colorb e^{+i\omega K^3}}\big|{\colorc\alpha_{\rm in}\big>}}_{\mbox{\scriptsize boosted state}}
\end{equation}
where ${\colorb K^3}$ is the generator of Lorentz boosts in the $+z$
direction.

Before doing any calculation, a simple argument can help understand
what happens in this limit. Quite generally, scattering amplitudes are
proportional to the duration of the overlap between the wavefunctions
of the two colliding objects. In the present case, it should scale as
the time spent by the incoming state in region occupied by the target
field. This time is inversely proportional to the energy of the
incoming state, and goes to zero in the limit $\omega\to +\infty$. If
the interaction between the projectile and the target field was via a
scalar exchange, then the conclusion would be that the scattering
amplitude vanishes in the high energy limit (in other words,
$S$-matrix elements would go to unity). However, interactions with a
color field are via a vector exchange, i.e. the target field couples
to a four-vector that represents the color current carried by the
projectile, by a term of the form $-ig{\cal A}_\mu J^\mu$. At high
energy, the longitudinal component of this four-vector increases
proportionally to the energy, and compensates the small time spent in
the interaction zone. Thus, for states that interact via a vector
exchange\footnote{By the same reasoning, gravitational interactions,
  that involve a spin two exchange, would lead to scattering amplitudes
  that grow linearly with energy.}, we expect that scattering
amplitudes have a finite high energy limit (nor zero, nor infinite).

This calculation is best done using {\sl light-cone coordinates}. For
any four-vector $a^\mu$, one defines
\begin{eqnarray}
{\colord a^+}\equiv \frac{a^0+a^3}{\sqrt{2}}\quad,\quad
{\colord a^-}\equiv \frac{a^0-a^3}{\sqrt{2}}\; .
\end{eqnarray}
The following formulas are often useful,
\begin{eqnarray}
&&
x\cdot y = x^+y^-+x^-y^+-\vec\x_\perp\cdot\vec\y_\perp
\nonumber\\
&&
d^4x=dx^+ dx^- d^2\vec\x_\perp
\nonumber\\
&&
\square=2\partial^+\partial^--\vec{\bs\nabla}_\perp^2
\qquad\mbox{with}\quad 
\partial^+\equiv\frac{\partial}{\partial x^-}\;,\;
\partial^-\equiv\frac{\partial}{\partial x^+}\; .
\end{eqnarray}
For a highly boosted projectile in the $+z$ direction, $x^+$ plays the
role of the time, and the Hamiltonian is the $P^-$ component of the
momentum. The generator of longitudinal boosts in light-cone
coordinates is
\begin{equation}
K^-\equiv M^{-+}=-K^3\; .
\end{equation}
In order to derive the eikonal approximation, the following identities
are also very useful,
\begin{eqnarray}
e^{i\omega K^-}\;{\colord P^-}\;e^{-i\omega K^-}&=& e^{-\omega} {\colord P^-}
\nonumber\\
e^{i\omega K^-}\;{\colord P^+}\;e^{-i\omega K^-}&=& e^{+\omega} {\colord P^+}
\nonumber\\
e^{i\omega K^-}\;{\colord P^j}\;e^{-i\omega K^-}&=& {\colord P^j}
\; .
\end{eqnarray}
They express the fact that, under longitudinal boosts, the components
$P^\pm$ of a four-vector are simply rescaled, while the transverse
components are left unchanged.

Under longitudinal boosts, states, creation operators and field
operators are transformed as follows,
\begin{eqnarray}
      &&
      {\colorb e^{-i\omega K^-}}\big|\vec\p\cdots{}_{\rm in}\big>
      =\big|({\colorb e^\omega} p^+,\vec\p_\perp)\cdots{}_{\rm in}\big>
      \nonumber\\
      && 
      {\colorb e^{-i\omega K^-}}a^\dagger_{\rm in}(q){\colorb e^{i\omega K^-}}
      =a^\dagger_{\rm in}({\colorb e^\omega} q^+,{\colorb e^{-\omega}}q^-,\vec\q_\perp)\nonumber\\
      &&\vphantom{\int}
      {\colorb e^{i\omega K^-}}\phi_{\rm in}(x){\colorb e^{-i\omega K^-}}
      =\phi_{\rm in}({\colorb e^{-\omega}} x^+,{\colorb e^{\omega}}x^-,\vec\x_\perp)
      \; .
\end{eqnarray}
Note that the last equation is valid only for a scalar field, or for
the transverse components of a vector field. In addition, the $\pm$
components of a vector field receive an overall rescaling by a factor
$e^{\pm\omega}$. Moreover, since a longitudinal boost does not alter
the time ordering, we can also write
\begin{equation}
      {\colorb e^{i\omega K^-}}U(+\infty,-\infty){\colorb e^{-i\omega K^-}}
      =
      T \exp i\int d^4x
      \;{\cal L}_{\rm int}({\colorb e^{i\omega K^-}}\phi_{\rm in}(x){\colorb e^{-i\omega K^-}})\; .
\end{equation}
Likewise, the components of the vector current that couples to the
target field transform as
\begin{eqnarray}
      &&
      {\colorb e^{i\omega K^-}}J^i(x){\colorb e^{-i\omega K^-}}
      =J^i({\colorb e^{-\omega}} x^+,{\colorb e^{\omega}}x^-,\vec\x_\perp)
      \nonumber\\
      &&
      {\colorb e^{i\omega K^-}}J^-(x){\colorb e^{-i\omega K^-}}
      ={\colorb e^{-\omega}}\,J^-({\colorb e^{-\omega}} x^+,{\colorb e^{\omega}}x^-,\vec\x_\perp)
      \nonumber\\
      &&
      {\colorb e^{i\omega K^-}}J^+(x){\colorb e^{-i\omega K^-}}
      ={\colorb e^\omega}\,J^+({\colorb e^{-\omega}} x^+,{\colorb e^{\omega}}x^-,\vec\x_\perp)
      \; .
\end{eqnarray}
Naturally, the target field ${\cal A}_\mu$ does not change when we
boost the projectile. For simplicity, let us assume that ${\cal
  A}_\mu$ is confined in the region $-L\le x^+ \le +L$. We can split
the evolution operator into three factors,
\begin{equation}
    U(+\infty,-\infty)=U(+\infty,+L)U(+L,-L)U(-L,-\infty)\; .
\end{equation}
The factors $U(+\infty,+L)$ and $U(-L,-\infty)$ do not contain the
external potential. For these two factors, the change of variables
{\colord$e^{-\omega} x^+\to x^+$}, {\colord$e^{\omega} x^-\to
  x^-$} leads to
\begin{eqnarray}
    \lim_{\omega\to+\infty}
    {\colorb e^{i\omega K^-}}
    U(+\infty,+L)
    {\colorb e^{-i\omega K^-}}
    &=&U_0(+\infty,0)\nonumber\\
    \lim_{\omega\to+\infty}
    {\colorb e^{i\omega K^-}}
    U(-L,-\infty)
    {\colorb e^{-i\omega K^-}}
    &=&U_0(0,-\infty)\; ,
\end{eqnarray}
where $U_0$ is the same as $U$, but with the self-interactions only
(since these two factors correspond to the evolution of the projectile
while outside of the target field).
For the factor $U(+L,-L)$, the change {\colord $e^{\omega}x^-\to x^-$} gives
\begin{equation}
    \lim_{\omega\to+\infty}
    {\colorb e^{i\omega K^-}}
    U(+L,-L)
    {\colorb e^{-i\omega K^-}}=
    \exp \Big[
      i {\colorc g}
    \int d^2\vec\x_\perp \;{\colord\chi(\vec\x_\perp)}\,{\colora\rho(\vec\x_\perp)}
    \Big]\; ,
\end{equation}
\begin{equation}
  \mbox{with}\qquad\left\{
  \begin{aligned}
    &
    {\colord\chi(\vec\x_\perp)}\equiv \int dx^+\;
    {\colord{\cal A}^-(x^+,0,\vec\x_\perp)}
    \\
    &
    {\colora\rho(\vec\x_\perp)}\equiv \int dx^-\;{\colora J^+(0,x^-,\vec\x_\perp)}
\; .
  \end{aligned}
  \right.
\end{equation}
Thus, the high-energy limit of the scattering amplitude is
\begin{equation}
    S_{\beta\alpha}^{(\infty)}=
    {\colorc\big<\beta_{\rm in}\big|}
    U_0(+\infty,0)
      \;\exp 
      \Big[i{\colorc g}
	\int d^2\x_\perp\;{\colord\chi(\vec\x_\perp)}
		   {\colora\rho(\vec\x_\perp)}
      \Big]
    U_0(0,-\infty)
    {\colorc\big|\alpha_{\rm in}\big>}\; .
\label{eq:eiko1}
\end{equation}
A few remarks are in order at this point
\begin{itemize}
\item Only the ${\cal A}^-$ component of the {\colora vector potential} matters.
\item The self-interactions and the interactions with the external
  potential are factorized into three separate factors -- this is a
  generic property of high energy scattering.
\item This is an exact result in the limit $\omega\to+\infty$.
\end{itemize}

Eq.~(\ref{eq:eiko1}) is an operator formula that still contains the
self-interactions of the fields to all orders. In order to evaluate
it, one must insert the identity operator written as a sum over a
complete set of states on each side of the exponential,
\begin{eqnarray}
 && 
  S_{\beta\alpha}^{(\infty)}=\smash{\sum_{{\colore\gamma},{\colore\delta}}}
  {\colorc\big<\beta_{\rm in}\big|}
  U_0(+\infty,0)
  {\colore\big|\gamma_{\rm in}\big>}
  \vphantom{\Big[}
  \nonumber\\
  &&\qquad\quad\times
  {\colore\big<\gamma_{\rm in}\big|}
    \exp \Big[
      i{\colorc g}
      {\int\limits_{\vec\x_\perp}}{\colord\chi(\vec\x_\perp)}
      {\colora\rho(\vec\x_\perp)}
    \Big]
  {\colore\big|\delta_{\rm in}\big>
  \big<\delta_{\rm in}\big|}
  U(0,-\infty)
  {\colorc\big|\alpha_{\rm in}\big>}\; .
\end{eqnarray}
The factor
\begin{equation}
      \sum_{\colore\delta}
	  {\colore \big|\delta_{\rm in}\big>\big<\delta_{\rm in}\big|}
  U(0,-\infty)
  {\colorc\big|\alpha_{\rm in}\big>} 
\end{equation}
is the {\colora Fock expansion} of the initial state: it accounts for
the fact that the state $\alpha$ prepared at $x^+=-\infty$ may have
fluctuated into another state $\delta$ before it interacts with the
external potential. The matrix elements of $U_0$ that appear in this
expansion can be calculated perturbatively to any desired
order. There is a similar factor for the final state evolution.

The interactions with the target field are contained in the central
factor, ${\colore\big<\gamma_{\rm in}\big|} \exp ...
{\colore\big|\delta_{\rm in}\big>}$. In order to rewrite it into a
more intuitive form, let us first rewrite the operator $\rho$ as
\begin{eqnarray}
      &&
      {\colora\rho^a(\vec\x_\perp)}={\colorc t^a_{ij}}
      \smash{\int \frac{dp^+}{4\pi p^+}\frac{d^2\vec\p_\perp}{(2\pi)^2}
      \frac{d^2\vec\q_\perp}{(2\pi)^2}}
      \Big\{
      {\colorb b^\dagger_{\rm in}(p^+,\vec\p_\perp;i)
	b_{\rm in}(p^+,\vec\q_\perp;j)}
      e^{i(\vec\p_\perp-\vec\q_\perp)\cdot\vec\x_\perp}
      \nonumber\\
      &&\qquad\qquad\qquad\qquad\qquad\qquad
      -
      {\colorb d^\dagger_{\rm in}(p^+,\vec\p_\perp;i)
	d_{\rm in}(p^+,\vec\q_\perp;j)}
      e^{-i(\vec\p_\perp-\vec\q_\perp)\cdot\vec\x_\perp}
      \Big\}\; ,
\end{eqnarray}
where the $t^a_{ij}$ are the generators of the fundamental
representation of the $SU(N)$ algebra and the $b,d,b^\dagger,d^\dagger$
the annihilation and creation operators for quarks and
antiquarks. $\rho^a$ also receives a contribution from gluons, not
written here, obtained with the generators in the adjoint
representation and the annihilation and creation operators for gluons
instead. This formula captures the essence of eikonal scattering:
\begin{itemize}
\item Each annihilation operator has a matching creation operator --
  therefore, the number of partons in the state does not change during
  the scattering, nor their flavor.
\item The $p^+$ component of the momenta are not affected by the scattering.
\item Only the colors and transverse momenta of the partons can change
  during the scattering.
\end{itemize}
Scattering amplitudes in the eikonal limit take a very simple form if
one trades transverse momentum for a transverse position, by doing a
Fourier transform. For each intermediate state ${\colore\big<\delta_{\rm
      in}\big|}\equiv {\colorb\big<\{k^+_i,\vec\k_{i\perp}\}\big|}$,
  define the corresponding {\sl light-cone wave function} by~:
\begin{equation}
{\colord\Psi_{\delta\alpha}(\{k^+_i,\vec\x_{i\perp}\})}\equiv 
\prod_{i\in\delta}\int \frac{d^2\vec\k_{i\perp}}{(2\pi)^2}
e^{-i\vec\k_{i\perp}\cdot\vec\x_{i\perp}}
{\colore\big<\delta_{\rm in}\big|}U(0,-\infty)
{\colore\big|\alpha_{\rm in}\big>}\; ,
\end{equation}
where the index $i$ runs over all the partons of the state
$\delta$. Then, each charged particle going through the external field
acquires a {\colora phase proportional to its charge} (antiparticles
get an opposite phase),
\begin{eqnarray}
&&
{\colord\Psi_{\delta\alpha}(\{k^+_i,\vec\x_{i\perp}\})}
\;
\longrightarrow
\;
{\colord\Psi_{\delta\alpha}(\{k^+_i,\vec\x_{i\perp}\})}
\prod_i {\colorb U_i(\vec\x_\perp)}
\nonumber\\
&&{\colorb U_i(\vec\x_\perp)}\equiv T\exp \Big[
i g\int dx^+\;
 {\colord{\cal A}^-_a(x^+,0,\vec\x_{i\perp})}{\colorc t^a}
\Big]
\; .
\end{eqnarray}
($U$ is replaced by $U^\dagger$ for antiquarks, and $t^a$ by an
adjoint generator for gluons.) The factors $U$ in this formula are
called {\sl Wilson lines}.

\subsection{Dipole scattering at Leading and Next to Leading Order}
Let us now assume that the initial and final states ${\colore \alpha}$
and ${\colore \beta}$ both contain a {\colora virtual photon}, as would be the
case in DIS at leading order. The simplest Fock state that contributes
to its wave function is a
$q\overline{q}$ pair in a color singlet state, and the bare scattering amplitude can be written as
\setbox1\hbox to
2cm{\hfil\resizebox*{2cm}{!}{\includegraphics{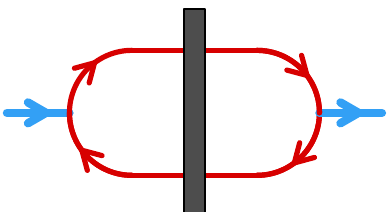}}\hfil}
\begin{eqnarray}
\raise -4mm\box1&\propto& 
{\colore\Psi^{(0)*}_{ij}(\vec\x_\perp,\vec\y_\perp)}
{\colore\Psi^{(0)}_{kl}(\vec\x_\perp,\vec\y_\perp)}
{\colorb U_{ik}(\vec\x_\perp)
U^\dagger_{lj}(\vec\y_\perp)}
\nonumber\\
&\propto&
{\colore\left|\Psi^{(0)}(\vec\x_\perp,\vec\y_\perp)\right|^2}
{\rm tr}\left[{\colorb U(\vec\x_\perp)
U^\dagger(\vec\y_\perp)}\right]
\; .
\end{eqnarray}
The color trace arises because the wavefunction of the photon is
diagonal in color, $\Psi_{ij}^{(0)}\sim \delta_{ij}$.  In the diagram
in the left hand side, the gray band represents the Lorentz contracted
target field.

It turns out that one-loop corrections due to the emission of a gluon
inside the dipole are enhanced by logarithms of the dipole
longitudinal momentum $p^+$, leading to possibly large corrections
proportional to $\alpha_s\log(p^+)$. In order to compute the dipole
amplitude at next to leading order, we need to evaluate the following
graphs \setbox1\hbox to
8cm{\hfil\resizebox*{8cm}{!}{\includegraphics{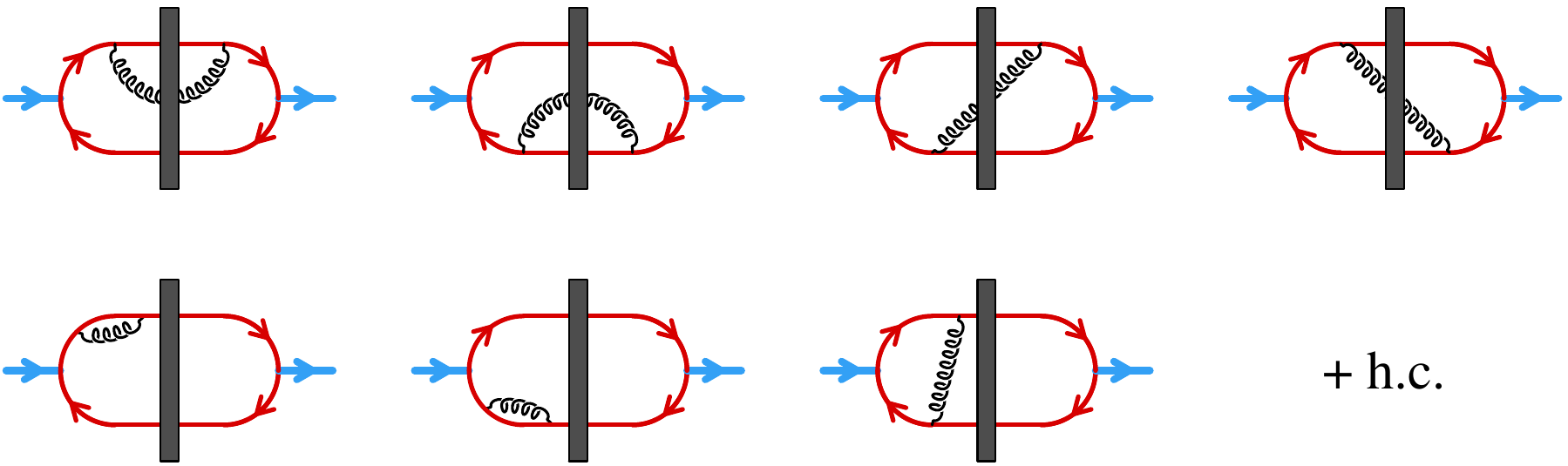}}\hfil}
\begin{equation*}
\box1
\end{equation*}
We will call {\sl real} the terms where the gluon traverses the target
field, and {\sl virtual} those where the gluon is a correction inside
the wavefunction of the incoming or outgoing dipole.  In the
light-cone gauge $A^+=0$, the emission of a gluon of momentum $k$ and
polarization $\lambda$ by a quark can be written as \setbox1\hbox to
15mm{\hfil\resizebox*{15mm}{!}{\includegraphics{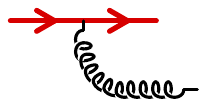}}\hfil}
\begin{equation}
\raise -3mm\box1\quad= \;2{\colorc g}t^a \; 
\frac{\vec{\bs \epsilon}_\lambda\cdot\vec\k_\perp}{k_\perp^2}\; ,
\end{equation}
where it is assumed that the gluon is soft compared to the quark and
where ${\bs\epsilon}_\lambda$ is the polarization vector of the
gluon. Trading transverse momentum in favor of transverse position,
this vertex reads
\begin{equation}
\int \frac{d^2\vec\k_\perp}{(2\pi)^2} \; e^{i\vec\k_\perp\cdot(\vec\x_\perp-\vec\z_\perp)} \; 2{\colorc g}t^a \; 
\frac{\vec{\bs \epsilon}_\lambda\cdot\vec\k_\perp}{k_\perp^2}
=
\frac{2i{\colorc g}}{2\pi}
t^a \frac{\vec{\bs \epsilon}_\lambda\cdot(\vec\x_\perp-\vec\z_\perp)}{(\vec\x_\perp-\vec\z_\perp)^2}\; .
\end{equation}
The other rule we need to compute these graphs is that when connecting
the gluons to form the loop, one must sum over their polarizations,
leading to a factor
\begin{equation}
\sum_{\lambda}\vec{\bs\epsilon}_\lambda^i \vec{\bs\epsilon}_\lambda^j = 
{\colorb \delta^{ij}}\; .
\end{equation}

Let us start with the virtual contributions. For instance, one gets
\setbox1\hbox to 2cm{
\hfil
\resizebox*{2cm}{!}{\includegraphics{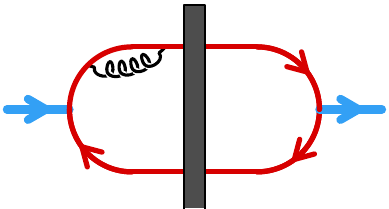}}
\hfil
}
\begin{eqnarray}
\raise -4mm\box1 &=& {\colore
\left|\Psi^{(0)}(\vec\x_\perp,\vec\y_\perp)\right|^2}
{\rm tr}\left[{\colord t^a t^a}
{\colorb U(\vec\x_\perp)U^\dagger(\vec\y_\perp)}\right]
\nonumber\\
&&
\times
-2{\colorc\alpha_s}{\colora\int \frac{dk^+}{k^+}}
\int\frac{d^2\vec\z_\perp}{(2\pi)^2}
\frac{(\vec\x_\perp-\vec\z_\perp)\cdot(\vec\x_\perp-\vec\z_\perp)}
{(\vec\x_\perp-\vec\z_\perp)^2 (\vec\x_\perp-\vec\z_\perp)^2}\; ,
\end{eqnarray}
and
\setbox1\hbox to 2cm{
\hfil
\resizebox*{2cm}{!}{\includegraphics{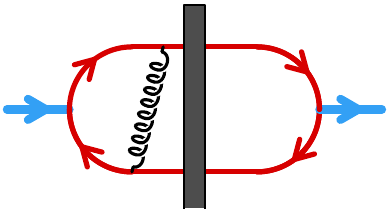}}
\hfil
}
\begin{eqnarray}
\raise -4mm\box1 &=& {\colore
\left|\Psi^{(0)}(\vec\x_\perp,\vec\y_\perp)\right|^2}
{\rm tr}\left[{\colord t^a} 
{\colorb U(\vec\x_\perp)U^\dagger(\vec\y_\perp)}{\colord t^a}\right]
\nonumber\\
&&\times
4{\colorc\alpha_s}
{\colora\int\frac{dk^+}{k^+}}\int\frac{d^2\vec\z_\perp}{(2\pi)^2} 
\frac{(\vec\x_\perp-\vec\z_\perp)\cdot(\vec\y_\perp-\vec\z_\perp)}
{(\vec\x_\perp-\vec\z_\perp)^2 (\vec\y_\perp-\vec\z_\perp)^2}\; .
\end{eqnarray}
Similar expressions can be obtained for the other virtual terms,
and the sum of all virtual corrections is
\begin{eqnarray}
&&
-\frac{{\colord C_{_F}}{\colorc\alpha_s}}{\pi^2}
{\colora\int\frac{dk^+}{k^+}}\int d^2\vec\z_\perp\;
\frac{(\vec\x_\perp-\vec\y_\perp)^2}
{(\vec\x_\perp-\vec\z_\perp)^2 (\vec\y_\perp-\vec\z_\perp)^2}
\nonumber\\
&&
\qquad\qquad\times\;{\colore
\left|\Psi^{(0)}(\vec\x_\perp,\vec\y_\perp)\right|^2}
{\rm tr}\left[
{\colorb U(\vec\x_\perp)U^\dagger(\vec\y_\perp)}\right]
\; ,
\end{eqnarray}
where we have introduced the Casimir of the fundamental representation
of $SU(N)$, ${\colord C_{_F}\equiv t^a t^a = (N^2-1)/2N}$. This
(incomplete, since the real terms are still missing) result exhibits
two pathologies~:
\begin{itemize}
\item The integral over ${\colora k^+}$ is divergent. It should have
  an upper bound at $p^+$, the longitudinal momentum of the quark or
  antiquark,
  \begin{equation}
    {\colora\int^{p^+}\frac{dk^+}{k^+}}=\log(p^+)\equiv Y
  \end{equation}
  When the rapidity $Y$ is large, ${\colorc \alpha_s} Y$ may not be
  small.  By differentiating with respect to $Y$, we will obtain an
  evolution equation in $Y$ whose solution resums all the powers
  $({\colorc \alpha_s} Y)^n$.
\item The integral over $\vec\z_\perp$ is divergent near
  $\vec\z_\perp=\vec\x_\perp$ or $\vec\y_\perp$. This is a collinear
  singularity, due to emission of the gluon at zero angle with respect
  to the quark or antiquark. We will see shortly that it is cancelled
  when we combine the virtual and real contributions.
\end{itemize}

Let us now turn to the real terms. For instance, one has
\setbox1\hbox to 2cm{
\hfil
\resizebox*{2cm}{!}{\includegraphics{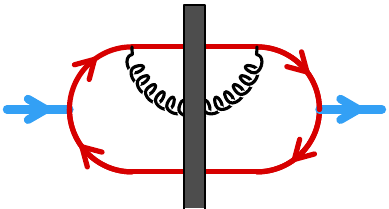}}
\hfil
}
\begin{eqnarray}
\raise -4mm\box1 &=& {\colore
\left|\Psi^{(0)}(\vec\x_\perp,\vec\y_\perp)\right|^2}
{\rm tr}\left[{\colord t^a}
{\colorb U(\vec\x_\perp){\colord t^b}U^\dagger(\vec\y_\perp)}\right]
\nonumber\\
&&
\times
4{\colorc\alpha_s}{\colora\int \frac{dk^+}{k^+}}
\int\frac{d^2\vec\z_\perp}{(2\pi)^2}{\colorb {\wt U}_{ab}(\vec\z_\perp)}
\frac{(\vec\x_\perp-\vec\z_\perp)\cdot(\vec\x_\perp-\vec\z_\perp)}
{(\vec\x_\perp-\vec\z_\perp)^2 (\vec\x_\perp-\vec\z_\perp)^2}\; ,
\end{eqnarray}
where $\wt{U}_{ab}$ is a Wilson line in the adjoint representation,
describing the eikonal scattering of the gluon off the target field.
In order to simplify the color structure, first
recall that
\begin{equation}
{\colord t^a}{\colorb {\wt U}_{ab}(\vec\z_\perp)}
=
{\colorb U(\vec\z_\perp)}{\colord t^b}{\colorb U^\dagger(\vec\z_\perp)}\; ,
\end{equation}
and then use the $SU(N)$ {\colora Fierz identity},
\begin{equation}
{\colord t^b_{ij} t^b_{kl}}={\colorc\frac{1}{2}} \delta_{il} \delta_{jk}
-{\colorc \frac{1}{2N_c} }\delta_{ij}\delta{kl}\; .
\end{equation}
Thanks to these identities, the Wilson lines can be rearranged into
\begin{eqnarray}
&&
{\rm tr}\left[{\colord t^a}
{\colorb U(\vec\x_\perp){\colord t^b}U^\dagger(\vec\y_\perp)}\right]
{\colorb {\wt U}_{ab}(\vec\z_\perp)}
=
{\colorc\frac{1}{2}}
{\rm tr}\left[{\colorb U^\dagger(\vec\z_\perp)U(\vec\x_\perp)}\right]
{\rm tr}\left[{\colorb U(\vec\z_\perp)U^\dagger(\vec\y_\perp)}\right]
\nonumber\\
&&
\qquad\qquad\qquad\qquad\qquad\qquad\qquad
-
{\colorc\frac{1}{2N_c}}
{\rm tr}\left[{\colorb U(\vec\x_\perp)U^\dagger(\vec\y_\perp)}\right]
\; .
\end{eqnarray}
Interestingly, the subleading term in the number of colors (i.e. in
${\colorc 1/2N_c}$) cancels exactly against a similar term in the
virtual contribution. Moreover, when we sum all the real terms, we
generate the same kernel as in the virtual terms,
\begin{equation}
\frac{(\vec\x_\perp-\vec\y_\perp)^2}
{(\vec\x_\perp-\vec\z_\perp)^2 (\vec\y_\perp-\vec\z_\perp)^2}\; .
\end{equation}

In order to make the following equations more compact, let us
introduce
\begin{equation}
{\colorb{\bs S}(\vec\x_\perp,\vec\y_\perp)}
\equiv
{\colorc \frac{1}{N_c}}
{\rm tr}\left[{\colorb U(\vec\x_\perp)U^\dagger(\vec\y_\perp)}\right]\; .
\end{equation}
In terms of this object, the sum of all the NLO contributions is
\begin{eqnarray}
&&
-\frac{{\colorc \alpha_s N_c^2}{\colora Y}}{2\pi^2}
{\colore\left|\Psi^{(0)}(\vec\x_\perp,\vec\y_\perp)\right|^2}
\int d^2\vec\z_\perp\;
\frac{(\vec\x_\perp-\vec\y_\perp)^2}
{(\vec\x_\perp-\vec\z_\perp)^2 (\vec\y_\perp-\vec\z_\perp)^2}
\nonumber\\
&&\qquad\qquad\qquad\qquad\qquad\times
\Big\{
{\colorb{\bs S}(\vec\x_\perp,\vec\y_\perp)}
-
{\colorb{\bs S}(\vec\x_\perp,\vec\z_\perp)}
{\colorb{\bs S}(\vec\z_\perp,\vec\y_\perp)}
\Big\}
\; ,
\end{eqnarray}
while the bare scattering amplitude was
$
{\colore\left|\Psi^{(0)}(\vec\x_\perp,\vec\y_\perp)\right|^2}
{\colorc N_c}\;
{\colorb{\bs S}(\vec\x_\perp,\vec\y_\perp)}\; .
$
By comparing these two formulas, we conclude that
\begin{eqnarray}
&&
\frac{\partial {\colorb{\bs S}(\vec\x_\perp,\vec\y_\perp)}}{\partial{\colora Y}}
=
-\frac{{\colorc \alpha_s N_c}}{2\pi^2}
\int d^2\vec\z_\perp\;
\frac{(\vec\x_\perp-\vec\y_\perp)^2}
{(\vec\x_\perp-\vec\z_\perp)^2 (\vec\y_\perp-\vec\z_\perp)^2}
\nonumber\\
&&
\qquad\qquad\qquad\qquad\times
\Big\{
{\colorb{\bs S}(\vec\x_\perp,\vec\y_\perp)}
-
{\colorb{\bs S}(\vec\x_\perp,\vec\z_\perp)}
{\colorb{\bs S}(\vec\z_\perp,\vec\y_\perp)}
\Big\}
\; .
\label{eq:BFKL+}
\end{eqnarray}
Note that, since ${\colorb {\bs S}(\vec\x_\perp,\vec\x_\perp)=1}$, the
integral over $\vec\z_\perp$ is now regular.

\subsection{BFKL equation}
In the derivation of eq.~(\ref{eq:BFKL+}), we have not made any
assumption about whether the amplitude $\bs{S}$ is large or small.
Let us now focus on the dilute regime, which corresponds to small
scattering amplitudes ${\colorb {\bs T}(\vec\x_\perp,\vec\y_\perp)}
\equiv 1-{\colorb{\bs S}(\vec\x_\perp,\vec\y_\perp)}$. By rewriting
eq.~(\ref{eq:BFKL+}) in terms of $\bs{T}$ and by keeping only the
linear terms, we obtain the {\sl Balitsky-Fadin-Kuraev-Lipatov
equation} (BFKL) \cite{KuraeLF1,BalitL1},
\begin{eqnarray}
      &&
      \frac{\partial\,{\colorb{\bs T}(\vec\x_\perp,\vec\y_\perp)}}
      {\partial{\colora Y}}
      =
      \frac{{\colorc \alpha_s N_c}}{2\pi^2}
      \int d^2\vec\z_\perp\;
      \frac{(\vec\x_\perp-\vec\y_\perp)^2}
      {(\vec\x_\perp-\vec\z_\perp)^2 (\vec\y_\perp-\vec\z_\perp)^2}
      \nonumber\\
      &&
      \qquad\qquad\qquad\qquad\qquad\times
      \Big\{
      {\colorb{\bs T}(\vec\x_\perp,\vec\z_\perp)}
      +
      {\colorb{\bs T}(\vec\z_\perp,\vec\y_\perp)}
      -
      {\colorb{\bs T}(\vec\x_\perp,\vec\y_\perp)}
      \Big\}\; .
\end{eqnarray}
The BFKL equation has a fixed point, ${\bs T}\equiv 0$, but it turns out
to be unstable. Moreover, one can see that it leads to violations
of unitarity~:
\begin{itemize}
\item The mapping ${\colorb\bs T}\to {\colorc\alpha_s
    N_c}\int_{\z}\cdots{\colorb\bs T}$ has a positive eigenvalue
  $\omega$.
\item Therefore, generic solutions of the BFKL equation grow
  exponentially without bound as $\exp(\omega {\colora Y})$ when
  ${\colora Y \to +\infty}$.  This leads to a violation of unitarity,
  since ${\bs T}$-matrix elements should be bounded.
\end{itemize}
This growth of the scattering amplitude with rapidity can be related
to the increase at small $x$ of the gluon distribution. Indeed, if one
is still in the dilute regime, the forward scattering amplitude
between a small dipole and a target made of gluons is proportional to
\begin{equation}
      {\colorb {\bs T}(\vec\x_\perp,\vec\y_\perp)}
      \propto {\colorb |\vec\x_\perp-\vec\y_\perp|^2}
      \;\; {\colora x}G({\colora x},{\colorb|\vec\x_\perp-\vec\y_\perp|^{-2}})
\end{equation}
where ${\colora Y}\equiv \ln({\colora 1/x})$. Therefore, the
exponential behavior of ${\colorb{\bs T}}$ translates into
\begin{equation}
{\colora x}G({\colora x},Q^2)\sim \frac{1}{{\colora x}^\omega}\; .
\end{equation}

\subsection{Gluon saturation and BK equation}
Interestingly, the original equation (\ref{eq:BFKL+}), did not have
this unitarity problem. When written in terms of ${\bs T}$, it reads
\begin{eqnarray}
      &&
      \frac{\partial\,{\colorb{\bs T}(\vec\x_\perp,\vec\y_\perp)}}
      {\partial{\colora Y}}
      =
      \frac{{\colorc \alpha_s N_c}}{2\pi^2}
      \int d^2\vec\z_\perp\;
      \frac{(\vec\x_\perp-\vec\y_\perp)^2}
      {(\vec\x_\perp-\vec\z_\perp)^2 (\vec\y_\perp-\vec\z_\perp)^2}
      \nonumber\\
      &&\!\!\!\!
      \times
      \Big\{
      {\colorb{\bs T}(\vec\x_\perp,\vec\z_\perp)}
      +
      {\colorb{\bs T}(\vec\z_\perp,\vec\y_\perp)}
      -
      {\colorb{\bs T}(\vec\x_\perp,\vec\y_\perp)}
	- {\colord{\bs T}(\vec\x_\perp,\vec\z_\perp)}
	{\colord{\bs T}(\vec\z_\perp,\vec\y_\perp)}
      \Big\}\; ,
\label{eq:BK2}
\end{eqnarray}
an equation known as the {\sl Balitsky-Kovchegov equation} (BK)
\cite{Kovch5,Kovch3}.  One can see that it has two fixed points, ${\bs
  T}\equiv 0$ and ${\bs T}\equiv 1$. The appearance of the second
fixed point at ${\bs T}=1$ is due to the nonlinear term in ${\bs T}$,
that we had neglected in deriving the BFKL equation. Moreover, this
new fixed point turns out to be stable. Therefore, the typical
behavior of solutions of the equations it that ${\bs T}$ starts at a
small value, grows exponentially with $Y$, and then saturates at ${\bs
  T}=1$.

\begin{figure}[htbp]
\begin{center}
\resizebox*{7cm}{!}{\includegraphics{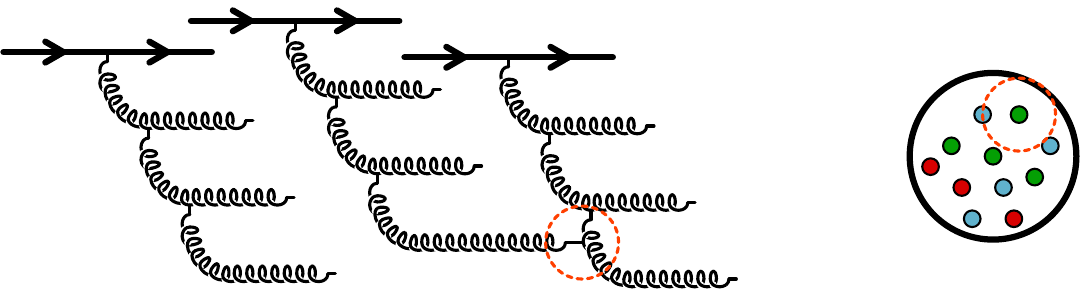}}\hfil
\resizebox*{5cm}{!}{\includegraphics{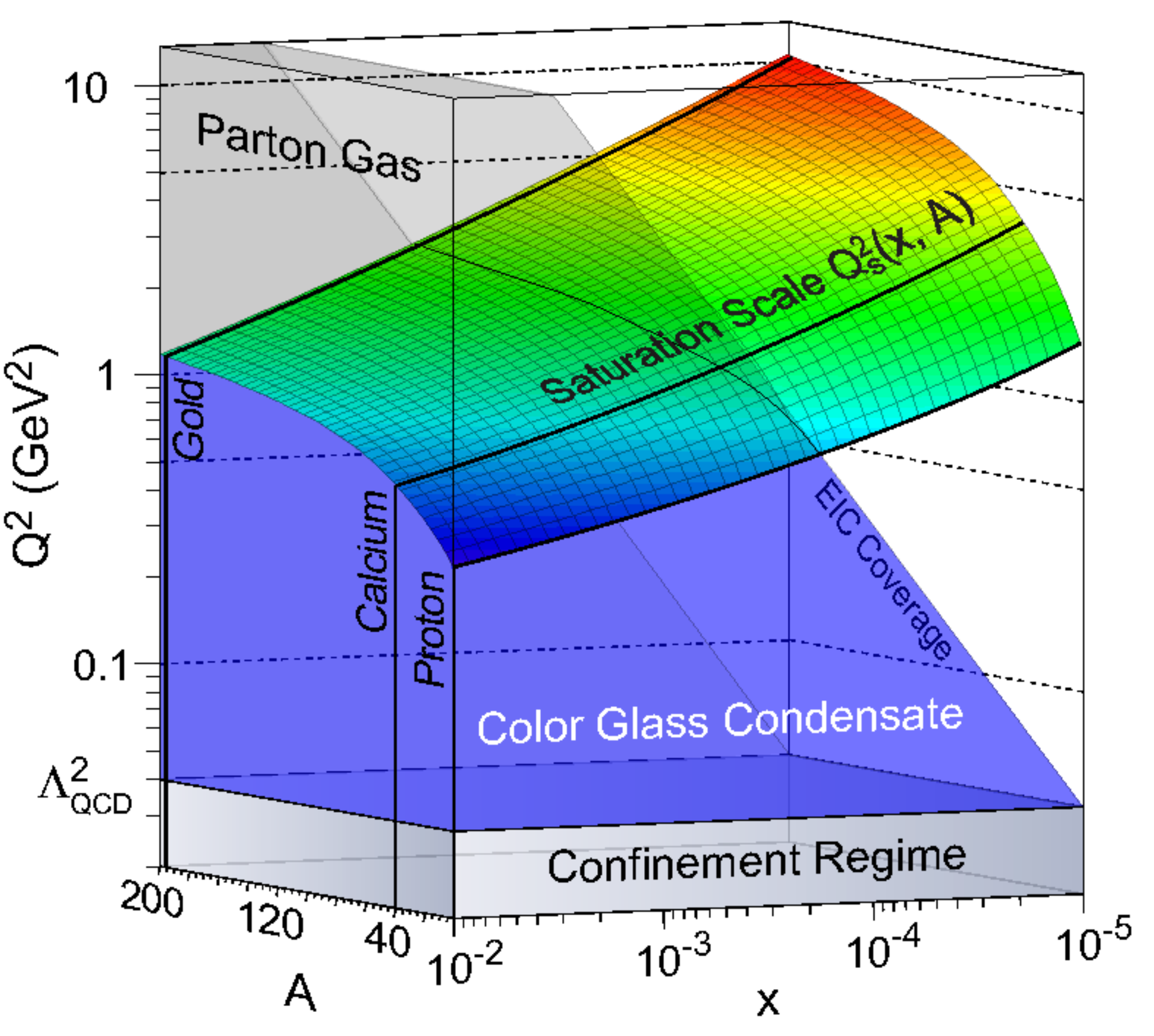}}
\end{center}
\caption{\label{fig:sat}Left: gluon cascades from a valence quark in a
  hadron. Right: saturation domain as a function of the value of $x$
  and the atomic number $A$ of the nucleus\cite{DeshpEM1}.}
\end{figure}

The physical interpretation of what makes the solutions of the
linearized BFKL equation grow exponentially, and why the solutions are
bounded when one keeps the nonlinear term in ${\bs T}$ is rather
transparent from the point of view of the gluon distribution in the
target. This is illustrated in the left part of the figure
\ref{fig:sat}.
At small $Y$, i.e. in the dilute regime where ${\bs T}$ is small, one
probes the valence quarks only. At larger values of $Y$, the dipole
can interact not only with the valence quarks, but also with gluons
radiated from the valence quarks. At even larger values of $Y$, these
gluons themselves emit more gluons, in ladder-like diagrams. In this
regime, the growth of the number of gluons is exponential in the
rapidity $Y$. This very fast growth continues until the gluon density
becomes too large. Roughly speaking, when the areal gluon density in
the transverse plane multiplied by the cross-section for the
recombination of two gluons by the process $gg\to g$ becomes of order
one, then the nonlinearities become important. Their effect is to tame
the growth of the gluon density, so that the scattering amplitude
${\bs T}$ remains bounded. This phenomenon is known as {\sl gluon
  saturation}.

The criterion for the onset of gluon saturation can be expressed in
terms of the resolution scale $Q$ at which the gluons are observed. It
reads
\begin{equation}
        \underbrace{\alpha_s Q^{-2}}_{\sigma_{gg\to g}} 
        \;\times\; 
        \underbrace{A^{-2/3} xG(x,Q^2)}_{\mbox{\scriptsize surface density}} 
        \;\ge\; 1\; ,
\end{equation}
where $A$ is the atomic number in the case the target is a nucleus
(otherwise it is unity). This inequality can be rearranged into
\begin{equation}
        Q^2 \le 
        \underbrace{{\colorb Q_s^2}\equiv 
        \frac{{\colord\alpha_s}{\colorb xG(x,Q_s^2)}}{{\colora A^{2/3}}}}_{\mbox{\scriptsize saturation momentum}}
        \;\sim\; {\colora A^{1/3}}{\colorb x^{-0.3}}\; .
\end{equation}
The right hand side of this inequality is called the {\sl saturation
  momentum}. The effects of gluon saturation are important below this
scale, and are therefore more visible when this scale is large. It is
easy to estimate the $A$ and $x$ dependence of the saturation
momentum. Because $Q_s^2$ is the ratio of a quantity proportional to
the volume of the target by an area, it should scale like
$A^{1/3}$. Its $x$ dependence can be inferred from the behavior of the
gluon distribution in nucleons at small $x$: it grows like a power of
$1/x$, the numerical value $\sim 0.3$ of the exponent being extracted
from HERA data for instance. The value of the saturation momentum as a
function of $x$ and $A$ is shown in the right plot of the figure
\ref{fig:sat}, and the dense/saturated regime comprises the volume
under the surface.

Note that the eq.~(\ref{eq:BK2}) derived here resums only the leading
log terms, since it is based only on the kernel at one loop. At this
order, there is no running of the coupling constant. This effect
arises via some parts of the two-loop kernel, that involve the beta
function. These corrections are known by
now\cite{Balit3,BalitC1,KovchW1,GardiKRW1} and play an important role
in reaching a good agreement with experimental
data\cite{AlbacAMSW1,AlbacK1,AlbacAMS1,AlbacAMS2}.

\subsection{Target average}
Until now, we have treated the target as a given patch of color field,
produced by the constituents of the target, that are static over the
timescale of the collision due to Lorentz time dilation. However, this
picture implies that details of this color field depend of the precise
configuration (position, color, ...) of the constituents of the target
at the time of the collisions, that of course we do not know. Since
this configuration changes from collision to collision, so does the
target field. Therefore, instead of a single target field, one should
instead have in mind an ensemble of such fields, corresponding to all
the possible configurations of the constituents of the target. This
means that all the quantities we have so far evaluated in a given
target field should in fact be averaged over this ensemble of target
fields,
\begin{equation}
{\colorb{\bs T}}\quad \to\quad\left<{\colorb{\bs T}}\right>\; .
\end{equation}
When performing this average at the level of the BK equation, we
naturally get
\begin{eqnarray}
      &&
      \frac{\partial\,\left<{\colorb{\bs T}(\vec\x_\perp,\vec\y_\perp)}\right>}
      {\partial{\colora Y}}
      =
      \frac{{\colorc \alpha_s N_c}}{2\pi^2}
      \int d^2\vec\z_\perp\;
      \frac{(\vec\x_\perp-\vec\y_\perp)^2}
      {(\vec\x_\perp-\vec\z_\perp)^2 (\vec\y_\perp-\vec\z_\perp)^2}
      \nonumber\\
      &&
      \times
      \Big\{\!\!   
      \left<{\colorb{\bs T}(\vec\x_\perp,\vec\z_\perp)}\right>
      +
      \left<{\colorb{\bs T}(\vec\z_\perp,\vec\y_\perp)}\right>
      -
      \left<{\colorb{\bs T}(\vec\x_\perp,\vec\y_\perp)}\right>
	- \left<{\colord{\bs T}(\vec\x_\perp,\vec\z_\perp)}
	{\colord{\bs T}(\vec\z_\perp,\vec\y_\perp)}\right>\!\! 
      \Big\}\, .
\end{eqnarray}
The trouble with this equation is that it is no longer a closed
equation, since the evolution with $Y$ of the quantity
$\left<{\colorb{\bs T}}\right>$ depends on the value of a new quantity
$\left<{\colord{\bs T}} {\colord{\bs T}}\right>$. By the same method
used to derive the evolution equation for $\left<{\colorb{\bs
      T}}\right>$, we could derive an evolution equation for
$\left<{\colord{\bs T}} {\colord{\bs T}}\right>$, but its right hand
side would contain the average value of an object containing three
factors ${\bs T}$, and so on.  Therefore, instead of a single closed
equation, we have now an infinite hierarchy of nested equations, known
as the {\sl Balitsky hierarchy}\cite{Balit1}.

There is an approximation of the above hierarchy of equations, that
leads us back to the original Balitsky-Kovchegov equation. This
approximation amounts to assuming that the average of a product of two
dipole operators factorizes into the product of their averages,
\begin{equation}
\left<{\colorb{\bs T}\,{\bs T}}\right>\approx 
\left<{\colorb{\bs T}}\right>\,
\left<{\colorb{\bs T}}\right>\; .
\end{equation}
Obviously, if this is true, then the first equation of the hierarchy
is nothing but the BK equation. This mean-field approximation is
believed to be valid for large targets such as nuclei, and in the
limit of a large number of color. Although subject to this
approximation, the BK equation is widely used in phenomenological
applications because of its simplicity.

\subsection{Geometrical scaling}
The BK equation leads to an interesting property of the DIS total
cross-section, called {\sl geometrical scaling}. Firstly, let us write
the $\gamma^*$-target cross-section as follows,
\begin{equation}
\sigma_{\gamma^*T}=\int_0^1 dz\int d^2\vec\r_\perp
\left|\psi(\q|z,\vec\r_\perp)\right|^2
\sigma_{\rm dipole}(\vec\r_\perp)\; ,
\end{equation}
with
\begin{equation}
\sigma_{\rm dipole}(\vec\r_\perp)\equiv {2}
\int d^2\vec\X_\perp\;{\bs T}(\vec\X_\perp+\frac{\vec\r_\perp}{2},\vec\X_\perp-\frac{\vec\r_\perp}{2})\; .
\label{eq:dipXsec}
\end{equation}
$\psi(\q|z,\vec\r_\perp)$ is the light-cone wavefunction for a virtual
photon of momentum $\q$ splitting into a $q\overline{q}$ pair or
transverse size $\r_\perp$, where the quarks carries the fraction $z$
of the longitudinal momentum of the photon. This object is a purely
electromagnetic quantity (at least at leading order), and is not
interesting from the point of view of strong interactions, that are
all contained in the dipole cross-section defined in
eq.~(\ref{eq:dipXsec}). This equation defines the total cross-section
between the $q\overline{q}$ pair and the target, expressed in terms of
the forward scattering amplitude thanks to the optical theorem. Thus,
the BK equation tells us about the rapidity dependence (i.e. the $x$
dependence since $Y\equiv \log(1/x)$) of the DIS cross-section.

For the sake of the argument, let us assume translation and rotation
invariance, and define
\begin{equation}
{\colord N(k_\perp)}\equiv
2\pi\int d^2\vec\x_\perp
\;e^{i\vec\k_\perp\cdot\vec\x_\perp}\;
\frac{\left<{\colorb {\bs T}(0,\vec\x_\perp)}\right>}{x_\perp^2}\; .
\end{equation}
From the Balitsky-Kovchegov equation for $\left<{\colorb {\bs
T}}\right>$, we obtain the following equation for ${\colord
N}$
\begin{equation}
      \frac{\partial {\colord N(k_\perp)}}
	   {\partial{\colora Y}}
	   =\frac{{\colorc
	       \alpha_s N_c}}{\pi} \Big[ {\colorc \chi(-\partial_L)}
	     {\colord N(k_\perp)}-{\colord N^2(k_\perp)} \Big]\; ,
\label{eq:BK1}
\end{equation}
where we denote 
\begin{eqnarray}
      &&{\colorb L}\equiv {\colorb\ln(k^2/k_0^2)}\nonumber\\
      &&{\colorc
      \chi(\gamma)}\equiv {\colorc
      2\psi(1)-\psi(\gamma)-\psi(1-\gamma)}\nonumber\\
    && \psi(z) \equiv \frac{d\ln\Gamma(z)}{dz}\; .
\end{eqnarray}
($\Gamma(z)$ is Euler's Gamma function.) The form (\ref{eq:BK1}) of
the BK equation is particularly simple, because its nonlinear term is
just the square of the function $N$.

The function ${\colorb \chi(\gamma)}$ has a minimum at
$\gamma=1/2$. Eq.~(\ref{eq:BK1}) can be approximated by expanding
${\colorb \chi(\gamma)}$ to quadratic order around the minimum. This
amounts to keeping only derivatives with respect to $L$ up to second
order, i.e. to a diffusion approximation. The physical content of this
equation becomes more transparent if one also introduces the following
variables,
\begin{eqnarray}
      &&{ t}\sim {\colora Y}\nonumber\\
      &&{z}\sim L +\frac{{\colorc\alpha_s N_c}}{2\pi}
      \chi^{\prime\prime}(1/2)\;{\colora Y}\; ,
\end{eqnarray}
in terms of which eq.~(\ref{eq:BK1}) becomes\cite{MunieP1,MunieP2,MunieP3}
\begin{equation}
{\partial_t}{\colord N} 
  ={\partial_z^2}{\colord N} +{\colord N}-{\colord N}^2\; .
\label{eq:FKPP}
\end{equation}
This equation is known as the {\colora
  Fisher-Kolmogorov-Petrov-Piscounov} (FKPP) equation, and it describes the
physics of {\sl reaction-diffusion processes}. These are processes
involving a certain entity $A$ that can do the following:
\begin{itemize}
\item An entity $A$ can hop from a location to neighboring
  locations. This is described by the diffusion term in the right hand
  side of eq.~(\ref{eq:FKPP}).
\item An entity $A$ can split into two, increasing the population by
  one unit. This is described by the $+N$ term.
\item Two of these entities can merge into a single one, reducing
  their population by one unit. This is described by the term $-N^2$.
\end{itemize}
The FKPP equation has two fixed points, $N=0$ which is unstable, and
$N=1$ which is a stable fixed point.  Note that the loss term, $-N^2$,
is essential for the existence of this stable fixed point. Moreover,
for rather generic initial conditions, the solutions of the {\colora
  FKPP equation} are known to behave like {\colorb traveling waves} at
asymptotic times
\begin{equation}
    {\colord N}(t,z)
    \empile{\sim}\over{t\to +\infty} 
	   {\colord N}(z-2t)\; .
\end{equation}
In other words, these solutions propagate in the $+z$ direction at the
constant velocity $dz/dt=2$, and instead of depending separately on
$z$ and $t$, they depend only on the combination $z-2t$. Going back to
the dipole scattering amplitude ${\bs T}$, we see that it does not
depend separately on $Y$ and on the dipole size $\r_\perp$, but on the
combination $Q_s(Y)\r_\perp$, where $Q_s(Y)$ has the following $Y$ dependence
\begin{equation}
{\colorb Q_s^2(Y)}
    =k_0^2 
    \;{\colorb Y}^{-\frac{3}{2(1-\bar\gamma)}}\;
    e^{{\overline{\alpha}_s\chi^{\prime\prime}(\frac{1}{2})(\frac{1}{2}-\bar\gamma)}{\colorb Y}}\; .
\end{equation}
This quantity is nothing but the saturation momentum introduced by a
simple semi-quantitative argument earlier. Here, we see it emerge
dynamically from the evolution equation itself. This is a major
feature of gluon saturation: the ability to generate a dimensionful
scale is a consequence of the nonlinear aspect of the saturation
phenomenon. Moreover, thanks to the fact that this scale {\sl
  increases} with $Y$, i.e. with the energy scale, one may hope that
gluon saturation falls in the realm of weakly coupled physics at
sufficiently high energy.

\begin{figure}[htbp]
\begin{center}
\resizebox*{5.5cm}{!}{\includegraphics{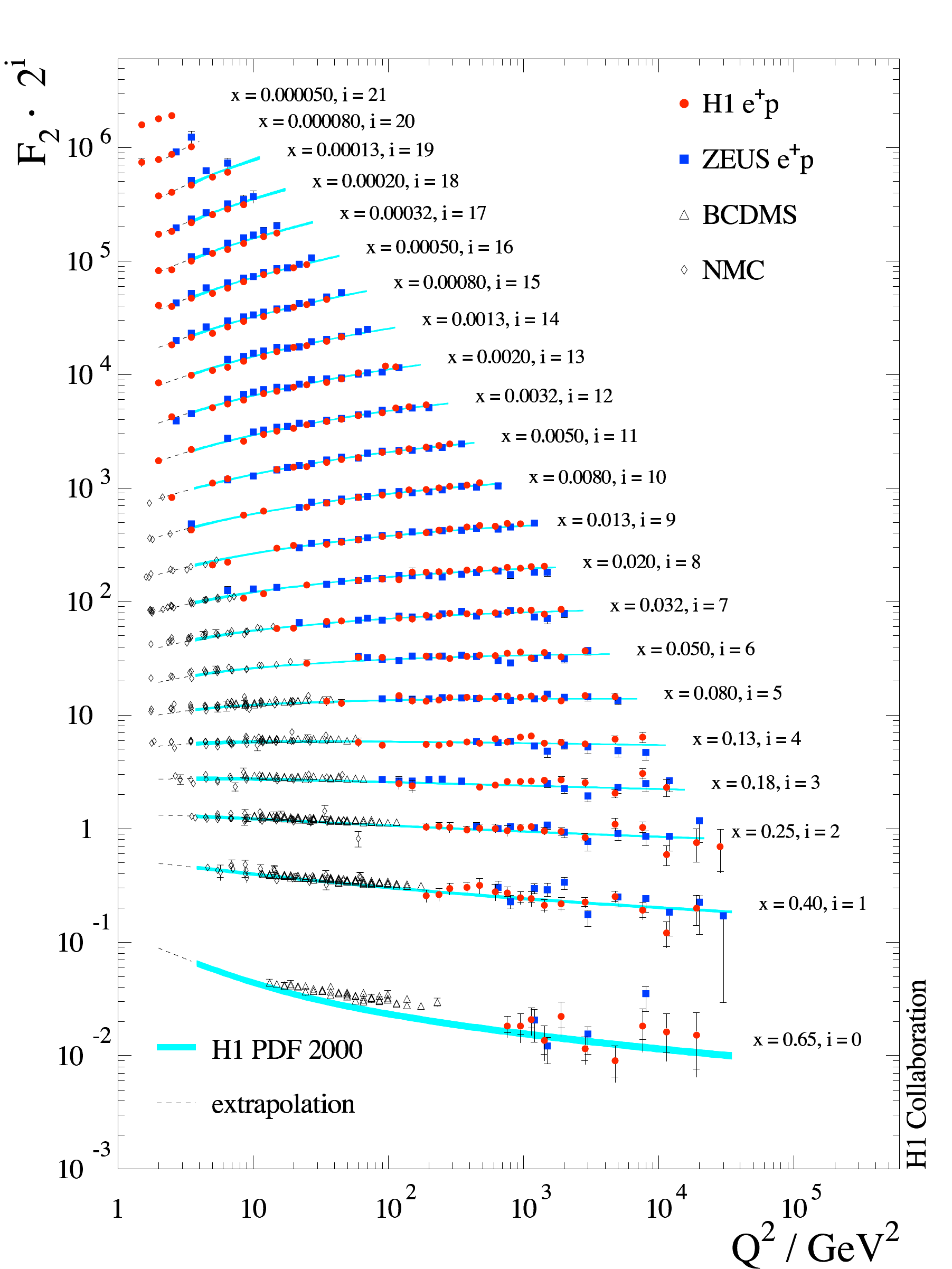}}\hfil
\resizebox*{7cm}{!}{\includegraphics{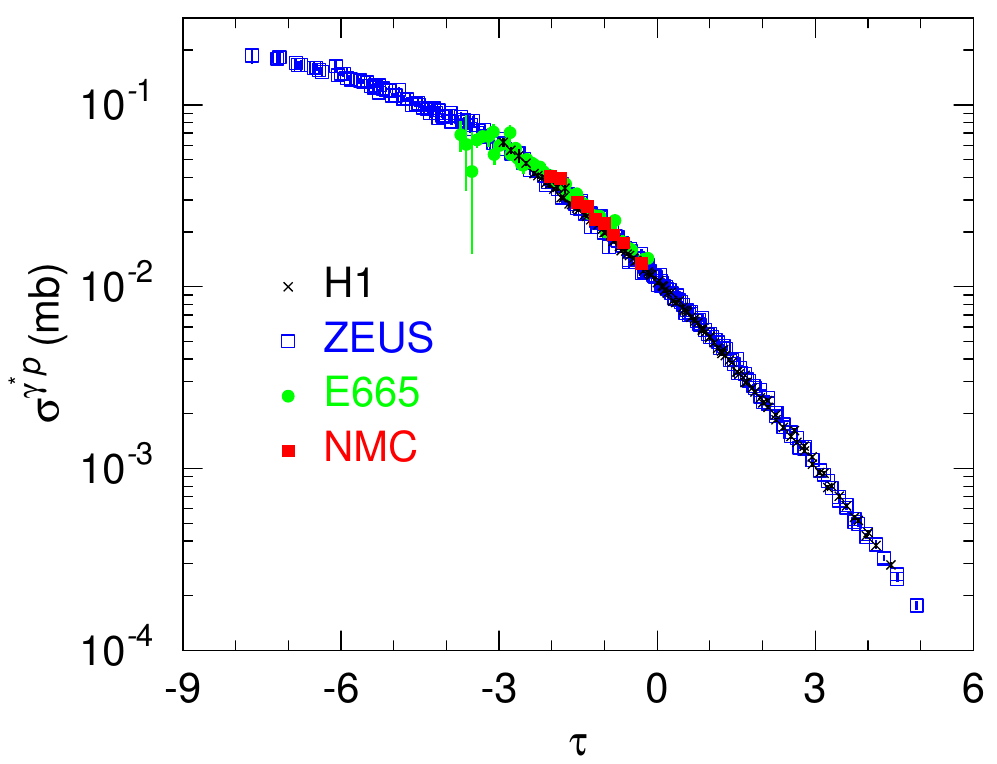}}
\end{center}
\caption{\label{fig:geom} DIS data at HERA. Left: displayed
  conventionally as a function of $x$ and $Q^2$. Right: displayed as a
  function of $\tau \equiv Q^2/Q_s^2(x)$ \cite{StastGK1,GelisPSS1}.}
\end{figure}

And at the level of the $\gamma^*$-target cross-section, this implies
that the cross-section depends only on $Q^2/Q_s^2(Y)$, rather than
$Q^2$ and $Y$ separately (this is strictly true assuming that we
neglect any other dimensionful parameters in the problem, such as the
quark masses that enter in the virtual photon wavefunction $\psi$ --
this is legitimate for the light $u,d$ and $s$ quarks).  This scaling
property of the DIS total cross-section has been observed in
experimental data, as shown in the right plot of the figure
\ref{fig:geom}.
What makes {\sl gluon saturation} a satisfying explanation of this
phenomenon is that geometrical scaling is a very robust consequence of
evolution equations with a nonlinear term that tames the growth of the
amplitude, since any realistic initial condition falls leads to
traveling wave solutions. Interestingly, the scaling shown in the
right plot of the figure \ref{fig:geom} also works for DIS off
nuclei\cite{FreunRWS1}, provided one scales the saturation momentum by
the appropriate factor $A^{1/3}$ (see the NMC data points). For large
nuclei like the ones employed in heavy ion collisions (Gold at RHIC
and Lead at the LHC), the nuclear enhancement of the saturation
momentum is $A^{1/3}\approx 6$. From fits to existing data on deep
inelastic scattering on various targets, one can get a more precise
idea of the numerical value of the saturation
momentum\cite{KowalT1,KowalLV1} (this is how $Q_s$ has been obtained
in the figure \ref{fig:sat}).

\section{Introduction to the Color Glass Condensate}
\label{sec:CGC}
\subsection{Other elementary reactions}
In the previous section, we have studied Deep Inelastic Scattering and
its energy evolution from the point of view of the projectile (in this
case, the virtual photon). When going to NLO, the correction was
applied to the $q\overline{q}$ dipole, while the target field was held
fixed. In other words, all the energy evolution was applied to the
dipole. This is perfectly legitimate: since cross-sections are Lorentz
invariant objects, they do not depend on the frame. The kind of
calculation we have performed in the previous section amounts to
observing the reaction in a frame where the target is nearly at rest,
and the boost to go to higher energy is applied entirely to the
virtual photon. In the case of DIS, the interest in doing so is
obvious: the target is very complicated (a nucleon or a nucleus),
while the projectile is a rather elementary object (a virtual
photon). Thus it seemed natural to look at the energy evolution by
boosting the object we understand best.

There is another situation where one can proceed in this way, that
involves colliding a dilute projectile on a dense target. This is the
case for instance in proton-nucleus collisions, at forward rapidities
in the direction of the proton beam. In this kinematical
configuration, one probes the dilute regime of the proton, and the
dense regime of the nucleus.
\begin{figure}[htbp]
\begin{center}
\resizebox*{!}{18mm}{\includegraphics{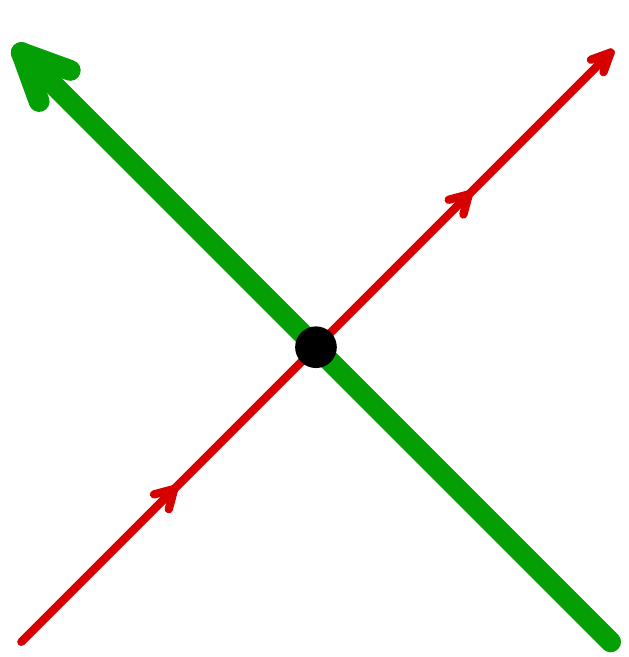}}\hfill
\resizebox*{!}{18mm}{\includegraphics{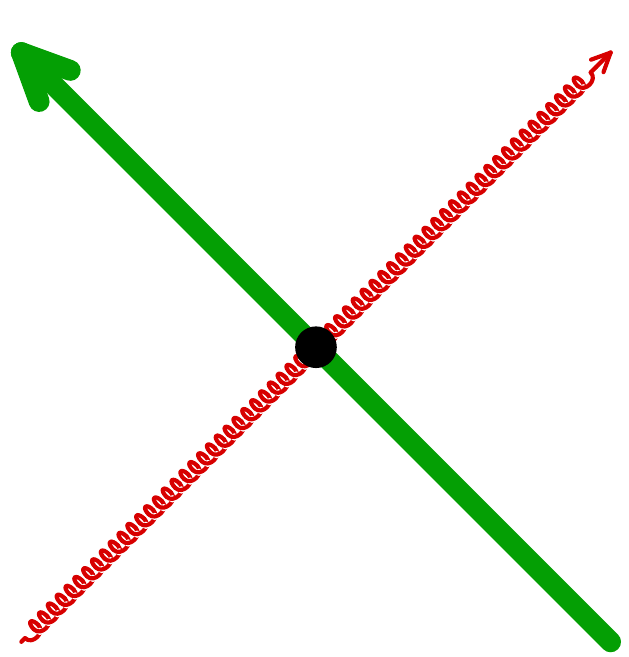}}\hfill
\resizebox*{!}{18mm}{\includegraphics{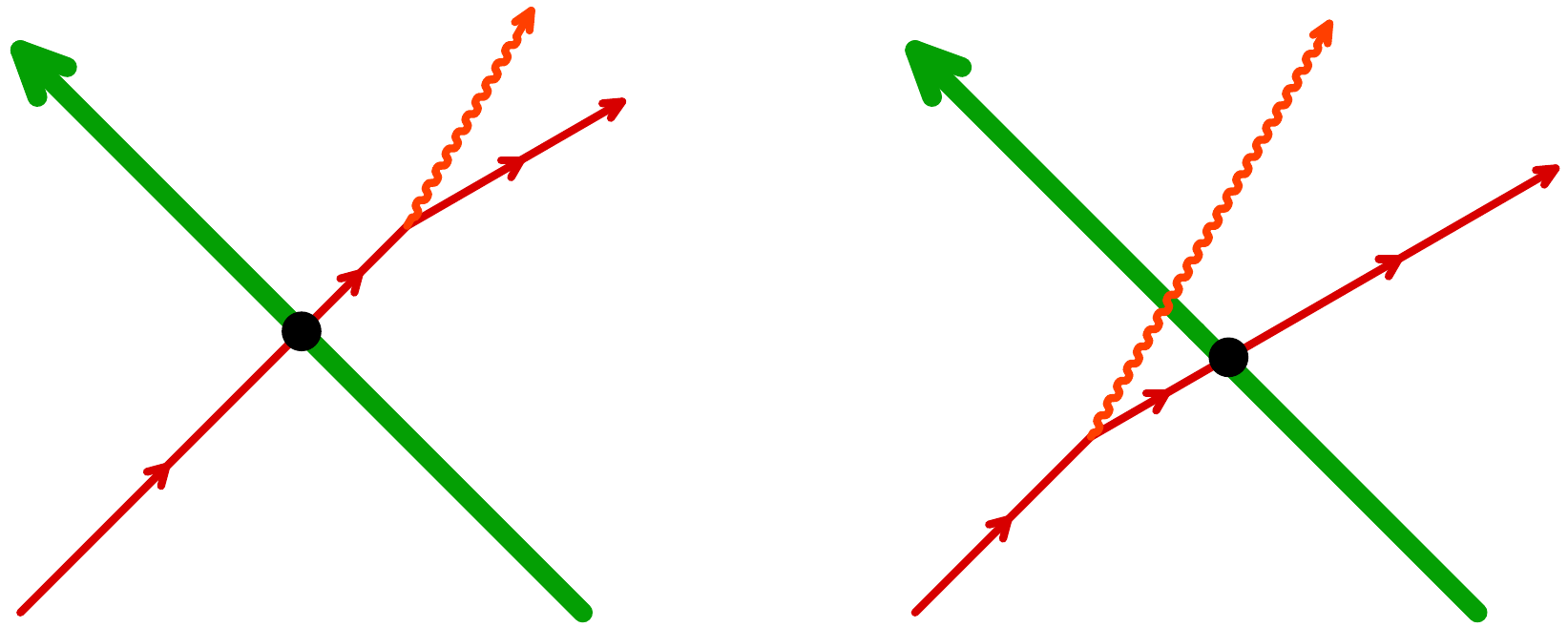}}
\end{center}
\begin{center}\hfill
\resizebox*{!}{18mm}{\includegraphics{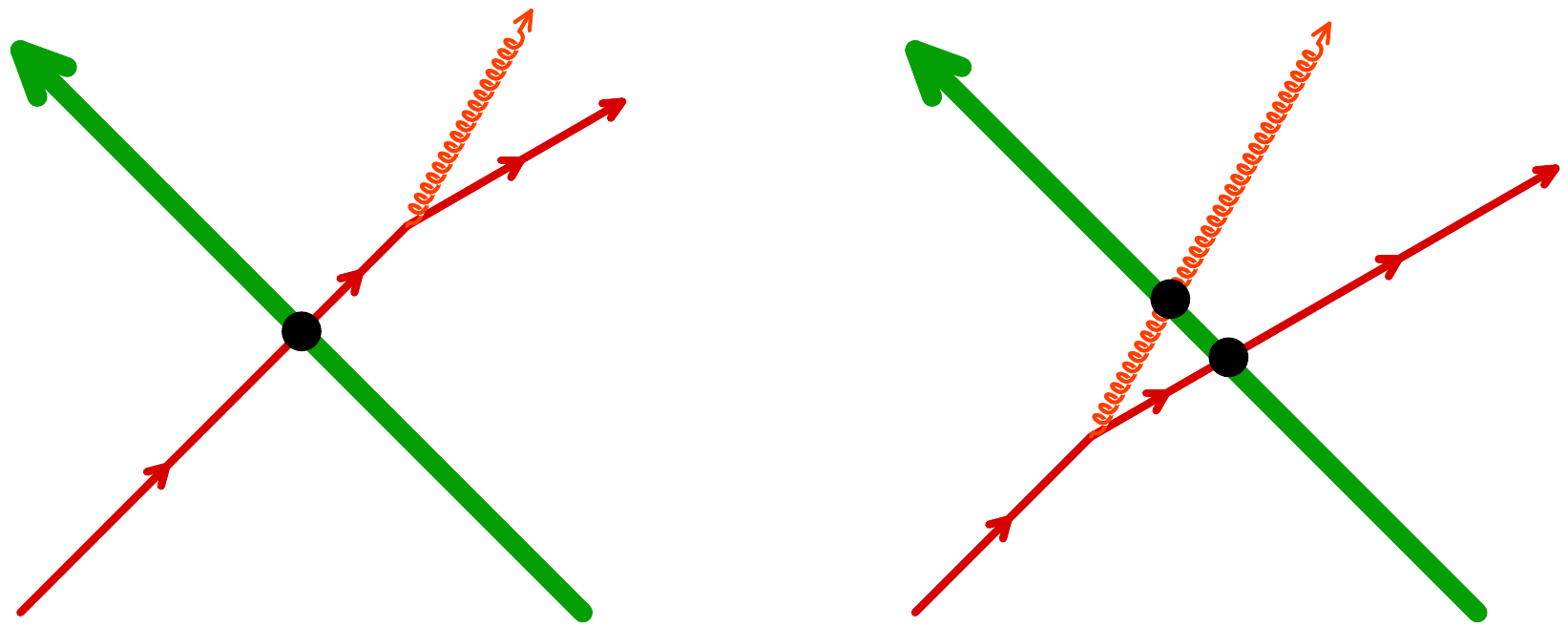}}\hfill
\resizebox*{!}{18mm}{\includegraphics{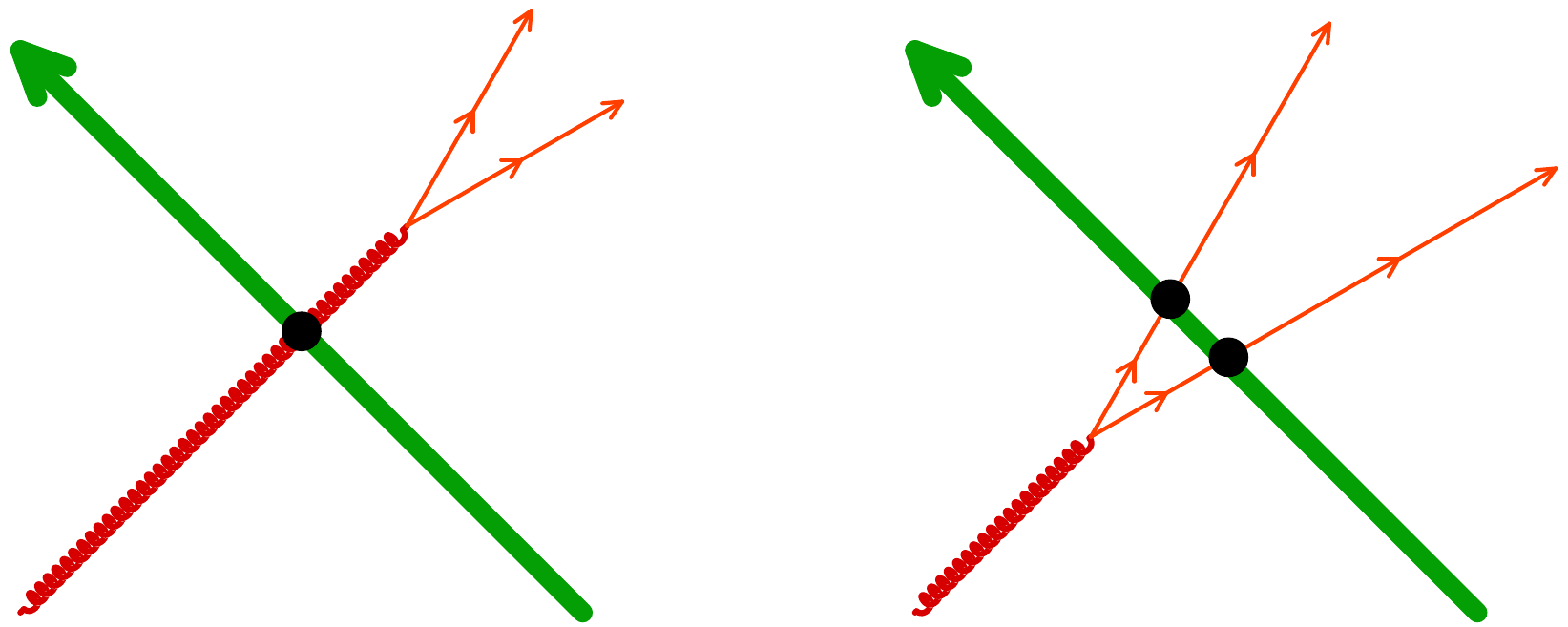}}\hfill
\end{center}
\caption{\label{fig:elem}Examples of elementary processes involving a
  projectile quark or gluon and the field of a target nucleus. Each
  black dot represents the Wilson line associated with the eikonal
  scattering of the corresponding parton.}
\end{figure}
In this case, the proton can be describe as a dilute beam of quarks or
gluons whose flux is given by the usual parton distributions, and the
elementary reactions we must consider are the interactions between a
quark or a gluon and the field of the target nucleus. Examples of
processes\cite{KovchT1,MarquXY1,KovchT2,DumitJ1,DumitJ2,GelisJ1,GelisJ2,GelisJ3,GelisJ4,DumitM1,BlaizGV1,BlaizGV2,IancuIT2,Jalil1,Jalil4,Jalil5,Jalil6,JalilK1,JalilK2,FujiiGV1,FujiiGV1,DumitHJ1,DumitHJ2,DumitJ3,Marqu1,DominXY1,ChiriXY1}
that have been evaluated in this way are represented in the figure
\ref{fig:elem}.

However, this approach suffers from a serious limitation, since it is
practical only when the projectile is much simpler than the
target. For symmetric collisions, such as nucleus-nucleus
collisions. In this case, there is no advantage in treating
differently the two nuclei. In fact, doing so is arguably rather
unnatural. As we shall see shortly, the Color Glass Condensate
description of gluon saturation allows a perfectly symmetric
description of the two nuclei in these collisions.

\subsection{Color sources and quantum fields}
Before we turn to the study of nucleus-nucleus collisions, let us
consider again DIS, but this time from the point of view of the
target, i.e. when we go to higher energy, we apply the boost to the
target while leaving the projectile wavefunction unchanged. In the
previous, we simply assumed that the target can be represented by a
certain ensemble of color field configurations, that we did not
specify. If we adopt the point of view that consists in applying the
boost to the target, then this ensemble of fields must change with
energy for the cross-section to have the correct energy dependence.

At the basis of the Color Glass Condensate framework is the fact that
the fast partons of a hadron or nucleus suffer Lorentz time dilation
and therefore do not evolve during the short duration of a collision.
They can thus be considered as time independent objects moving at the
speed of light along the light-cone, and the only relevant information
we need to know about them is the color charge they carry. This can be
encoded in a color current of the form
\begin{equation}
J^\mu_a={\colora\delta^{\mu+}}{\colorb\rho_a(x^-,\vec\x_\perp)}\; .
\label{eq:CGCcur}
\end{equation}
This formula keeps only the longitudinal component of the current (the
$+$ component in light-cone coordinates), and neglects all the other
components, because only this component is enhanced by the Lorentz
factor. The lack of $x^+$ dependence stems from the assumption that
these color charges are time independent. The function $\rho_a$
describes the spatial distribution of these color charges, both in
$x^-$ and in $\x_\perp$. Note that the support of its $x^-$ dependence
is very narrow, due to Lorentz contraction.  At leading order, the
target field can be obtained by solving the Yang-Mills equation with
the source $J^\mu$,
\begin{equation}
\big[D_\mu, F^{\mu\nu}\big]=J^\nu\; .
\end{equation}
In the Lorenz gauge, defined by $\partial_\mu A^\mu=0$, this equation
can be solved analytically for the above current, and one obtains
\begin{equation}
A^-=A^i=0\quad,\quad 
A^+(x) =-\frac{1}{{\bs\nabla}_\perp^2}\;\rho(x^-,\x_\perp)\; . 
\end{equation}
Therefore, specifying the ensemble of target fields is equivalent to
specifying the ensemble of the functions $\rho$. Thus, the CGC
framework must be supplemented by a probability distribution $W[\rho]$.

The ideas behind the Color Glass Condensate originate in the
McLerran-Venugopalan model\cite{McLerV1,McLerV2,McLerV3} that, in
addition to the description of a dense object in terms of classical
color sources, also argued that their distribution $W[\rho]$ should be
nearly Gaussian, with only local correlations,
\begin{equation} 
\left<\rho_a(x^-,\x_\perp)\rho_b(y^-,\y_\perp)\right>=\mu^2\delta^{ab}\delta(x^--y^-)
\delta(\x_\perp-\y_\perp)\; .
\end{equation}
The justification of this model is that, in a highly Lorentz
contracted nucleus, there is a large density of color charges at each
impact parameter. The charges from different nucleons are
uncorrelated, and thus by the central limit theorem, the resulting
distribution should be approximately Gaussian for a large nucleus.
The locality of the correlations is also a consequence of the fact
that charges in different nucleus are not correlated.

\begin{figure}[htbp]
\begin{center}
\resizebox*{4cm}{!}{\includegraphics{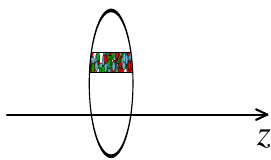}}\hfil
\resizebox*{6cm}{!}{\includegraphics{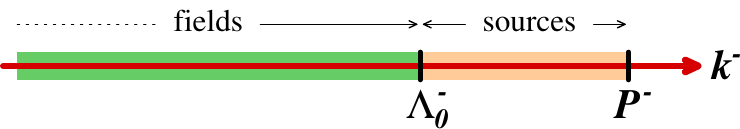}}
\end{center}
\caption{\label{fig:MV}Left: illustration of the McLerran-Venugopalan
  model for a large nucleus. Right: separation in longitudinal
  momentum between fields and sources (here for a target moving in the
  $-z$ direction -- hence the cutoff on the $k^-$ component of the
  momentum).}
\end{figure}

\subsection{Cutoff dependence and renormalization group evolution}
The justification for treating the constituents of the target as
static color sources is Lorentz time dilation. Arguably, this is only
justified if these partons have a large enough longitudinal
momentum. The partons that are too slow must be treated in terms of
the usual gauge fields (if they are gluons). Therefore, the CGC should
be viewed as an effective theory with a cutoff $\Lambda^+$, such that
\begin{itemize}
\item Partons with $k^+>\Lambda^+$ are treated as static sources, via
  eq.~(\ref{eq:CGCcur}).
\item Partons with $k^+<\Lambda^+$ are treated as standard quantum fields.
\end{itemize}
Because these two types of degrees of freedom have well separated
longitudinal momenta, the interactions between them can be
approximated by an eikonal coupling $J_\mu A^\mu$, and the action that
describes the complete effective theory is
\begin{equation}
{\cal S}=\underbrace{-\frac{1}{4}\int F_{\mu\nu}F^{\mu\nu}}_{\mbox{Yang-Mills}}
+\underbrace{\int{\colord J^\mu}{\colorb A_\mu}}_{\mbox{fast partons}}.
\end{equation}
In the CGC framework, the expectation value of an observable is obtained
by first computing the observable for a fixed configuration $\rho$ of
the color sources of the target, and then by averaging it with the
distribution $W[\rho]$,
\begin{equation}
\left<{\colorb{\bs{\cal O}}}\right>=
{\int \left[D\rho\right]\;
{\colord W[\rho]}\;
{\colorb{\bs{\cal O}}}[\rho]}\; .
\end{equation}

The cutoff $\Lambda^+$ that has been introduced to separate the fast
and the slow partons is arbitrary, and observable quantities should
not depend on it. At leading order, this is always the case. But when
one computes loop corrections, the cutoff enters in the calculation as
an upper limit on the longitudinal momentum that runs in the loop, in
order to prevent the quantum modes that are integrated in the loop to
duplicate modes that are already introduced via the static source
$\rho_a$. In general, this leads to a logarithmic dependence of the
loop corrections on the cutoff $\Lambda^+$~\cite{AyalaJMV2}. The only
way to have $\Lambda$-independent observables at the end of the day is
to let the distribution $W[\rho]$ depend itself on $\Lambda$,
precisely in such a way that the two $\Lambda$ dependences cancel,
\begin{equation}
W[\rho]\to W_{_\Lambda}[\rho]\; .
\end{equation}
Moreover, it turns out that this can be achieved with the same
distribution $W_{_\Lambda}[\rho]$ for a wide range of different
observables, with a $\Lambda$ dependence governed by the so-called
JIMWLK equation\cite{JalilKMW1,JalilKLW1,JalilKLW2,JalilKLW3,JalilKLW4,IancuLM1,IancuLM2,FerreILM1}, of the form 
\begin{equation}
\frac{\partial W_{_\Lambda}}{\partial\log\Lambda}
={\cal H}\,
W_{_\Lambda}\; ,
\label{eq:JIMWLK}
\end{equation}
where ${\cal H}$ (the JIMWLK {\sl Hamiltonian}) is a quadratic
operator in functional derivatives with respect to $\rho$.  Note that
the $\Lambda$ dependence is often traded for a rapidity dependence, by
using $Y\equiv\log(\Lambda)$. The evolution equation (\ref{eq:JIMWLK})
can be seen as a renormalization group equation, that describes how
the color charge content of the target changes as one changes the
longitudinal momentum $\Lambda$ down to which they are
considered. When the cutoff $\Lambda$ is large, and close to the
longitudinal momentum of the target, only its valence partons are
included in the description in terms of $\rho$. Lowering $\Lambda$
means that one includes in this effective description more and more
soft modes ({\sl sea partons}).

The JIMWLK equation does not predict by itself what the distribution
$W_{_\Lambda}[\rho]$ is for a given target, only how it changes when
on lowers the cutoff. It must be supplemented by an initial condition
at some $\Lambda_0$ in order to lead to definite results. This initial
condition is by essence non-perturbative, and must be
modelled\footnote{In this sense, the JIMWLK equation has exactly the
  same status as the DGLAP equation for the $Q^2$ dependence of the
  ordinary parton distributions.}. When the evolution of the
distribution $W_{_\Lambda}[\rho]$ is considered, the status of the
McLerran-Venugopalan model changes a bit, because the Gaussian
distribution proposed by the MV model is not a fixed point of the
JIMWLK equation. Instead, the MV model is often viewed as a plausible
initial condition for a large nucleus at large cutoff (i.e. close to
the valence region of the target).

Although it is necessary to specify an initial condition in order to
solve the JIMWLK equation, it should be stressed that its solutions
have a universal scaling property when evolved far enough from the
initial cutoff scale: the equation generates an intrinsic momentum
scale, the saturation momentum $Q_s$, and in this scaling regime all
the correlation functions of Wilson lines depend on the transverse
coordinates only via combinations such as $Q_s \x_\perp$. This scaling
of the solutions of the JIMWLK equation is what led to asymptotic
traveling wave solutions for the BK equation. This behavior of the
solutions mean that they tend to forget the details of their initial
condition, if evolved sufficiently far from the starting scale. This
suggests that predictions of the CGC should be less sensitive to the
model employed for the initial condition when applied to the study of
collisions at very high energy.

\subsection{Relationship between the Balitsky's and CGC formulations}
The correspondence between Balitsky's hierarchy and the CGC
formulation can be summarized by the following identity,
\begin{equation}
\left<{\colorb{\bs{\cal O}}}\right>_{_Y}=
\underbrace{\int \left[D\rho\right]\;
{\colord W_{0}[\rho]}\;
{\colorb{\bs{\cal O}}}_{_Y}[\rho]}_{\mbox{\scriptsize Balitsky's description}}
=
\underbrace{\int \left[D\rho\right]\;
{\colord W_{_Y}[\rho]}\;
{\colorb{\bs{\cal O}}}_{0}[\rho]}_{\mbox{\scriptsize CGC description}}\; .
\end{equation}
In the Balitsky's approach, the rapidity dependence is obtained by
applying a boost to the observable, while the distribution of sources
is kept unchanged. In the CGC approach, the observable is a fixed
functional of the sources, and the boost leads to a change in the
distribution of the sources. In the simple case of collisions between
an elementary projectile and a dense target, where observables can be
expressed in terms of Wilson lines of the target field, the
equivalence between these two points of view follows from the existence
of a universal operator ${\cal H}$ such that
\begin{equation}
\frac{\partial{\colorb{\bs{\cal O}}}_{_Y}[\rho]}{\partial{\colora Y}}
={\cal H}\left[\rho,\frac{\delta}{\delta\rho}\right]\;\;
{\colorb{\bs{\cal O}}}_{_Y}[\rho]\; .
\end{equation}
Moreover, this operator is self-adjoint, which means that one can
``integrate by parts'' and transpose its action from the observable
to the distribution $W[\rho]$, leading to the JIMWLK evolution equation.

\subsection{Example~: DIS in the CGC framework}
\begin{figure}[htbp]
\begin{center}
\resizebox*{3.5cm}{!}{\includegraphics{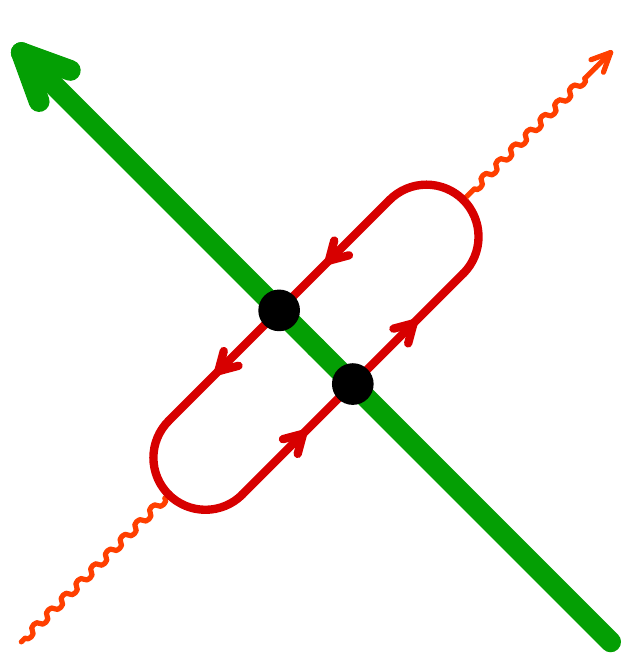}}\hfil
\resizebox*{3.5cm}{!}{\includegraphics{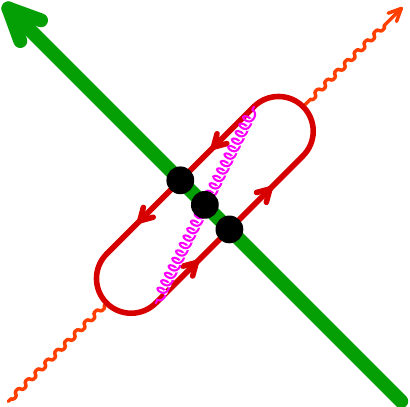}}
\end{center}
\caption{\label{fig:DIS1} Deep inelastic scattering in the CGC
  framework. The green arrow represents the trajectory of the target,
  and the support of its color sources. Left: Leading Order: the
  virtual photon fluctuates into a $q\overline{q}$ dipole that
  scatters off the color field of the target. Right: Next to Leading
  Order: the photon fluctuates into a $q\overline{q}g$ state.}
\end{figure}

As an illustration of the use of the CGC, let us reconsider Deep
Inelastic Scattering in this framework (see the figure
\ref{fig:DIS1}). We present here only a sketch of the calculation, but
not the details\cite{McLerV4,McLerV5}.
At LO, the photon must fluctuate into a $q\overline{q}$ dipole in
order to interact with the field of the target. As with the BK
approach, the forward scattering amplitude of this dipole off the
target can be expressed in terms of Wilson lines,
\begin{eqnarray}
{\colorb {\bs T}_{_{\rm LO}}(\vec\x_\perp,\vec\y_\perp)}
&=&
1-\frac{1}{N_c}\,{\rm tr}\,({{\colord U(\vec\x_\perp)U^\dagger(\vec\y_\perp)}})
\nonumber\\
U(\vec\x_\perp)&=& {\rm P}\,\exp i{\colord g}\int
 dz^+\,{\colorb{\cal A}^-(z^+,\vec\x_\perp)}\; ,
\end{eqnarray}
and the main difference resides in the fact that the target field is
not arbitrary but obtained by solving the Yang-Mills equation
\begin{equation}
\left[{\colorb{\cal D}_\mu},{\colorb{\cal F}^{\mu\nu}}\right]={\colora\delta^{\nu-}}\,{\colord\rho(x^+,\vec\x_\perp)}\; .
\end{equation}
Then the cross-section is obtained by averaging over all the
configurations of the color source $\rho$, with the distribution
$W[\rho]$. At LO, the difference with the BK point of view is purely
semantic, since we are simply trading the (arbitrary) distribution of
the target fields for the (arbitrary) distribution of the color sources.

\begin{figure}[htbp]
\begin{center}
\resizebox*{7cm}{!}{\includegraphics{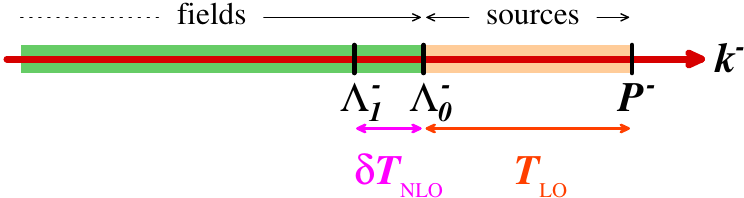}}
\end{center}
\caption{\label{fig:DIS2} Infinitesimal step in the cutoff evolution,
  from $\Lambda_0$ to $\Lambda_1$.}
\end{figure}
At NLO, one must now consider corrections such as the one represented
in the right part of the figure \ref{fig:DIS1} (there are several
graphs, only one of them is represented there). The longitudinal
momentum of the gluon in the calculation of these corrections should
be limited at the upper value $\Lambda_0^-$, to avoid over-counting
modes that are already included via the sources $\rho$. In practice,
it is more convenient to compute only the correction $\delta T_{_{\rm
    NLO}}$ to the scattering amplitude due to the field modes in a
small strip $\Lambda_1^-< k^- < \Lambda_0^-$. The important part in
this discussion are the terms that have logarithm divergences when
$\Lambda_1^-\ll \Lambda_0^-$. One can show that they take the
following form,
\begin{equation}
  {\colorc\delta {\bs T}_{_{\rm NLO}}(\vec\x_\perp,\vec\y_\perp)}
  =
  \ln\left(\frac{\Lambda_0^-}{\Lambda_1^-}\right)\;
  {\colorb{\cal H}}\;
  {\colord{\bs T}_{_{\rm LO}}(\vec\x_\perp,\vec\y_\perp)}\; ,
\end{equation}
where ${\cal H}$ is an operator that acts on the $\rho$'s. Besides the
fact that they depend on an unphysical cutoff, these logarithms
question the validity of the perturbative expansion. Indeed, the NLO
corrections are suppressed by one power of the coupling constant
$\alpha_s$ compared to the LO result, but the appearance of these
possibly large logarithms can compensate the smallness of the
coupling. This is why one should consider only a small slice of
longitudinal momenta at a time. It turns out that these logarithms can
be hidden by averaging over the color sources, thanks to the identity
\begin{equation}
\Big<{\colord{\bs T}_{_{\rm LO}}}+{\colorc\delta{\bs T}_{_{\rm NLO}}}\Big>_{\Lambda_0^-}
=
\Big<{\colord{\bs T}_{_{\rm LO}}}\Big>_{\Lambda_1^-}\; ,
\label{eq:eff-th}
\end{equation}
where we use the shorthand
\begin{equation}
\big<\cdots\big>_{\Lambda}\equiv \int[D\rho]\;W_{_\Lambda}[\rho]\;\cdots\; ,
\end{equation}
provided the distributions of sources at the scales $\Lambda_0^-$ and
$\Lambda_1^-$ are related by
\begin{equation}
W_{\Lambda_1^-}\equiv \Big[1+\ln\left(\frac{\Lambda_0^-}{\Lambda_1^-}\right)\;
    {\colorb{\cal H}}\Big]\;W_{\Lambda_0^-}\; .
\label{eq:jimwlk1}
\end{equation}
The meaning of eq.~(\ref{eq:eff-th}) is that the sum of the LO+NLO
contributions in the original effective theory (that has its cutoff at
the scale $\Lambda_0^-$) is equivalent to the LO only in a new
effective theory that has its cutoff at the lower scale $\Lambda_1^-$,
and a new distribution of sources given by
eq.~(\ref{eq:jimwlk1}). Note that eq.~(\ref{eq:jimwlk1}) is nothing
but the JIMWLK equation for an infinitesimal change of the cutoff
scale. Moreover, the right hand side of eq.~(\ref{eq:eff-th}) has no
dependence on the former scale $\Lambda_0^-$. This means that in the
left hand side, this dependence must cancel between the NLO
contribution and the scale dependence of the distribution $W[\rho]$.

The previous process, by which we integrated out the logarithms coming
from a small slice of field modes, can be repeated indefinitely by
considering a sequence of lower and lower cutoffs,
\begin{equation}
\cdots< \Lambda_n^-<\cdots < \Lambda_1^- <\Lambda_0^-\; .
\end{equation}
At each step, new logarithms are generated, that can be absorbed by
defining a new effective theory at the lower scale $\Lambda_{n+1}^-$
and a new distribution of sources at this scale. The dependence on the
cutoff at the previous scale $\Lambda_n^-$ disappears in this
process. We must repeat this procedure until we reach a cutoff scale
which is lower than all the physically relevant longitudinal momentum
scales: at this point, lowering the cutoff further only introduces new
sources that are too slow to be relevant in the observable under
consideration -- the observable does not depend on these sources, and
thus becomes independent of the cutoff\footnote{In the case of DIS,
  the dependence on the cutoff disappears once it becomes smaller than
  $xP^-$, where $P^-$ is the longitudinal momentum of the target and
  $x\equiv Q^2/(2P\cdot Q)$.}.

\section{Factorization in high-energy hadronic collisions}
\label{sec:factorization}
\subsection{Introduction}
In the previous section, we have introduced the Color Glass Condensate
effective theory, and illustrated it by considering again the example
of Deep Inelastic Scattering. In this context, the CGC is a mere
rephrasing of the physics that was already present in Balitsky's
hierarchy. Roughly speaking, the CGC puts the emphasis on the energy
evolution of the distribution that describes the target, rather than
on radiative corrections to the operator that is being measured, but
the end result is  exactly the same.

The situation is qualitatively different in the case of
nucleus-nucleus collisions. In this case, it would be highly desirable
to have a framework in which the projectile and target can be treated
on the same footing, and the CGC appears as a good candidate for
achieving this. In the CGC framework, we expect that each of the
nuclei will be described by static color sources $\rho_1$ and
$\rho_2$, moving in opposite directions along the light-cone, with
their respective distributions $W_1[\rho_1]$ and $W_2[\rho_2]$.

\begin{figure}[htbp]
\begin{center}
\resizebox*{5.5cm}{!}{\includegraphics{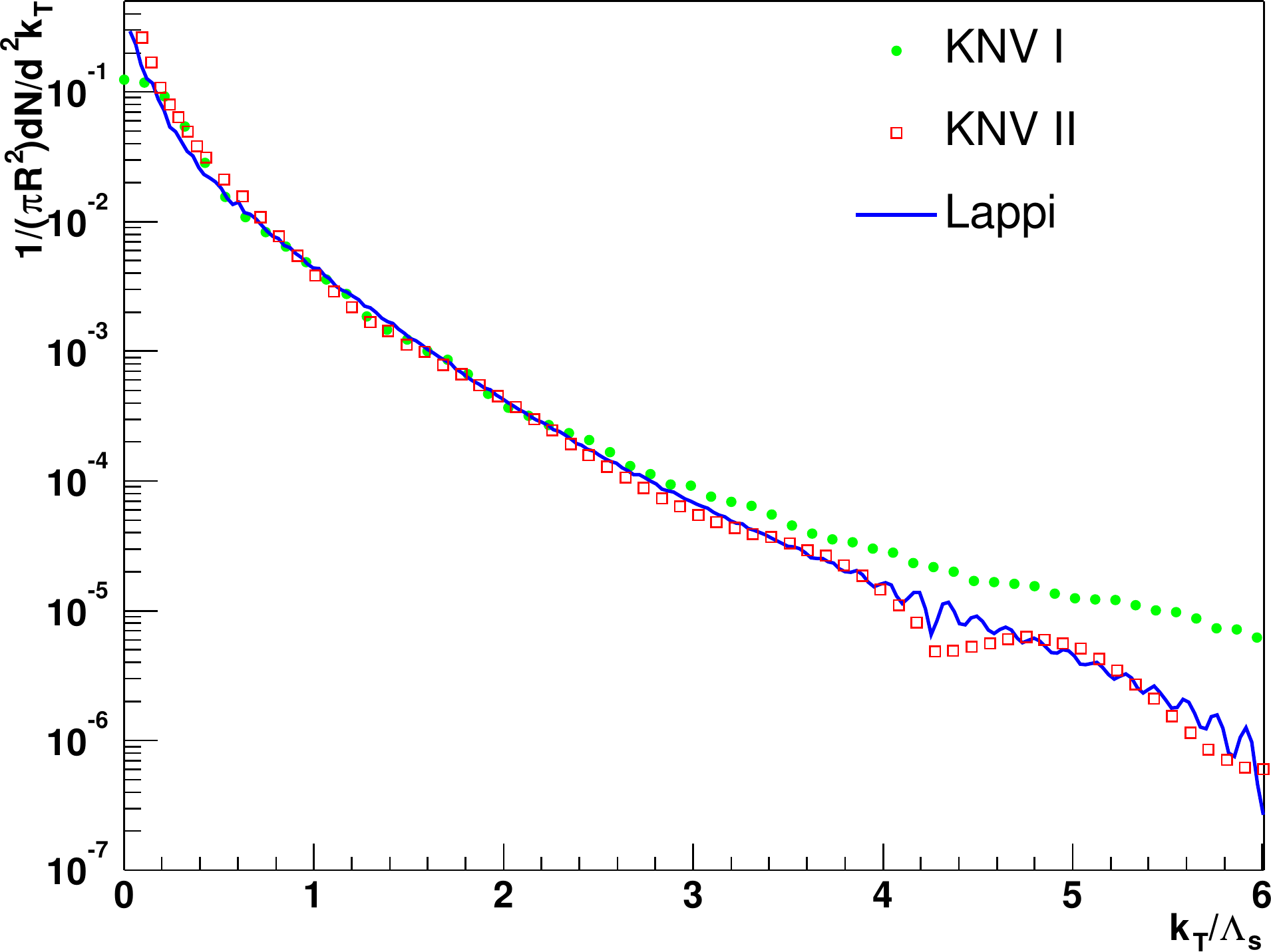}}
\end{center}
\caption{\label{fig:MV1}
  Transverse momentum dependence of the single inclusive gluon
  spectrum\cite{KrasnNV1}.}
\end{figure}

In several
works\cite{KrasnV3,KrasnV1,KrasnV2,KrasnNV2,KrasnNV1,Lappi1,Lappi3,KrasnNV3,KrasnNV4,LappiV1},
the CGC has been used as follows in order to compute observables in
heavy ion collisions~:
\begin{itemize}
\item Pick randomly two color sources $\rho_1$ and $\rho_2$ (one for
  each nucleus), for instance by using the Gaussian distributions
  encountered in the MV model.
\item Solve (numerically) the classical Yang-Mills equation with two
  sources,
\begin{equation}
\left[{\colorb{\cal D}_\mu},{\colorb{\cal F}^{\mu\nu}}\right]={\colora\delta^{\nu-}}\,{\colord\rho_1}+\delta^{\mu+}\,\rho_2\; ,
\label{eq:YM1}
\end{equation}
with an initial condition such that the color field vanishes at
$x^0\to -\infty$.
\item Evaluate the observable of interest on the classical color field
  ${\cal A}^\mu$ obtained by solving the previous equation. For
  instance, the gluon spectrum is obtained as
\begin{equation*}
\left.
\frac{d{{\colora{N_1}}}}{dY d^2\vec\p_\perp}\right|_{_{\rm LO}}=
\frac{1}{16\pi^3}\int_{x,y}\; {\colorc e^{ip\cdot (x-y)}}\;
\square_x\square_y\;
\sum_\lambda \epsilon^\mu_\lambda \epsilon^\nu_\lambda\;\;
{\colord {\cal A}_\mu(x)}{\colord {\cal A}_\nu(y)}\; .
\end{equation*}
\item Repeat these steps in order to perform a Monte-Carlo average
  over the distributions for $\rho_1$ and $\rho_2$.
\end{itemize}
As an illustration, we show in the figure \ref{fig:MV1} some numerical
results for the single inclusive spectrum of the gluons produced in a
nucleus-nucleus collision, obtained by following this procedure.

The goal of this section is to provide a theoretical justification for
the above procedure, and in particular to answer the following questions~:
\begin{itemize}
\item Does it correspond to the leading term in an expansion in powers
  of the coupling constant?
\item What is the justification for the use of retarded boundary
  conditions?
\item In the case of a collision between two nuclei, can one still
  absorb the logarithms that arise in loop corrections by letting the
  distributions $W_1[\rho_1]$ and $W_2[\rho_2]$ evolve according to
  the JIMWLK equation?
\end{itemize}

\subsection{Power counting}
\begin{figure}[htbp]
\begin{center}
\resizebox*{5.5cm}{!}{\includegraphics{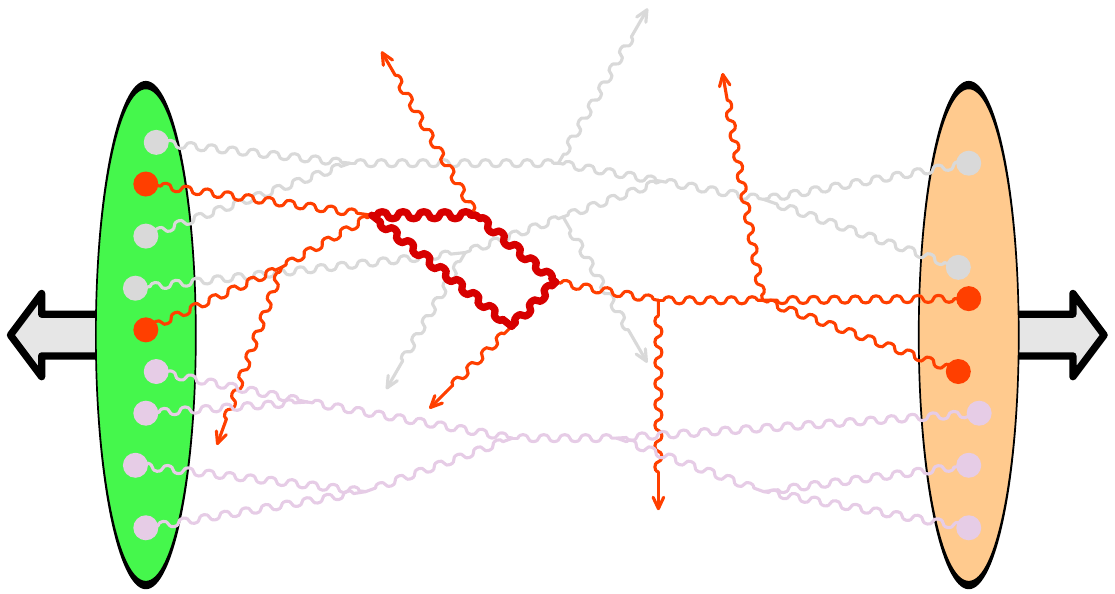}}
\end{center}
\caption{\label{fig:PC}Typical connected graph in a nucleus-nucleus
  collision described in the CGC framework. In this representation,
  the circular dots represent insertions of the sources $\rho_1$ or
  $\rho_2$.}
\end{figure}
Before we can start answering these questions, we need a way to assess
the order of magnitude of a given graph\cite{GelisV2,GelisV3}. Typical graphs in a nucleus-nucleus collisions
are made of several disconnected components (see the graph on the
right of the figure \ref{fig:coll}). In order to perform this power
counting, it is sufficient to consider only one of these connected
components, as illustrated in the figure \ref{fig:PC}. In this
representation, the circular dots represent insertions of the sources
$\rho_1$ or $\rho_2$. In the saturated regime, their contribution to
the power counting is $g^{-1}$, since the gluon occupation number,
proportional to $\rho^2$, should be of order $g^{-2}$. In addition,
three-gluon vertices bring one power of $g$ and four-gluon vertices a
factor $g^2$.

With these rules, one obtains the following order of magnitude for a
connected graph such as the one represented in the figure \ref{fig:PC}~:
\begin{equation}
\frac{1}{g^2}\,g^{n_{_E}+2n_{_L}}\; ,
\label{eq:PC}
\end{equation}
where $n_{_E}$ is the number of external gluons (the gluon lines
terminated by an arrow in the figure \ref{fig:PC}) and $n_{_L}$ the
number of independent loops in the graph. 

Interestingly, the order of the graph does not depend on the number of
internal lines and on the number of sources $\rho_{1,2}$ attached to
the graph. The latter property is specific of the dense regime, where
the sources are of order $g^{-1}$. Indeed, each insertion of a source
necessarily brings a factor $g$ for the vertex at which this source is
attached to the graph; this vertex cancels the $g^{-1}$ coming from
the source and therefore inserting a source in the graph does not cost
any power of $g$. This means that, for graphs with a fixed number of
external lines and a fixed number of loops, there is an infinity of
topologies of the same order in $g$ that differ in the number of
sources they contain. Thus, computing any observable in the CGC
framework, even at Leading Order, requires that one resums this
infinite series of terms.

For instance, the inclusive gluon spectrum has the following expansion
\begin{equation}
\frac{d{\colora N_1}}{d^3\vec\p}
=
\frac{1}{{\colorb g^2}}\;\Big[
c_0+c_1\,{\colorb g^2}+c_2\,{\colorb g^4}+\cdots
\Big]\; ,
\label{eq:Nexp}
\end{equation}
where the coefficients $c_0, c_1,\cdots$ are themselves series that
resum all orders in $(g\rho_{_{1,2}})^n$. For instance,
\begin{equation}
c_0=\sum_{n=0}^\infty c_{0,n}\,({\colorb g}{\colorc\rho_{_{1,2}}})^n\; .
\end{equation}
Our goal in the rest of this section is to calculate the complete
zeroth order term, ${\colorb c_0/g^2}$, and a subset of the
higher order terms.

\subsection{Bookkeeping}
Among the observables that one could possibly consider, the {\sl
  inclusive observables} are especially simple because they do not
veto any final state. This is the case for instance of the single
inclusive gluon spectrum\footnote{By using the completeness of the final states,
  one can check that this spectrum is also the expectation value of
  the number operator,
\begin{eqnarray*}
    \smash{\frac{dN_1}{d^3{\colorb\vec\p}}}
    \sim
    \big<0_{\rm in}\big|{\colorb a^\dagger_{\rm out}(\p)a_{\rm out}(\p)}\big|0_{\rm in}\big>\; .
  \end{eqnarray*}}, defined in terms of the transition amplitudes
as
\begin{equation}
    \frac{dN_1}{d^3{\colorb\vec\p}}
    \sim
    \sum_{n=0}^\infty (n+1)
    \int \frac{1}{(n+1)!}\Big[\underbrace{d\Phi_1\cdots d\Phi_n}_{n\mbox{\scriptsize\ part. phase-space}}\Big]\;
    \Big|\big<{\colorb\p}\p_1\cdots\p_n{}_{\rm out}\big|0{}_{\rm in}\big>\Big|^2\; ,
\end{equation}
where we use the shorthand $d\Phi\equiv d^3\p/(2\pi)^3 2p$ to denote
the invariant phase-space of a final state particle.

\begin{figure}[htbp]
\begin{center}
\resizebox*{5cm}{!}{\includegraphics{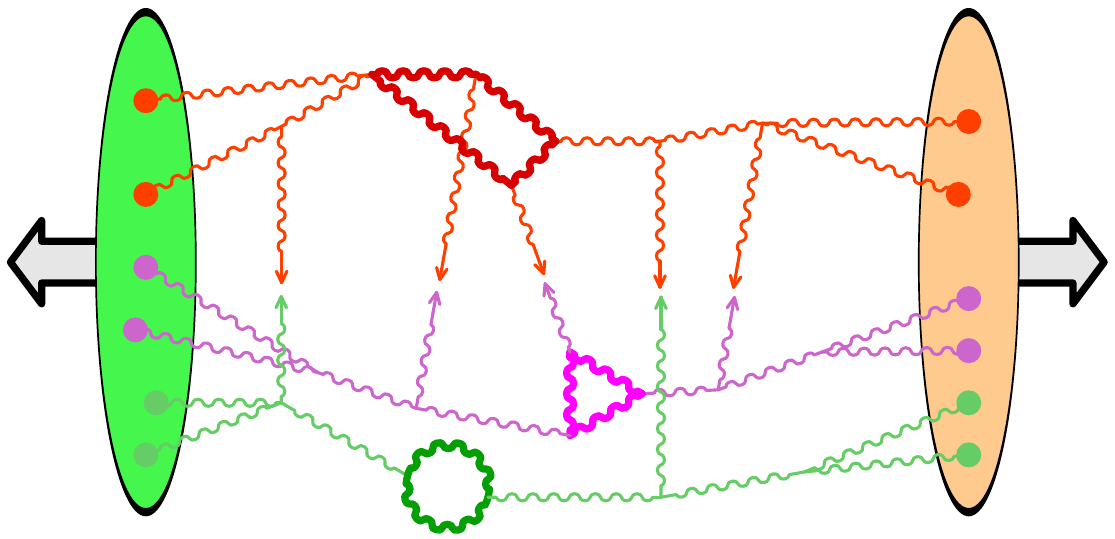}}\hfil
\resizebox*{5cm}{!}{\includegraphics{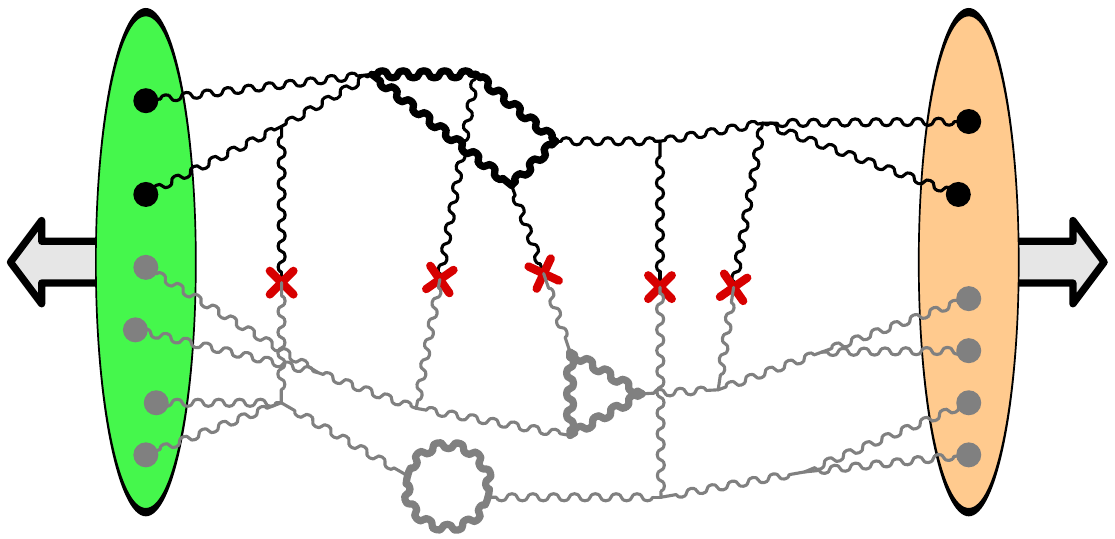}}
\end{center}
\caption{\label{fig:VAC}Left: product of a term of the amplitude with
  a term from the complex conjugate amplitude. Right: interpretation
  of this product as a cut vacuum graph.}
\end{figure}

In order to sum all the relevant graphs at a given order in $g^2$, we
need tools to list and manipulate them. For the sake of th discussion
in this section, we will consider observables related to particle
spectra, such as the inclusive gluon spectrum, correlations,
etc... All these observables can be obtained from a generating
functional
\begin{equation}
F[{\colord z}]
\equiv \sum_n \frac{1}{n!}\int\Big[d\Phi_1\cdots d\Phi_n\Big]\;
{\colord z(\p_1)}\cdots {\colord z(\p_n)}
\;
\Big|\big<\p_1\cdots\p_n{}_{\rm out}\big|0{}_{\rm in}\big>\Big|^2\; ,
\end{equation}
obtained by weighting each final state particle by an arbitrary
function $z(\p)$. The diagrammatic interpretation of this generating
functional is illustrated in the figure \ref{fig:VAC}.
The product of an amplitude by its complex conjugate can be
interpreted as performing a cut through a vacuum graph (i.e. a graph
without any external gluon)\cite{Cutko1,t'HooV1}. To construct the
generating functional $F[z]$, one simply needs to weight each cut
propagator (the propagators with a cross in the right part of the
figure \ref{fig:VAC}) by the appropriate function $z(\p)$. From this
generating functional, the single inclusive gluon spectrum introduced
above is given by
\begin{equation}
\frac{dN_1}{d^3{\colorb\vec\p}}
=
\left.\frac{\delta F[z]}{\delta z(\p)}\right|_{z=1}\; .
\end{equation}
In fact, quite generally, inclusive observables are derivatives of $F[z]$
at $z=1$, while exclusive observables are obtained via derivatives at
other values of $z$.

A crucial property in the CGC effective theory is {\sl unitarity},
i.e. at a basic level the fact that the sum of the probabilities for
all the possible final states in a collision should equal unity.  In
terms of the generating functional $F[z]$, unitarity is encoded in a
very simple way, $F[1]=1$. The fact that inclusive observables are
given by derivatives at $z\equiv 1$, combined to this property due to
unitarity, is the reason why they are much simpler than exclusive
observables. In more physical terms, inclusive observables are simpler
because many graphs cancel thanks to unitarity.

By using reduction formulas for the transition amplitudes and their
complex conjugate, one can write the generating functional $F[z]$ as
follows
\begin{equation}
  \displaystyle{{\colora F[z]}=e^{\colorb{\cal C}[z]}
    \left.\; Z[j_+]\;
    Z^*[j_-]
    \right|_{j_+=j_-=j}}\; ,
\label{eq:Fgen}
\end{equation}
where $Z[j]$ is the usual generating functional for the time-ordered
Green's functions that enter in the reduction formulas\cite{ItzykZ1},
and where
\begin{equation}
  \begin{aligned}
      &
      {\colorb{\cal C}[z]}\equiv
      \int d^4x d^4y\; {\colorc G_{+-}^0(x,y)}\; 
      \square_x\square_y\;
      \frac{\delta}{\delta {\colord j_+(x)}}
      \frac{\delta}{\delta {\colord j_-(y)}}&&&
      \nonumber\\
      &
      {\colorc G_{+-}^0(x,y)}\equiv
      \int \frac{d^4p}{(2\pi)^4}
      \; e^{ip\cdot(x-y)}\;{\colorb z(\vec\p)}\;
      {\colorc 2\pi\theta(-p^0)\delta(p^2)}&&&
      \; .
    \end{aligned}
\end{equation}
(Note that here we have simplified the notations by not writing color,
Lorentz and polarization indices -- when dealing with gluons, these
indices are of course necessary.)  Eq.~(\ref{eq:Fgen}), albeit being
rather formal, is useful to obtain more explicit expressions for
observables than can be derived from the generating functional. At
this point, it is crucial to note that
\begin{equation}
e^{\colorb{\cal C}[1]}\; Z[j_+]\;Z^*[j_-] = {\cal Z}[j_+,j_-]\; ,
\end{equation}
where ${\cal Z}[j_+,j_-]$ is the generating functional for
path-ordered Green's functions in the {\sl Schwinger-Keldysh
  formalism}\cite{Schwi1,Keldy1}.

\begin{figure}[htbp]
\begin{center}
\resizebox*{5cm}{!}{\includegraphics{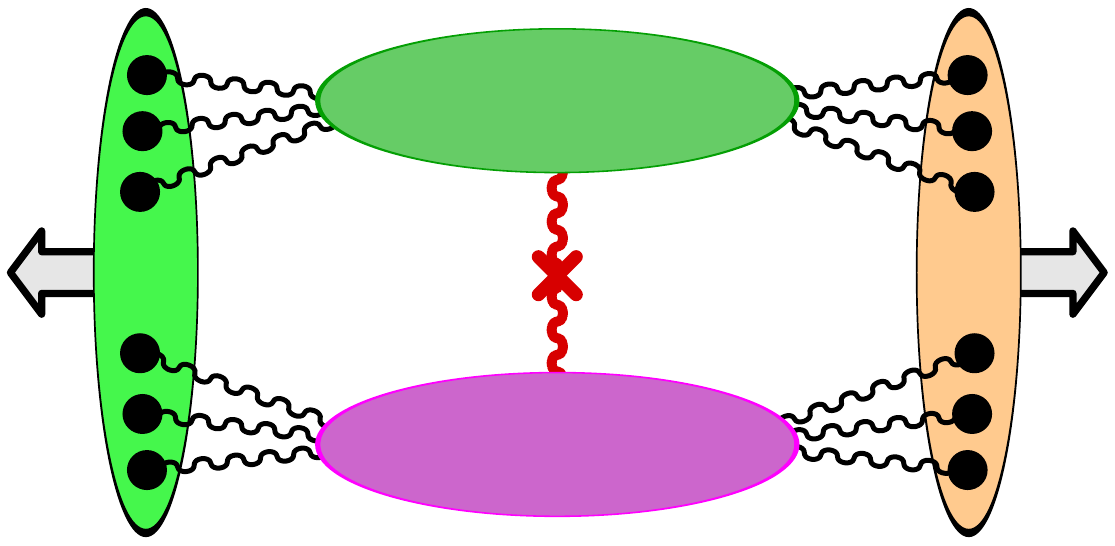}}\hfil
\resizebox*{5cm}{!}{\includegraphics{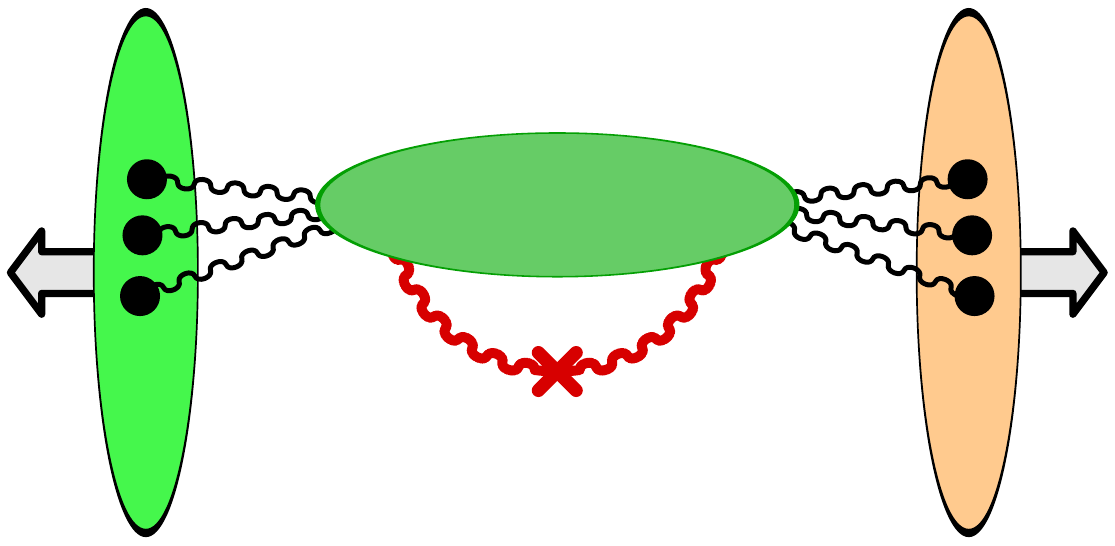}}
\end{center}
\caption{\label{fig:mult}Diagrammatic representation of the two terms
  in the right hand side of eq.~(\ref{eq:N11}). Left: term in
  $A_+A_-$. Right: term in $G_{+-}$. The crossed wavy line represents
  the on-shell momentum $\p$.}
\end{figure}
For instance, one can use eq.~(\ref{eq:Fgen}) in order to obtain the
following expression for the single inclusive spectrum,
\begin{equation}
\frac{dN_1}{d^3\p}
=
\left.\frac{\delta F[z]}{\delta (\p)}\right|_{z=1}
=
\int d^4x d^4y\; e^{ip\cdot(x-y)}\;\square_x \square_y
\Big[{\colorb A_+(x)A_-(y)}+{\colorb G_{+-}(x,y)}\Big]\; ,
\label{eq:N11}
\end{equation}
where $A_\pm$ and $G_{+-}$ are respectively the one-point and
two-point Green's functions in the Schwinger-Keldysh formalism.  The
two terms in this formula are illustrated in the figure
\ref{fig:mult}. This formula is true to all orders, and evaluating the
spectrum at a given order amounts to evaluated these Green's functions
to the desired order (the power counting rule of eq.~(\ref{eq:PC}) can
be applied to these functions, respectively with $n_{_E}=1$ and
$n_{_E}=2$).

\subsection{Leading order and retarded classical fields}
\begin{figure}[htbp]
\begin{center}
\resizebox*{5cm}{!}{\includegraphics{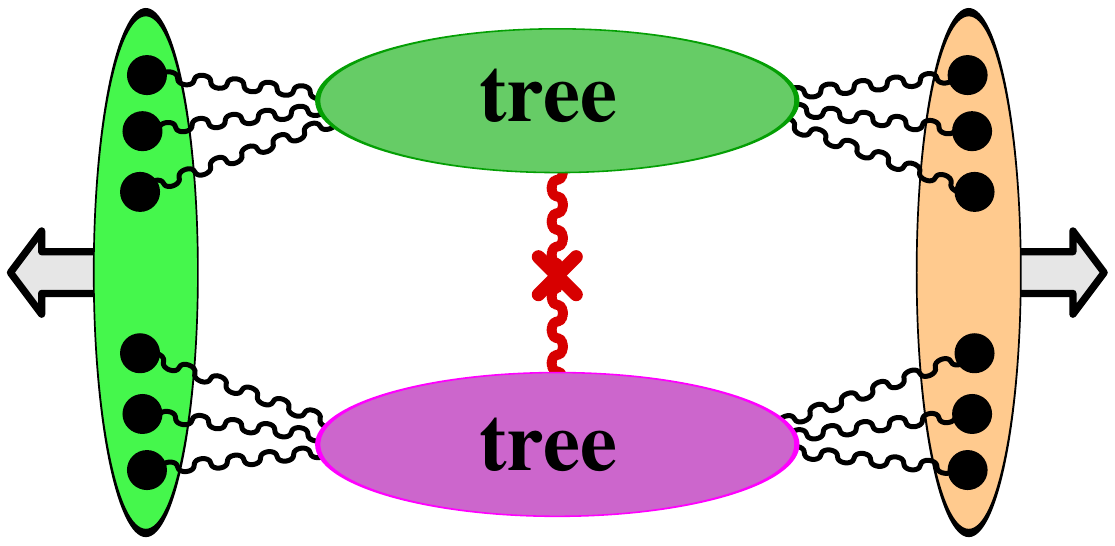}}\hfil
\resizebox*{5cm}{!}{\includegraphics{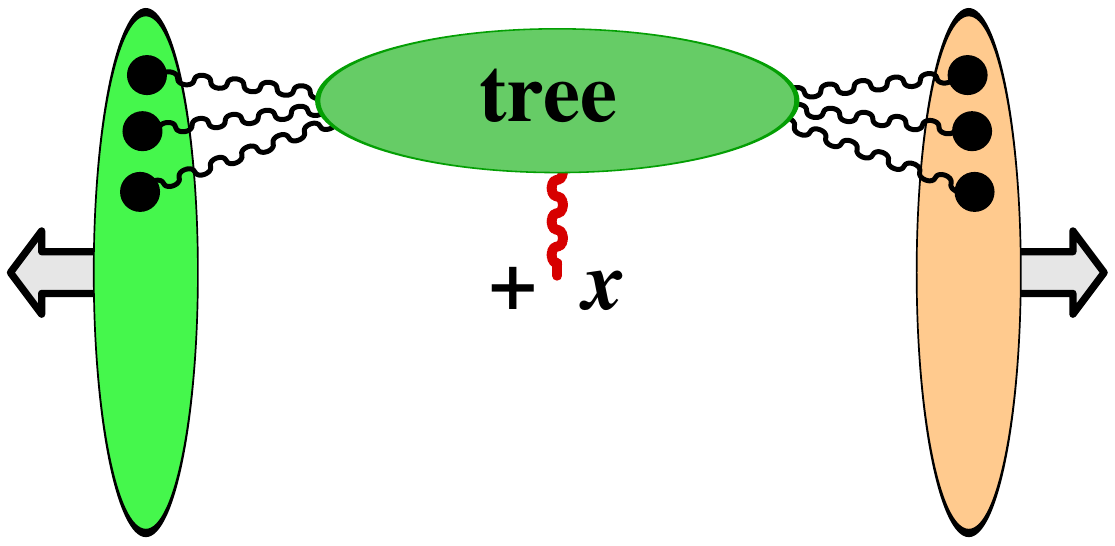}}
\end{center}
\caption{\label{fig:mult-LO}Left: only contribution to the single
  inclusive spectrum at leading order. Right: $A_+(x)$ at leading order.}
\end{figure}

\begin{figure}[htbp]
\begin{center}
\resizebox*{10cm}{!}{\includegraphics{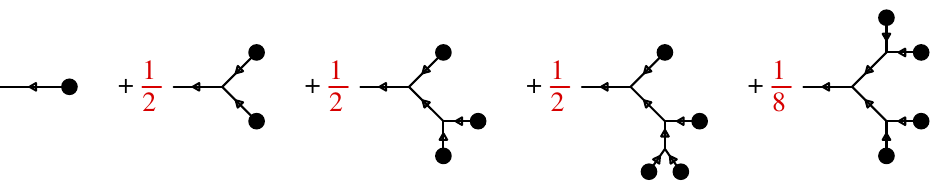}}
\end{center}
\caption{\label{fig:class}Beginning of the series of tree diagrams
  that contribute to $A_\pm$ at LO (only cubic vertices are
  included here, but of course in QCD one would also have four-gluon
  vertices). The black dots denote the color sources $\rho_1$ or
  $\rho_2$, and the arrows on the lines indicate that the propagators
  are retarded.}
\end{figure}

When we apply the power counting rule (\ref{eq:PC}) to $A_\pm$ and
$G_{+-}$, we see that they have the following expansion in $g^2$
\begin{eqnarray}
A_\pm&=& \frac{a_0}{g}+a_1\,g+a_2\,g^3+\cdots\nonumber\\
G_{+-}&=& b_0+b_1\,g^2+b_2\,g^4+\cdots\; ,
\end{eqnarray}
where the $a_i$ and $b_i$ are coefficients of order one. Therefore, to
evaluate the spectrum at leading order, we need only the LO of
$A_\pm$, and we can disregard completely the term with $G_{+-}$, as
illustrated in the left figure \ref{fig:mult-LO}.
Despite this simplification, one still needs to sum an infinite set of
graphs. Indeed, all the tree topologies contribute to $A_{\pm}$ at the
order $1/g$, and for all the internal vertices, we must sum over all
the possible assignments of the types $+$ or $-$, per the rules of the
Schwinger-Keldysh formalism\cite{Schwi1,Keldy1}. Since the color
sources are identical on the two branches of the Schwinger-Keldysh
closed time contour, this sum only gives the following combinations of
propagators
\begin{eqnarray}
  G_{++}^0(p)-G_{+-}^0(p)
  =
  G_{-+}^0(p)-G_{--}^0(p)
  {\colorb = G_{_R}^0(p)}\; ,
\end{eqnarray}
i.e. the retarded propagator. Thus, we only need to sum all the tree
diagrams built with retarded propagators, as illustrated in the figure
\ref{fig:class}.
The result of this sum of tree graphs is well known: it is the solution of
the classical field equations of motion that vanishes in the remote
past. In the case of QCD, these are the {\sl Yang-Mills equations},
\begin{equation}
\left[{\cal D}_\mu ,{\cal F}^{\mu\nu}\right] =
\delta^{\nu-}\rho_1+\delta^{\nu+}\rho_2\quad,\qquad
\lim_{x^0\to-\infty}{\cal A}^\mu(x)=0\; .
\end{equation}
This result is what justifies the procedure described in the
introduction of this section in order to compute the inclusive gluon
spectrum.

Interestingly, at LO, the $n$-gluon inclusive spectrum
factorizes into products of single gluon spectra,
\begin{equation}
\left.
\frac{d{{\colora{N_n}}}}{d^3\p_1\cdots d^3\p_n}\right|_{_{\rm LO}}=
\left.
\frac{d{{\colora{N_1}}}}{d^3\p_1}\right|_{_{\rm LO}}
\times\cdots\times
\left.
\frac{d{{\colora{N_1}}}}{d^3\p_n}\right|_{_{\rm LO}}\; ,
\label{eq:Nn}
\end{equation}
as can be checked by simple power counting arguments.

\subsection{Next to Leading Order, logarithms and factorization}
The previous discussion justifies partly the procedure employed to
compute observables in the CGC framework. But since it is based solely
on LO considerations, it says nothing about the logarithmic cutoff
dependence that may arise in loop corrections.  Indeed, in
eq.~(\ref{eq:Nexp}), it is implicitly assumed that the coefficients
$c_0,c_1,\cdots$ are of order one. But these coefficients in fact
contain logarithms of the cutoff, and should be more accurately written
as
\begin{align*}
  c_1
  &=&
  & &
  c_{10}\;
  &+&
  {\colord c_{11}\,\ln\Lambda^\pm}\;\;\;\;
  \\
  c_2
  &=&
  c_{20}\;
  &+&
  c_{21}\,\log\Lambda^\pm\;
  &+&
\underbrace{{\colord c_{22}\,\log^2\Lambda^\pm}}\;\;\;\;
\\
  &&
  &&
  &&\mbox{Leading Log terms}
\end{align*}
At each order, there can be powers of logarithms up to the number
of loops. The terms that maximize the number of logarithms are
called the {\sl Leading Log terms}, and are the most important.

On a more phenomenological side, the above LO result for the gluon
spectrum is insufficient because at this order the spectrum has no
dependence on the rapidity of the produced gluon. The two issues
are in fact closely related~: logarithms of the cutoff appear at
NLO, and in order to cancel them one must let the distributions of
sources $W_1[\rho_1]$ and $W_2[\rho_2]$ evolve according to the JIMWLK
equation. It is this evolution that gives the spectrum its rapidity
dependence.

\begin{figure}[htbp]
\begin{center}
\resizebox*{6cm}{!}{\includegraphics{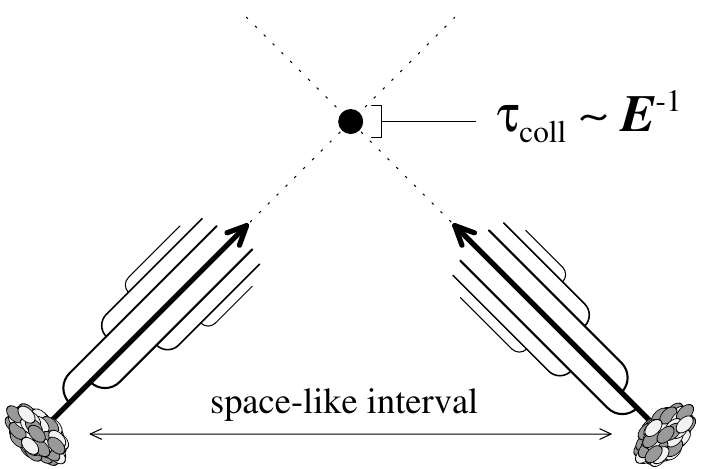}}
\end{center}
\caption{\label{fig:tau-eta}Causality argument explaining why the
  distributions $W[\rho]$ are universal in nucleus-nucleus
  collisions.}
\end{figure}
Let us first start with a qualitative argument to explain why it
should be possible to absorb the logarithms of the cutoff into
universal distributions $W[\rho]$ in the case of nucleus-nucleus
collisions (in fact the same distributions as those encountered in
DIS). This argument, based on causality, is illustrated in the figure
\ref{fig:tau-eta}, and goes as follows
\begin{itemize}
\item The duration of the collision is very short, and decreases as
  the inverse of the collision energy.
\item In contrast, the radiation of the soft gluons responsible for
  the logarithms takes a much longer time. Therefore, this radiation
  cannot happen during the collision itself -- these gluons must be
  emitted before the collision.
\item Before the collision, the two nuclei are not in causal
  contact. Therefore, what happens inside the first nucleus cannot be
  influenced by the second nucleus, and conversely. It should also be
  independent of the quantities that an observer may measure in the
  final state (provided that this measurement does not lead to
  discarding some class of events -- hence the special role of
  inclusive observables). Therefore, each nucleus should be described
  by a universal distribution $W[\rho]$.
\end{itemize}

\begin{figure}[htbp]
\begin{center}
\resizebox*{3.2cm}{!}{\includegraphics{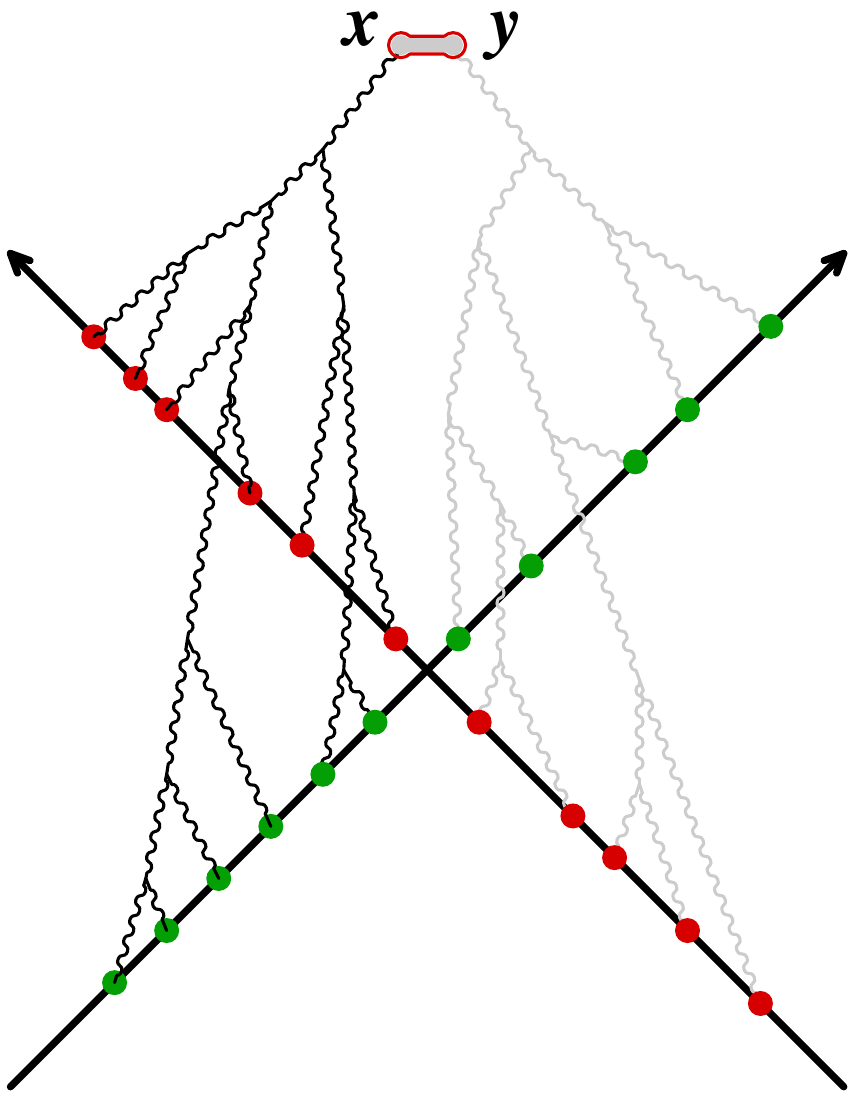}}\hfil
\resizebox*{3.2cm}{!}{\includegraphics{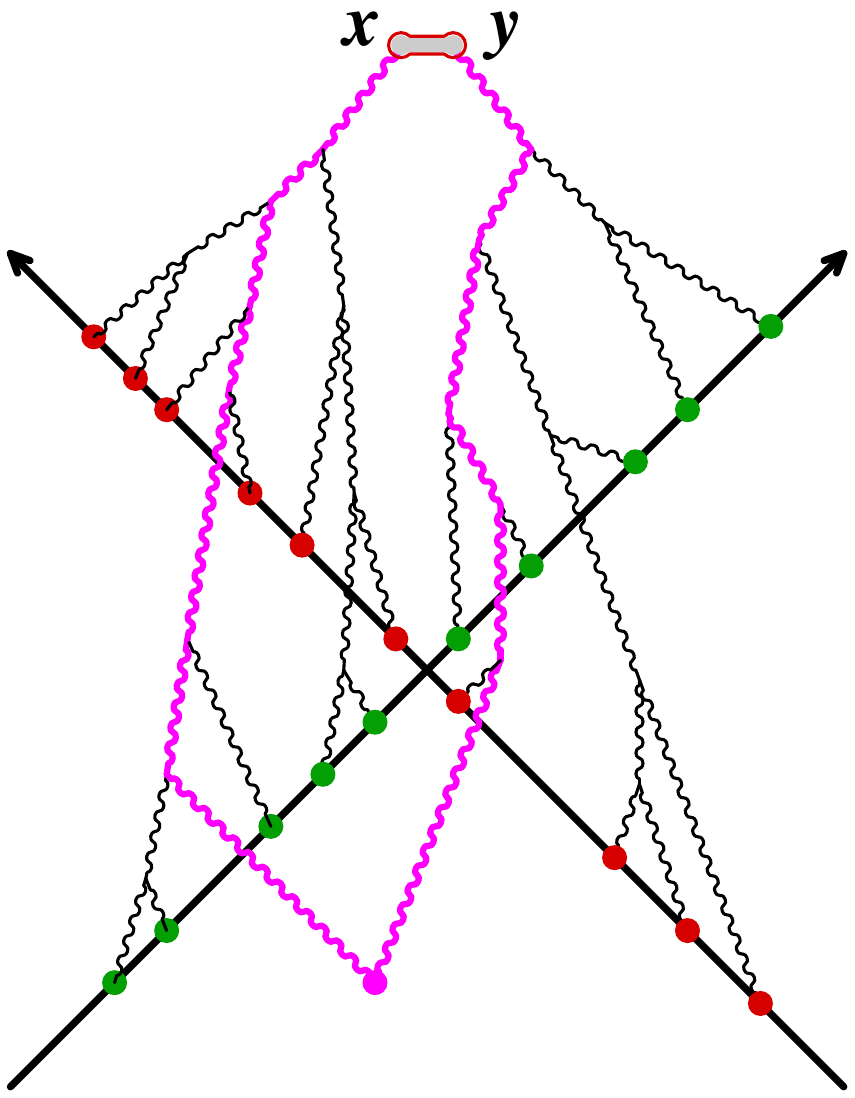}}\hfil
\resizebox*{3.2cm}{!}{\includegraphics{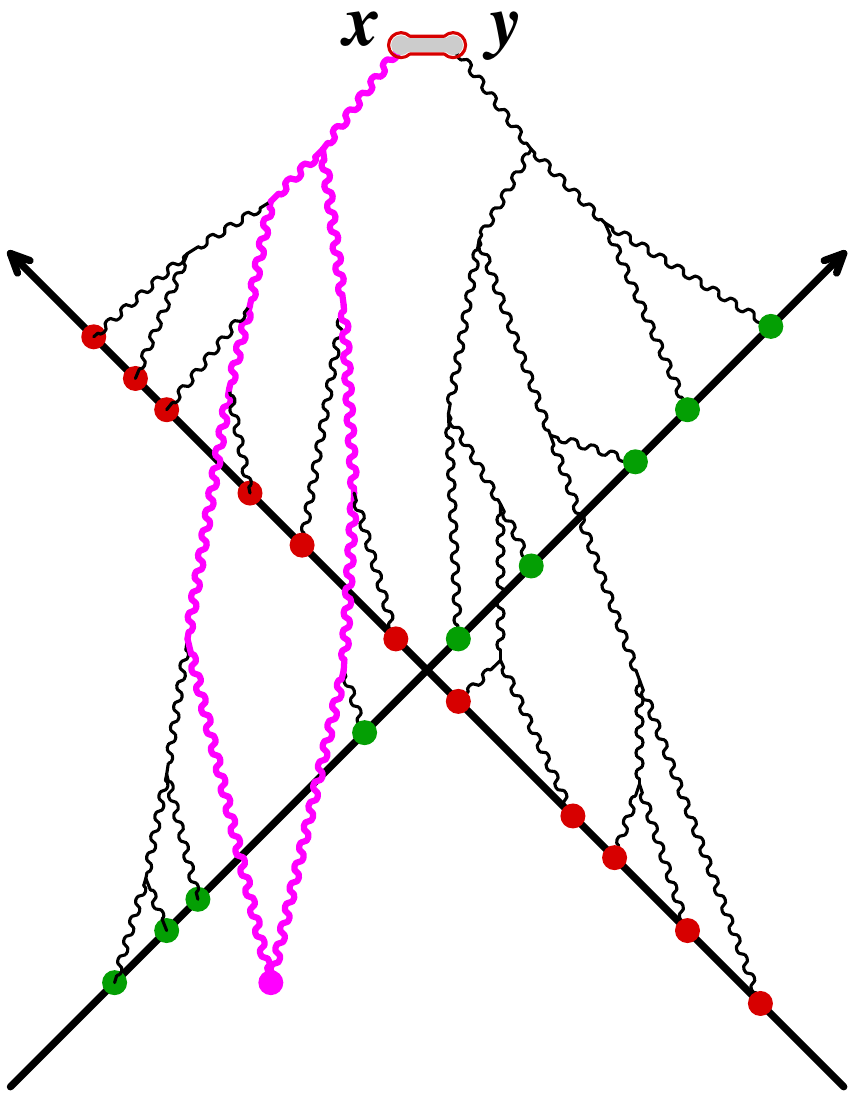}}
\end{center}
\caption{\label{fig:N1}Left: diagrams in the gluon spectrum at
  LO. Middle and Right: diagrams in the gluon spectrum at NLO. Here,
  the graphs are represented in coordinate space (with the two
  light-cones figuring the trajectories of the colliding nuclei). The
  green and red dots are the color sources carried by the two nuclei.}
\end{figure}
Now, let us consider the NLO corrections to the single inclusive gluon
spectrum, in order to see more precisely the structure of the
logarithms. There are two types of contributions at NLO, represented
in the middle and right graphs of the figure \ref{fig:N1} (for
comparison, we have represented in the left graph a typical
contribution at LO). The graph in the middle involves the two-point
Green's function $G_{+-}(x,y)$, that makes a first appearance at NLO,
while the graph on the right contains a one-loop correction to the
factor $A_+(x)$ (there is a third class of graphs, not shown in the
figure \ref{fig:N1}, that contain a one-loop correction to $A_-(y)$).

The calculation of the NLO correction to the gluon spectrum is rather
involved and will not be detailed here. Instead, we just sketch
the main steps in this study\cite{GelisLV3,GelisLV4,GelisLV5},

\begin{itemize}
\item[{\bf i.}] Express the single gluon spectrum at LO and NLO in
  terms of classical fields and small field fluctuations. A crucial
  property is that these objects all obey retarded boundary
  conditions.

\item[{\bf ii.}] Write the NLO terms as a perturbation of the initial
  value of the classical fields. If ${\cal A}^\mu(x)$ is a solution of
  the classical Yang-Mills equations, and $a^\mu(x)$ a small (in the
  sense that $a^\mu\ll {\cal A}^\mu$, so that the equation of motion
  of $a^\mu$ can be linearized) perturbation around this solution,
  they are formally related by
  \begin{equation}
    {\colord{\bs a}^\mu(x)} \equiv
    \int\limits_{\u\in\Sigma}\big[{\colorb\alpha}(\u)\,{\mathbbm T}_\u\big]
    \;{\cal A}^\mu(x)\; ,
    \label{eq:aTA}
  \end{equation}
  where ${\mathbbm T}_\u$ is the generator of the shifts of the initial
  value of ${\cal A}^\mu$ on some surface $\Sigma$ (roughly speaking,
  one can see ${\mathbbm T}_\u$ as a derivative with respect to the
  initial value of the field at the point $\u\in\Sigma$).  Thanks to
  this identity, one can obtain the following formula for the gluon
  spectrum at NLO
  \begin{equation}
    \left.\frac{dN_1}{d^3\vec\p}\right|_{_{\rm NLO}}
    =
    \Big[
    \frac{1}{2}\int\limits_{_{\vec\u,\vec\v\in\Sigma}}
    {\colord{\cal G}(\vec\u,\vec\v)}\,{\mathbbm T}_\u{\mathbbm T}_\v
    +\int\limits_{_{\vec\u\in\Sigma}}
    {\colord{\bs\beta}(\vec\u)}\,{\mathbbm T}_\u
    \Big]\;
    \left.\frac{dN_1}{d^3\vec\p}\right|_{_{\rm LO}}\; ,
    \label{eq:N1-LO-NLO}
  \end{equation}
  where ${\bs\beta}$ and ${\cal G}$ are one-point and two-point
  functions that can be computed analytically. The main interest of
  this formula is that it provides a {\sl factorization in time} of the NLO
  spectrum: the operator inside the square brackets depends only on
  the fields under the surface $\Sigma$, while the factor
  $dN_1/d^3\p|_{_{\rm LO}}$ on which it acts depends only on what
  happens above $\Sigma$.  Therefore, by choosing appropriately the
  surface $\Sigma$, this formula separates what happens in the two
  nuclei before the collision from the collision itself. Note also
  that in this formula, the first factor can be calculated
  analytically, while the second cannot because it involves solutions
  of the classical Yang-Mills equations that are not known
  analytically.

  \begin{figure}[htbp]
    \begin{center}
      \resizebox*{6cm}{!}{\includegraphics{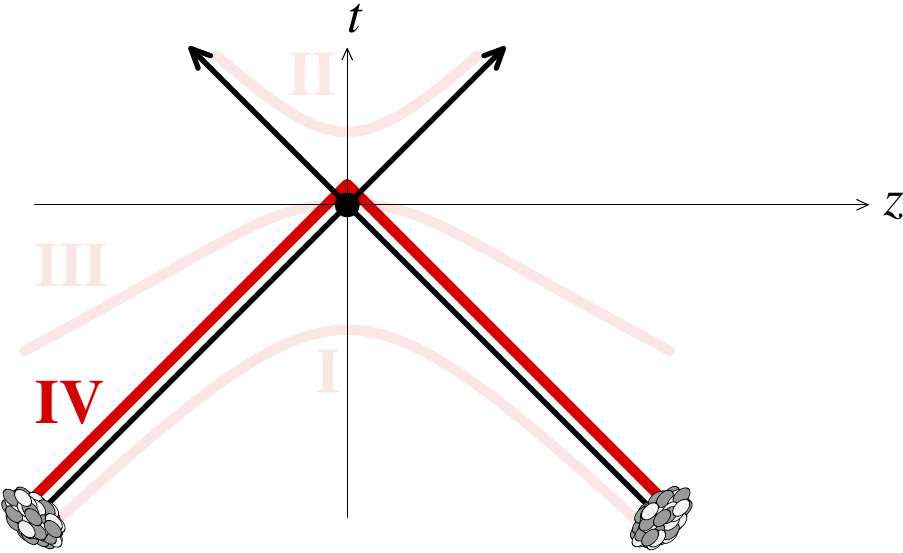}}
    \end{center}
    \caption{\label{fig:sigma}Choice of the surface $\Sigma$ to
      extract the initial state logarithms.}
  \end{figure}
\item[{\bf iii.}] Choose $\Sigma$ as in the figure
  \ref{fig:sigma}. When $\vec\u,\vec\v$ are both on the $t=z$ branch
  of the light-cone, one has~:
  \begin{equation}
    \frac{1}{2}\int\limits_{_{\vec\u,\vec\v\in\Sigma}}
    {\colord{\cal G}(\vec\u,\vec\v)}\,{\mathbbm T}_\u{\mathbbm T}_\v
    +\int\limits_{_{\vec\u\Sigma}}
    {\colord{\bs\beta}(\vec\u)}\,{\mathbbm T}_\u
    =\log{{\colorb\Lambda^+}}\,\times\,{\colorb{\cal H}_1}
    \;+\;\mbox{finite terms}\; ,
  \end{equation}
  where ${\cal H}_1$ is the JIMWLK Hamiltonian of the nucleus that
  moves in the $+z$ direction. If the two points $\u,\v$ are on the
  other branch of $\Sigma$, we obtain a logarithm of $\Lambda^-$,
  whose coefficient is the JIMWLK Hamiltonian ${\cal H}_2$ of the
  other nucleus. If the points $\u,\v$ are on different branches of
  the light-cone, there is no logarithm\footnote{This property is
    crucial~: if this were not true, there would be a mixing of the
    sources of the two nuclei in the coefficients of the logarithms,
    preventing their factorization.}. Therefore, at leading log
  accuracy, we have
  \begin{equation}
    \left.\frac{dN_1}{d^3\vec\p}\right|_{_{\rm NLO}}
    \;\empile{=}\over{\scriptsize\mbox{Leading Log}}\;
    \Big[
    \log\left(\Lambda^+\right){\colorb{\cal H}_1}
    +
    \log\left(\Lambda^-\right){\colorb{\cal H}_2}
    \Big]\;
    \left.\frac{dN_1}{d^3\vec\p}\right|_{_{\rm LO}}\; .
  \end{equation}

\item[{\bf iv.}] Like in the case of DIS, the logarithms can be hidden
  by integrating over the sources $\rho_1$ and $\rho_2$, with
  distributions $W_1[\rho_1]$ and $W_2[\rho_2]$ that obey the JIMWLK
  equation, with the JIMWLK Hamiltonian of the corresponding
  nucleus. To see this, one simply uses the self-adjointness of the
  JIMWLK Hamiltonian in order to transpose its action from the
  observable on the distributions $W[\rho]$. The final formula at
  Leading Log for the gluon spectrum is
  \begin{equation}
    \frac{dN_1}{d^3\vec\p}
    \;\empile{=}\over{\scriptsize\mbox{Leading Log}}\;
    \int 
    \big[D{\colora\rho_{_1}}\,D{\colorb\rho_{_2}}\big]
    \;
    {\colora W_1\big[\rho_{_1}\big]}\;
    {\colorb W_2\big[\rho_{_2}\big]}
    \;
    \underbrace{\left.\frac{dN_1}{d^3\vec\p}\right|_{_{\rm LO}}}_{\rm fixed\ \rho_{1,2}}
    \; .
    \label{eq:N1fact}
  \end{equation}
\end{itemize}
The factorization formula (\ref{eq:N1fact}) ends the justification of
the procedure employed in the literature in order to evaluate the
gluon spectrum in the CGC framework. Similar factorized formulas also
exist for $n$-gluon inclusive spectra, or for observables like the
energy-momentum tensor. In each of these extensions, it is the same
distributions $W[\rho]$ that enter in the formula, therefore providing
examples of their universality. In order to extend this factorization
to other observables, the central step is to prove that
eq.~(\ref{eq:N1-LO-NLO}) is true for the observable under
consideration -- with the same operator in the square brackets.

\section{Glasma fields, Long range rapidity correlations}
\label{sec:pheno}
\subsection{Glasma fields}
\begin{figure}[htbp]
\begin{center}
\resizebox*{5cm}{!}{\includegraphics{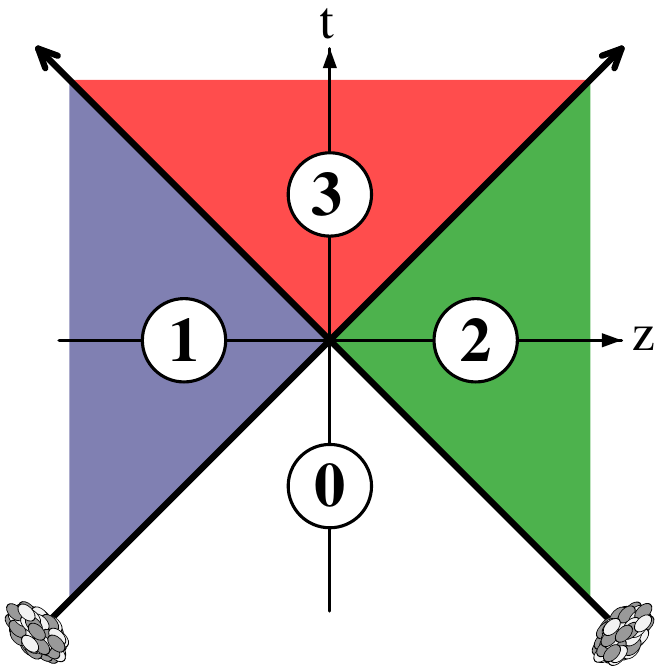}}\hfil
\resizebox*{6cm}{!}{\includegraphics{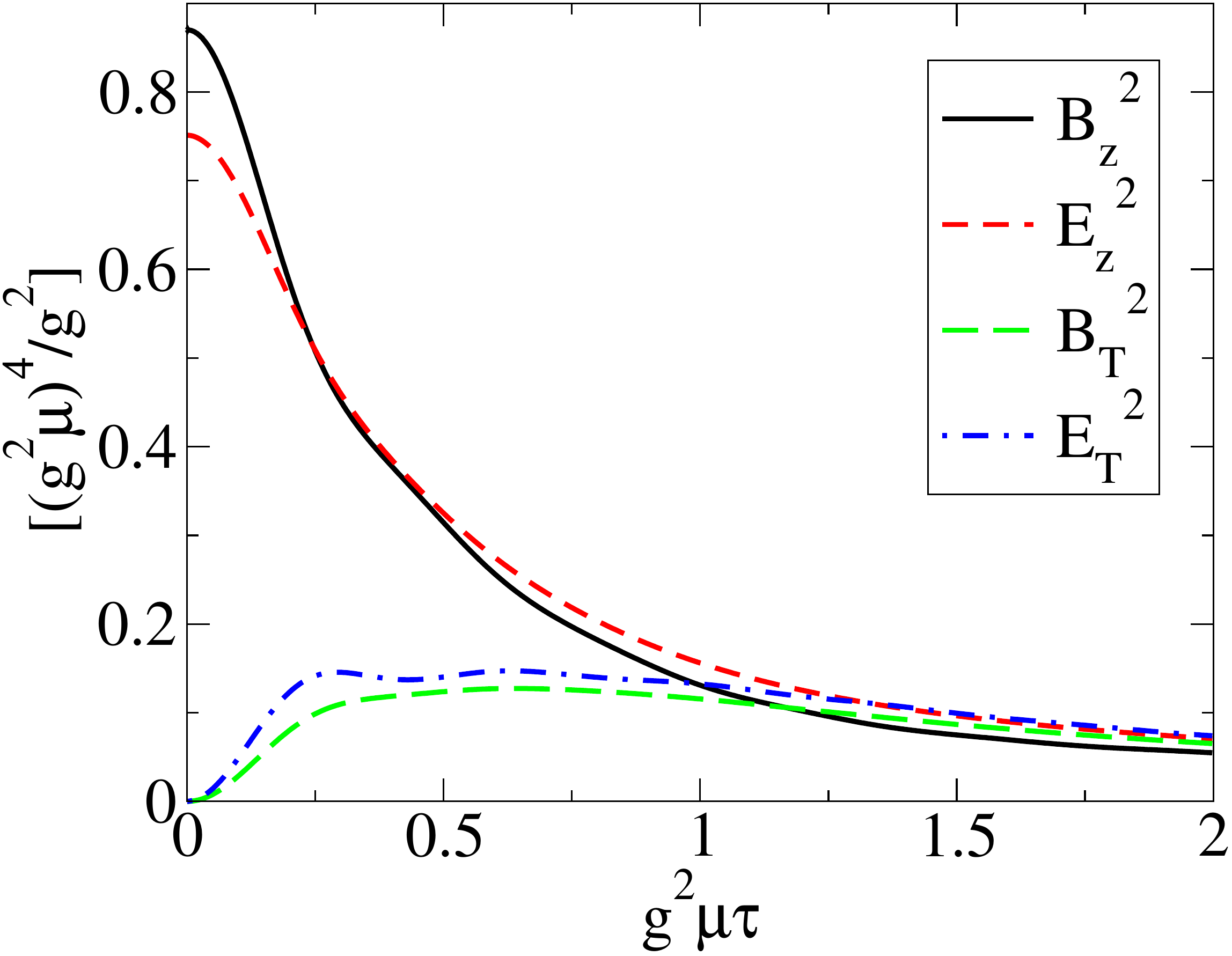}}
\end{center}
\caption{\label{fig:glasma}Right: transverse and longitudinal
  components of the chromo-electric and chromo-magnetic fields in the
  region three (see the left part of the figure), obtained by solving
  numerically the classical Yang-Mills equations
  (\ref{eq:YM1})\cite{LappiM1}.}
\end{figure}
The factorization formula~(\ref{eq:N1fact}) indicates that in
order to compute the single inclusive spectrum (or any other inclusive
quantity, for which the same factorization holds) it is sufficient to
compute the gluon spectrum in a classical field obtained by solving
the Yang-Mills equation (\ref{eq:YM1}), and then to average over all
the configurations of the color sources $\rho_{1,2}$. This classical
field therefore plays a crucial role in the CGC description of heavy
ion collisions, and it is worth spending some time discussing its
properties.

First of all, thanks to causality, one can divide space-time in four
regions (see the left part of the figure \ref{fig:glasma}). The region
labeled zero is completely trivial: an observer in this region sees
only the vacuum, and in this region the classical field is zero.
Observers in the regions one or two see only one of the nuclei,
but not the collision itself. In these regions, the solution of the
classical Yang-Mills equations is also very simple. In light-cone
gauge, it reads\cite{Kovch1},
\begin{equation}
{\cal A}^\pm=0\quad,\quad {\cal A}^i_{1,2} = \frac{i}{g}U_{1,2}^\dagger\partial^i U_{1,2}\; ,
\end{equation}
where $U_{1,2}$ is a Wilson line constructed with the source $\rho_1$
or $\rho_2$ respectively. In the regions one and two, the chromo-electric
and chromo-magnetic fields are transverse to the collision axis.  Note
also that, since these fields are pure gauges, there is no field
strength in the regions one and two. In other words, no energy is
deposited in these regions\footnote{This is of course expected, since
  all we have in these regions is a single (stable) nucleus moving at
  a constant speed.}.

The most interesting region is the third one, because it causally
connected to two nuclei. This region is the locus of all the events
that happen after the collision. The gauge fields can be obtained
analytically at $\tau=0^+$~\cite{KovneMW2}, but one has to resort to
numerical methods beyond that. The problem is best formulated in the
Fock-Schwinger gauge\footnote{In this gauge, the constraint of
  covariant current conservation $[{\cal D}_\mu,J^\mu]=0$ becomes
  trivial.}\cite{KovneMW1,KovneMW2},
\begin{equation}
x^+ {\colord {\cal A}^-}+x^- {\colord {\cal A}^+} = 0\; ,
\end{equation}
and by exploiting the invariance under longitudinal boosts in
collisions at very high energy. If we use the above gauge condition to
parameterize the fields ${\cal A}^\pm$ as ${\cal A}^\pm \equiv \pm
x^{\pm}\beta$, then the function $\beta$ and the transverse fields
${\cal A}^i$ are independent of the rapidity $\eta$. Solving
numerically the Yang-Mills equations in the forward light-cone leads
to the chromo-electric and chromo-magnetic fields shown in the right
plot of the figure \ref{fig:glasma}. At $\tau=0^+$, i.e. just after
the collision, these fields have vanishing transverse components: the
field lines are all parallel to the collision axis, and form elongated
tubular structures. Later on, as time increases, the transverse
components of the fields become comparable to the longitudinal
components. The {\sl glasma}\footnote{The word ``glasma'' is a
  contraction of glass (from colored glass condensate) and plasma
  (from quark-gluon plasma).}  designates these strong color fields
that populate the system at early times\cite{LappiM1}.

In the longitudinal direction, the correlation length of the fields is
infinite at leading order (since the system is boost invariant). When
the leading log corrections are resummed, the distributions of sources
evolve with rapidity, thus breaking the boost invariance. However,
since the resummed terms are powers of $\alpha_s \Delta \eta$, it
takes at least a rapidity shift $\Delta\eta\sim \alpha_s^{-1}$ in
order to see an appreciable variation of the sources. Therefore, even
after summing the leading logs, the glasma fields remain coherent over
rapidity intervals of order $\alpha_s^{-1}$. In the transverse
direction, the correlation length of the glasma fields is controlled
by the saturation momentum\cite{IancuIM2}, and is therefore of order
$Q_s^{-1}$.

The fact that the field lines are parallel to the collision axis
immediately after the collision leads to a very peculiar form for the
energy-momentum tensor,
\begin{equation}
  T^{\mu\nu}
  ={\rm diag}\,(\epsilon,\epsilon,\epsilon,{\colorb -\epsilon})
\; .
\end{equation}
The most striking feature is that the longitudinal pressure is
negative, which is reminiscent of strings stretching in the
longitudinal direction. As we shall see in the section
\ref{sec:glasma}, this result is not the end of the story though, and
further resummations are necessary due to the presence of
instabilities in the solutions of the classical Yang-Mills equations.

\subsection{Long range rapidity correlations}
The correlation properties of the glasma fields have been proposed as
the source of the long range rapidity correlations observed among
pairs of hadrons in heavy ion
collisions\cite{Daugh1,Abelea1,Adamsa4,Adamsa5,Wang1,AdareA1,Wosie1,CMS1}.
\begin{figure}[htbp]
\begin{center}
\resizebox*{5.5cm}{!}{\includegraphics{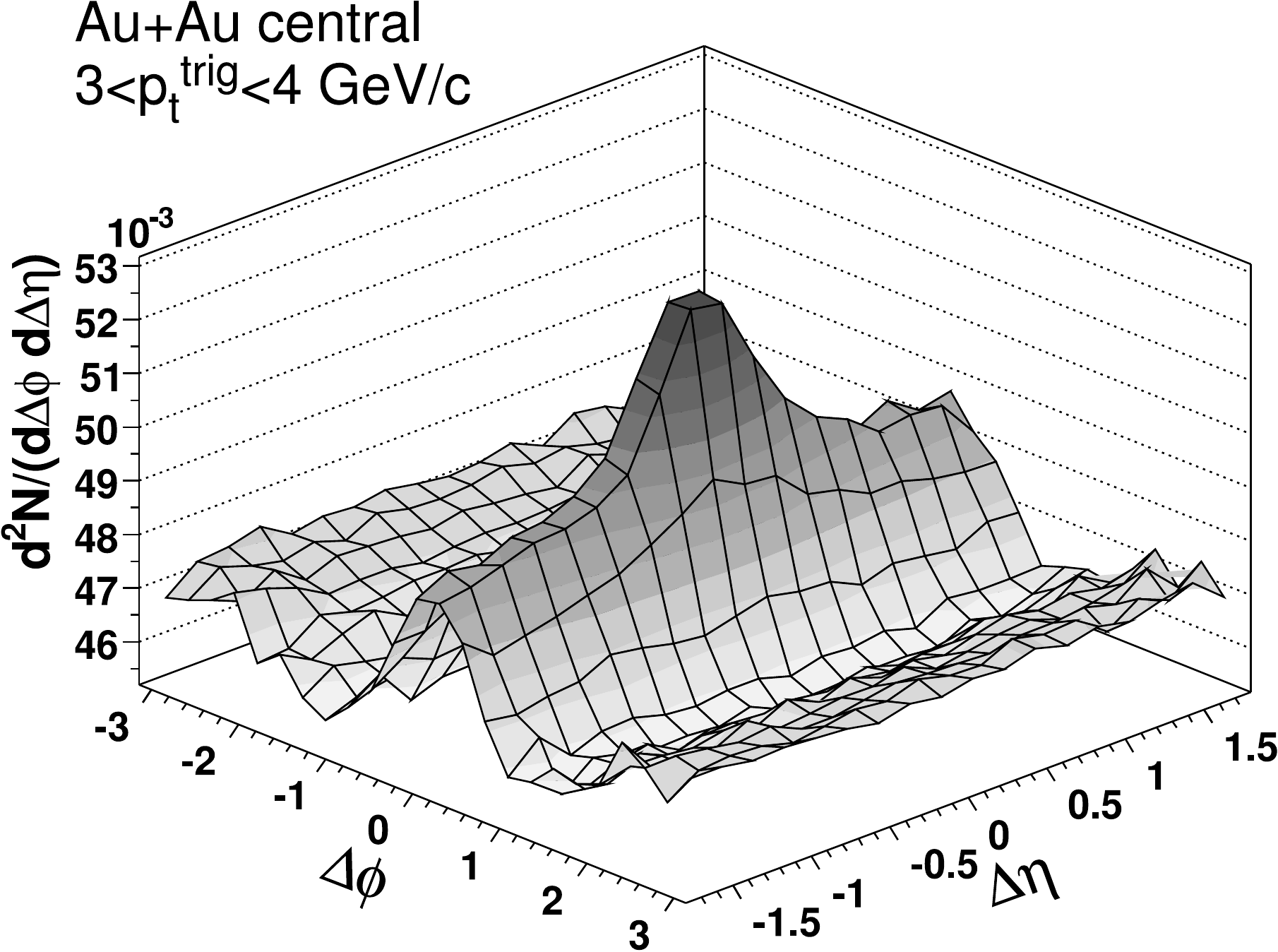}}\hfill
\resizebox*{7cm}{!}{\includegraphics{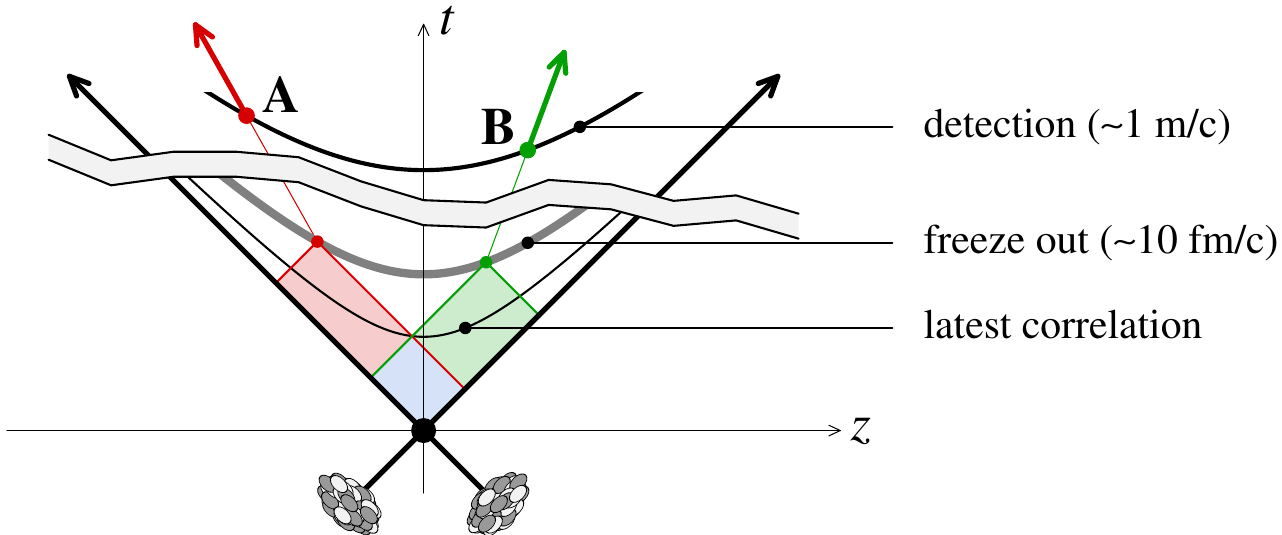}}
\end{center}
\caption{\label{fig:ridge} Left: measured two-hadron correlation, as a
  function of the rapidity difference and azimuthal angle difference
  between the two hadrons\cite{Abelea1}. Right: causal
  structure of the two-particle correlations.}
\end{figure}
The data is shown in the left part of the figure \ref{fig:ridge}. It
displays several features: a central peak (centered at
$\Delta\eta=\Delta\varphi=0$) that can be interpreted as due to
collinear fragmentation of a fast particle, and a {\sl ridge}-like
structure, narrow in $\Delta\varphi$ and very elongated in
$\Delta\eta$. The rest of the discussion will focus on the latter.

Before trying to interpret this correlation, one can reach a very
general conclusion, based on causality, regarding the nature of the
phenomena that may produce it. This is illustrated in the right part
of the figure \ref{fig:ridge}. Consider two particles A and B, and
assume that they are correlated. They are detected long after the
collision (compared to the strong interaction timescales), and their
last interaction occurred on the {\sl freeze-out surface}, at a proper
time of order $\tau_f\approx 10~$fm/c. Between their last interaction
and their detection, they traveled on straight lines, with an angle
determined by their rapidity. From the point at which they crossed the
freeze-out surface, draw a light-cone pointing in the past direction:
these light-cones (in red and green in the figure) are the locus of
the events that may have influenced A or B respectively. Any event
outside these light-cones cannot possibly have any influence, by
causality. A correlation between A and B means that some event had an
influence on both A and B; therefore this event must lie in the
overlap of the two light-cones described above, i.e. in the region in
blue in the figure.  From this figure, we see that this overlap region
extends only to a maximal time,
\begin{equation}
\tau_{\rm correlation}\;\le\;\tau_f\;\;e^{-|\Delta y|/2}\; .
\end{equation}
This upper bound depends exponentially on the rapidity separation
between the two particles, and therefore becomes very small for long
range correlations in rapidity. This simple argument tells
us that, regardless of their precise nature, the effects that produce
these correlations must happen very shortly after the
collision\footnote{This causality argument is very similar to the
  reason why, in cosmology, the observation of angular correlations in
  the Cosmic Microwave Background provides informations about the
  physics that prevailed long before the CMB was emitted.}, or
pre-exist in the wavefunctions of the colliding nuclei.

The leading log factorization formula that generalizes
eq.~(\ref{eq:N1fact}) to the two-particle correlation
is\cite{GelisLV4,GelisLV5}
\begin{equation}
  \frac{dN_2}{d^3\vec\p_{1}d^3\vec\p_{2}}
  \;\empile{=}\over{\scriptsize\mbox{Leading Log}}
  \int 
  \big[D{\colora\rho_{_1}}\,D{\colorb\rho_{_2}}\big]
  \;
  {\colora W_1\big[\rho_{_1}\big]}\;
  {\colorb W_2\big[\rho_{_2}\big]}
  \;
  \left.\frac{dN_1}{d^3\vec\p_{1}}\;
    \frac{dN_1}{d^3\vec\p_{2}}\right|_{_{\rm LO}}\; .
\label{eq:N2fact}
\end{equation}
Note that the integrand is the product of the single inclusive spectra
for gluons of momenta $\p_1$ and $\p_2$ respectively, a consequence of
eq.~(\ref{eq:Nn}). This indicates that, at this order, all the
correlations between the two particles result from the averaging over
the sources $\rho_{1,2}$, and therefore must come from the
distributions $W_1[\rho_1]$ and $W_2[\rho_2]$ themselves.

\begin{figure}[htbp]
\begin{center}
\resizebox*{5cm}{!}{\includegraphics{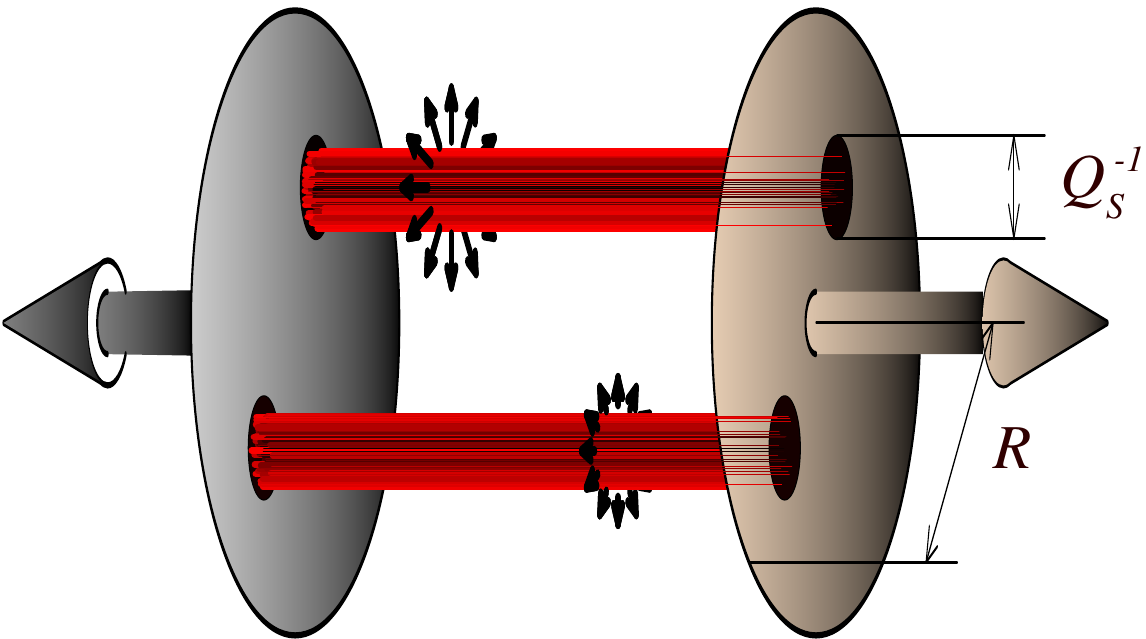}}\hfil
\resizebox*{5cm}{!}{\includegraphics{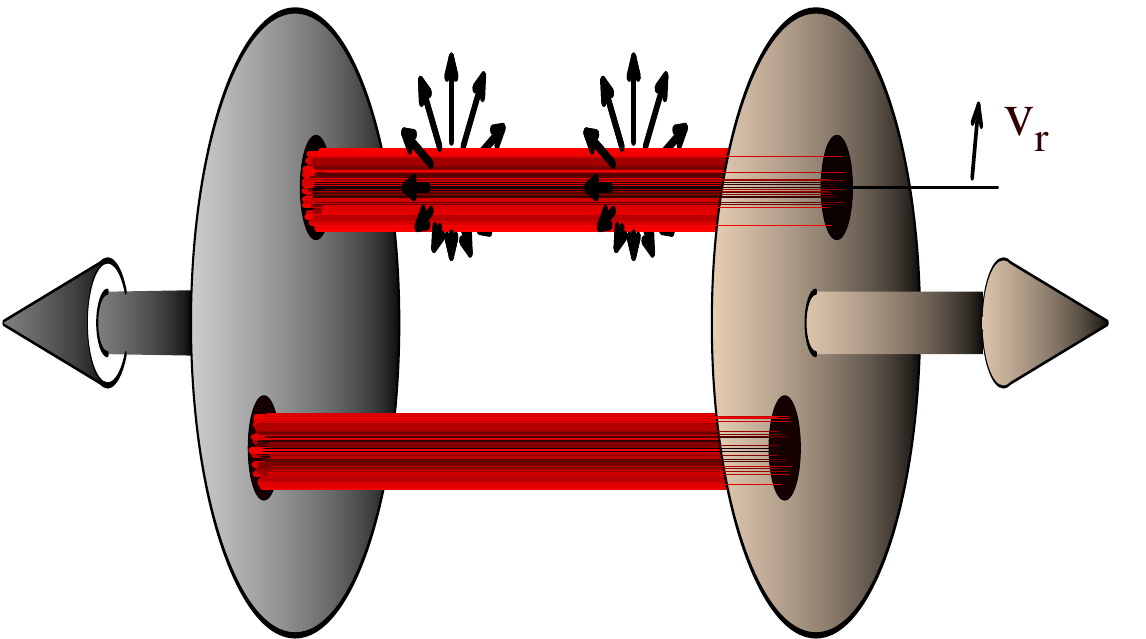}}
\end{center}
\caption{\label{fig:tubes}Left: particle emission from two distinct
  glasma flux tubes. Right: effect of radial flow on the angular
  distribution of the emitted particles.}
\end{figure}

A natural candidate for the formation of long range correlations in
rapidity is the glasma color fields\cite{DumitGMV1,LappiSV1}, since they are
coherent over rapidity intervals that extend to at least $\Delta
\eta\sim\alpha_s^{-1}$ (in turn, the approximate boost invariance of
the glasma fields is a consequence of the slow JIMWLK evolution of the
distributions $W_{1,2}$, which is an effect of order $\alpha_s$).
Note that the peculiar structure of the glasma field lines, that form
longitudinal tubes at early times, is not essential to this argument.
Another important parameter in the argument is the transverse size
over which the glasma fields are coherent; this size is of order
$Q_s^{-1}$. Particles emitted from two distinct glasma flux tubes
(i.e. separated by more than $Q_s^{-1}$ in the transverse direction)
are not correlated (see the left figure \ref{fig:tubes}) since they
are produced by incoherent fields.
The transverse size of the tubes controls the strength of the
correlation. Indeed, two particles are correlated only if they come
from the same tube. The probability for that is of order $(Q_s
R)^{-2}$, where $R$ is the typical size of the transverse overlap in
the collision (it coincides with the radius of the nuclei in a
collision at zero impact parameter).

\begin{figure}[htbp]
\begin{center}
\resizebox*{6cm}{!}{\includegraphics{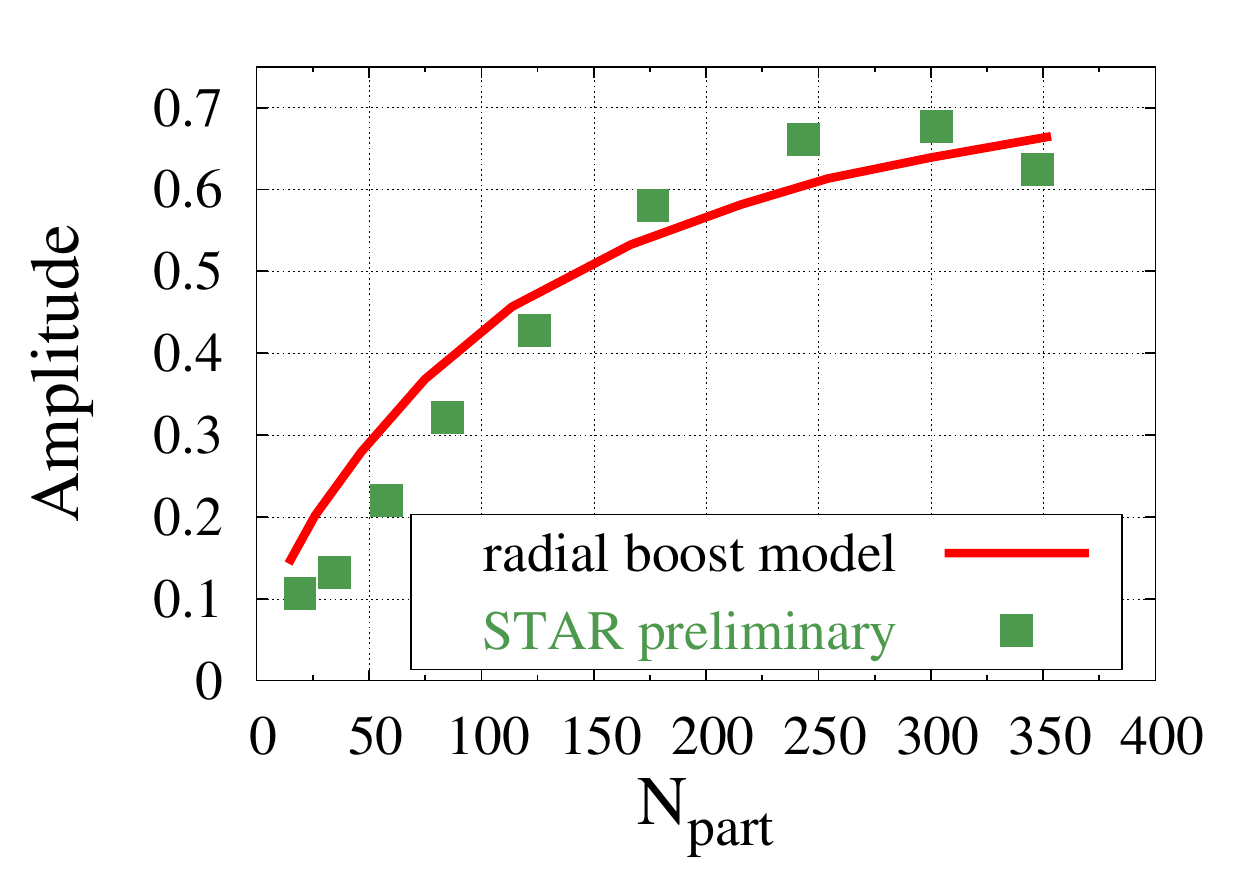}}\hfil
\resizebox*{6cm}{!}{\includegraphics{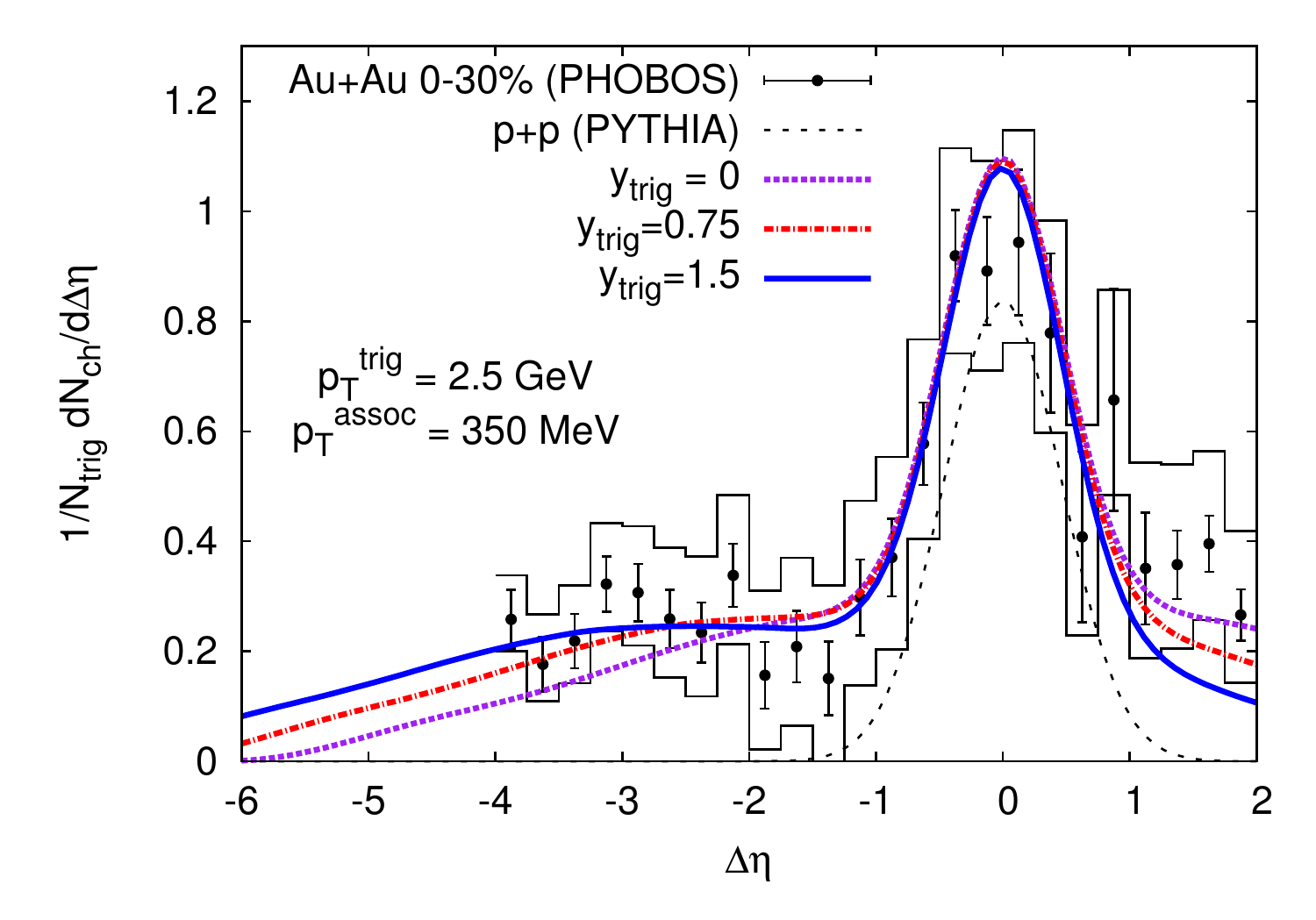}}
\end{center}
\caption{\label{fig:ridge1}Left: amplitude of the peak in
  $\Delta\varphi$, compared to data. Right: estimate of the rapidity
  dependence of the correlation in the CGC (note: the central peak is
  due to jet fragmentation and has been superimposed by hand) and
  comparison with data from PHOBOS\cite{Alvera1}.}
\end{figure}

However, the two-gluon correlation one gets from the glasma fields at
early times has no angular dependence: the two particles are not
correlated in azimuth. Note that there is no causality argument that
tells that the azimuthal correlation must be created early; it can be
produced later on by {\sl radial flow}\cite{Volos1,PruneGV1}. The
transverse pressure of the matter created in the collision sets it in
motion radially, reaching speeds that are a sizable fraction of the
speed of light. This collective radial motion transforms an
azimuthally flat two-particle spectrum in a spectrum that has a peak
around $\Delta\varphi=0$ (see the right figure \ref{fig:ridge}). The
width and height of this peak depend on the velocity of the radial
flow. In the left plot of the figure \ref{fig:ridge1} is shown a crude
calculation of the amplitude of this peak\cite{DumitGMV1} (several
works\cite{Shury2,GavinMM1} have performed more detailed studies of
this effect), based on radial velocities extracted from the data
itself\cite{Kiyom1}.

From the rapidity dependence of the distributions $W_1[\rho_1]$ and
$W_2[\rho2]$, due to the JIMWLK evolution, it is possible to determine
the dependence of the two-gluon spectrum on the rapidities of the two
gluons, from eq.~(\ref{eq:N2fact}). Using some shortcuts (such as
solving the mean field BK equation\footnote{The JIMWLK equation is
  considerably harder to solve numerically. The main idea is a
  reformulation of the equation as a Langevin equation\cite{BlaizIW1},
  that allows a numerical study of the JIMWLK equation on the
  lattice\cite{RummuW1}. This study has shown that the
  Balitsky-Kovchegov equation is indeed a rather good approximation,
  at least as far as 2-point correlators are concerned. Recently, this
  direct numerical approach has started to find its way into more
  phenomenological applications\cite{Lappi7,DumitJLSV1}.}, rather than
the much more complicated JIMWLK equation), one obtains the rapidity
dependence\cite{DusliGLV1} shown in the right plot of the figure
\ref{fig:ridge1}.

\section{Glasma evolution: isotropization and thermalization}
\label{sec:glasma}

\subsection{Resummation of the leading secular terms}
As we have seen in the previous section, the early stages of heavy ion
collisions are dominated by strong color fields, the
glasma. Initially, these chromo-electric and chromo-magnetic fields
are purely longitudinal.  A consequence of this is that the
longitudinal pressure is negative (in fact, exactly the opposite of
the energy density at $\tau=0^+$). This is problematic in view of the
many successes of the hydrodynamical description of the evolution of
the matter produced in heavy ion
collisions\cite{Adamsa3,Adcoxa1,Arsena2,Backa2,HuoviR1,Romat1,Teane1,RomatR1}, because a large and
negative pressure poses problems if used as initial condition for
hydrodynamics.

However, there is another fact that suggests that the CGC description
of nucleus-nucleus collisions that we have described so far is still
incomplete. It has been observed in several studies that classical
solutions of the Yang-Mills equations suffer from
instabilities\cite{RomatV1,RomatV2,RomatV3,FukusG1,BiroGMT1,HeinzHLMM1,BolteMS1,FujiiI1,FujiiII1,KunihMOST1},
that make them extremely sensitive to their initial condition.
\begin{figure}[htbp]
\begin{center}
\resizebox*{6cm}{!}{\rotatebox{-90}{\includegraphics{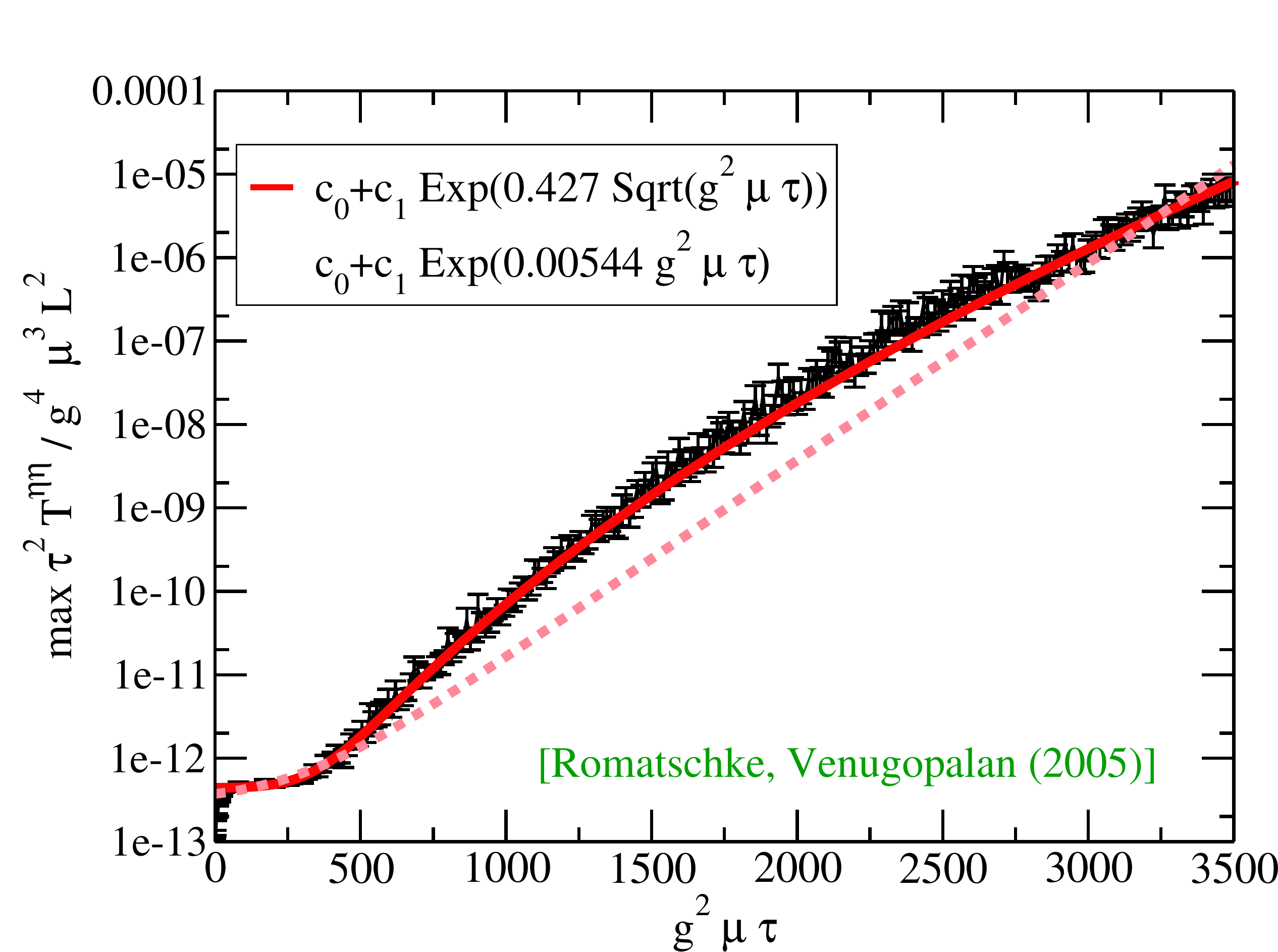}}}
\end{center}
\caption{\label{fig:insta}Growth of some instabilities in classical
  Yang-Mills equations\cite{RomatV1}.}
\end{figure}
The plot of the figure \ref{fig:insta} shows the growth of such an
unstable mode, when one disturbs a boost invariant classical solution
by a small rapidity dependent perturbation. These instabilities appear
to be related to the well known {\sl Weibel
  instability}\cite{Mrowc3,Mrowc4,RebhaRS1,RebhaRS2,MrowcRS1,RomatS1,RomatS2},
or {\sl filamentation instability}, in plasma physics. Many works have
already investigated the possible role of these instabilities in the
thermalization of the quark-gluon
plasma\cite{RebhaS1,RebhaSA1,ArnolLM1,ArnolLMY1,ArnolM1,ArnolM2,ArnolM3,ArnolMY4,DumitNS1,BodekR1,BergeGSS1,BergeSS2,KurkeM1,KurkeM2}.

In the CGC framework, these instabilities question the validity of the
power counting that was the very basis for the organization of the
expansion in power of $g^2$. Indeed, this power counting implicitly
assumes that if a classical field ${\cal A}^\mu$ is of order $g^{-1}$
and a small perturbation around it is initially of order $1$, then the
perturbation remains negligible compared to the background at all
times. The existence of instabilities invalidates this assumption.  It
has been argued that the instabilities grow exponentially\footnote{The
  square root inside the exponential is due to the longitudinal
  expansion of the system, that reduces the growth of the instability
  compared to a system that has a fixed volume.} as
\begin{equation}
a(\tau)\sim e^{\sqrt{\gamma\tau}}\; ,
\end{equation}
where the instability growth rate $\gamma$  is of the order of the
saturation momentum. This means that it will become comparable in
magnitude to the classical background field at a time
\begin{equation}
\tau_{\rm max}\sim \gamma^{-1}\,\log^2(g^{-1})\; .
\end{equation}
$\tau_{\rm max}$ is the time where the power counting rule in
eq.~(\ref{eq:PC}) completely breaks down, and the one-loop corrections
become as large as the leading order results. These terms, that are
formally suppressed by powers of the coupling $g$, but with prefactors
that grow with time, are called {\sl secular terms}.

In order to study the evolution of the system beyond this time, it is
necessary to perform a resummation of the secular terms that have the
fastest growth in time. A small amendment to the power counting rule
of eq.~(\ref{eq:PC}) is sufficient to track these terms,
\begin{eqnarray}
  \mbox{loop}\sim g^2\quad,\qquad {\mathbbm T}_\u \sim e^{\sqrt{\gamma\tau}}\; .
\end{eqnarray}
The new rule is to assign a factor $e^{\sqrt{\gamma\tau}}$ to each
occurrence of the operator ${\mathbbm T}$. The reason for this new
rule is eq.~(\ref{eq:aTA}), that shows that each power of
${\mathbbm T}$ generates a perturbation on top of the classical
background.

\begin{figure}[htbp]
\begin{center}
\resizebox*{3.5cm}{!}{\includegraphics{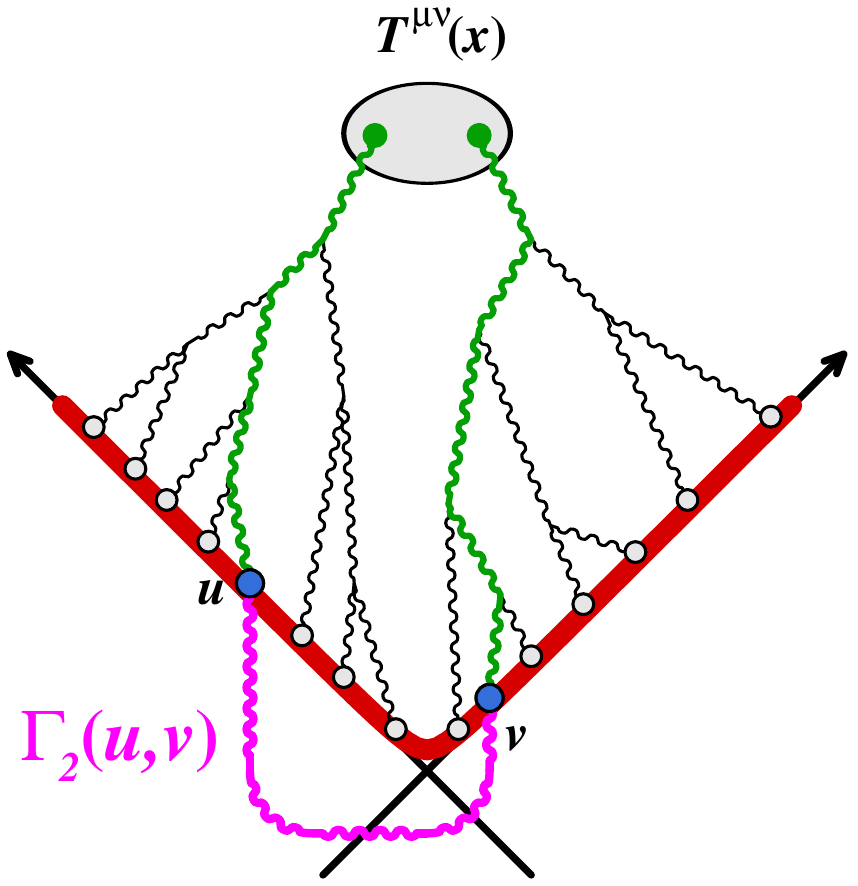}}\hfil
\resizebox*{3.5cm}{!}{\includegraphics{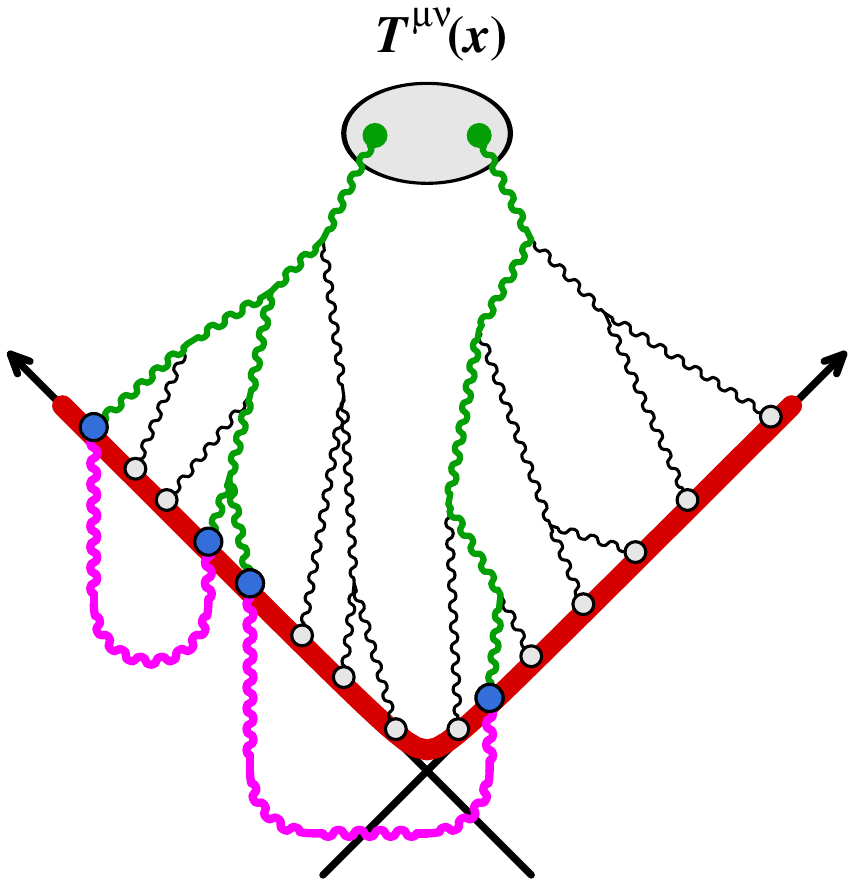}}\hfil
\resizebox*{3.5cm}{!}{\includegraphics{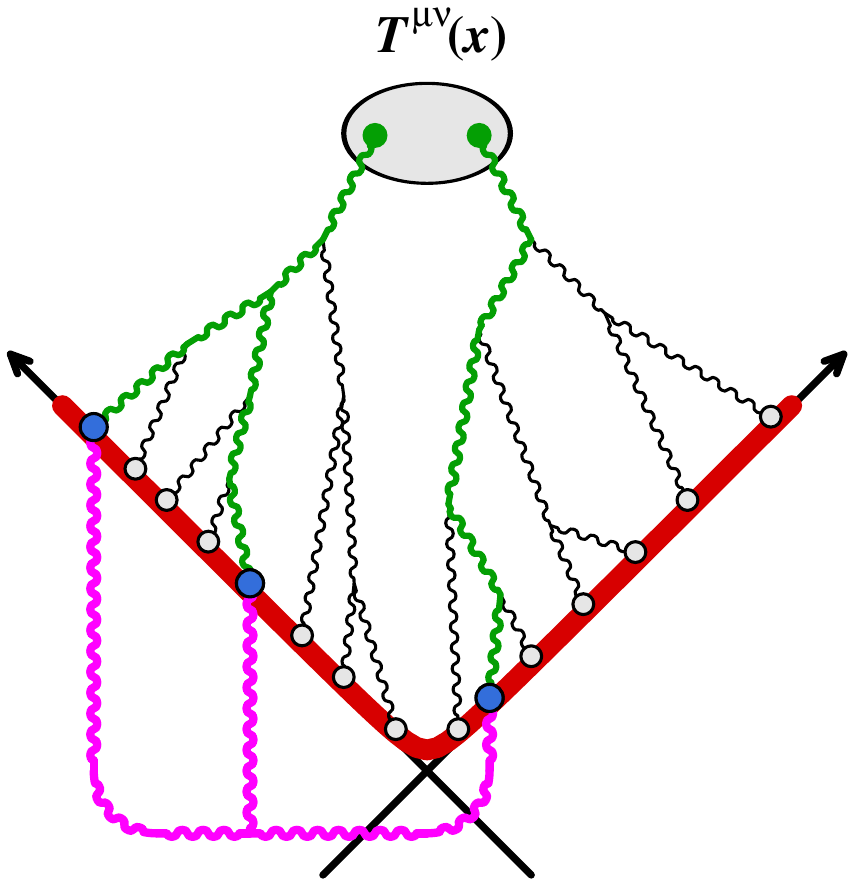}}
\end{center}
\caption{\label{fig:resum}Examples of one-loop and two-loop
  contributions to the energy momentum tensor.  The operators
  ${\mathbbm T}$ are represented as blue dots.}
\end{figure}
In order to illustrate these rules, we have represented in the figure
\ref{fig:resum} some one-loop and two-loop contributions to the energy
momentum tensor. With these rules, the one-loop graph on the left
would bring a factor
\begin{equation}
\big(g e^{\sqrt{\gamma\tau}}\big)^2\; ,
\end{equation}
and the two-loop graph in the middle a factor
\begin{equation}
\big(g e^{\sqrt{\gamma\tau}}\big)^4\; ,
\end{equation}
while the two-loop graph on the right gives only
\begin{equation}
g\big(g e^{\sqrt{\gamma\tau}}\big)^3\; .
\end{equation}
(These factors are all relative to the tree level result.)  At the
time $\tau_{\rm max}$, the first two contributions have the same order
of magnitude as the leading order result, while the third contribution
we have considered in this example is still suppressed.  It is easy to
resum all the terms that are leading at the time $\tau_{\rm max}$:
this corresponds to all the graphs that have the same structure as the
first two on the left of the figure \ref{fig:resum}, with an
arbitrary number of loops provided that these loops are not nested
below the light-cone. The sum of these graphs can be obtained by
exponentiating the operator in the square
brackets in eq.~(\ref{eq:N1-LO-NLO}). For a generic inclusive
observable, this reads
\begin{equation}
  {\cal O}_{\rm resum}
  =
  \exp\Big[
  \frac{1}{2}\int\limits_{_{\vec\u,\vec\v\in\Sigma}}
  {\colord{\cal G}(\vec\u,\vec\v)}\,{\mathbbm T}_\u{\mathbbm T}_\v
  +\int\limits_{_{\vec\u\in\Sigma}}
  {\colord{\bs\beta}(\vec\u)}\,{\mathbbm T}_\u
  \Big]\;
  {\cal O}_{_{\rm LO}}\; ,
  \label{eq:resum}
\end{equation}
By construction, this resummation contains the full LO and NLO
results, and a subset of all the higher order terms,
\begin{equation}
{\cal O}_{\rm resum}
  =
\underbrace{{\cal O}_{_{\rm LO}}+{\cal O}_{_{\rm NLO}}}_{\scriptsize\mbox{in full}}
+
\underbrace{{\cal O}_{_{\rm NNLO}}+\cdots}_{\scriptsize\mbox{partially}}
\end{equation}
Eq.~(\ref{eq:resum}) can be turned in a much more practical expression. By
noticing that the exponential of ${\mathbbm T}$ is a translation
operator,
\begin{equation}
    {\cal F}[{\colord{\cal A}_{\rm initial}}+{\colorc \alpha}]\equiv
\exp\Bigg[\int_{_{\vec\u\in{\Sigma}}}
\big[{\colorc\alpha}\cdot{\mathbbm T}\big]_\u\Bigg]\;\;{\cal F}[{\colord{\cal A}_{\rm initial}}]\; ,
\end{equation}
one can rewrite eq.~(\ref{eq:resum}) as\footnote{In order to prove
  this formula, we need also
  \begin{equation*}
    e^{\frac{{\colord\alpha}}{2}\partial_x^2}\;{\colorc f(x)}
  =
  \int\limits_{-\infty}^{+\infty}{\colora dz}\;
  \frac{e^{-{\colora z^2}/2{\colord\alpha}}}{\sqrt{2\pi\alpha}}\;
  {\colorc f(x+{\colora z})}\; .
  \end{equation*}}
\begin{equation}
{\cal O}_{\rm resum}
  =
\int[D{\colord\chi}]\,
\exp\Bigg[-\frac{1}{2}
\int\limits_{_{\u,\v}\in\Sigma}
{\colord\chi(\u)}\;{\colorc{\cal G}^{-1}(\u,\v)}\;{\colord\chi(\v)}\Bigg]\;
{\cal O}_{_{\rm LO}}[{\cal A}_{\rm init}+{\colord\chi}+{\colorb\bs\beta}]\; .
\label{eq:resum1}
\end{equation}
In words, the resummed observable is obtained by shifting the initial
value on $\Sigma$ (in practice a surface of constant proper time
$\tau_0$) by a constant shift ${\bs\beta}$ and by a fluctuating shift
$\chi$ that has a Gaussian distribution\footnote{One can check that
  this Gaussian fluctuation of the initial field amounts to promoting
  the purely classical state ${\cal A}_{\rm init}$ into a quantum
  mechanical coherent state centered at ${\cal A}_{\rm
    init}$. Coherent states are states that have the minimal extent in
  phase-space allowed by the uncertainty principle, and this extent is
  symmetrical between the coordinates and their conjugate
  momenta. Another equivalent point of view is that these fluctuations
  amount to filling each mode with $1/2$ particle, thereby reproducing
  the ground state of a quantum oscillator. }. The variance of this
Gaussian distribution can be calculated easily, and the functional
integration over $\chi$ in eq.~(\ref{eq:resum1}) can be evaluated by a
Monte-Carlo sampling. This resummation is sometimes described in the
literature as {\sl classical statistical field theory}. The same
Gaussian average was also obtained in different
approaches\cite{PolarS1,Son1,KhlebT1,MichaT1,FukusGM1}.  A similar
method has also been applied to cold atom physics, in problems
related to Bose-Einstein condensation\cite{Norri1,NorriBG1}.

In the rest of this section, we will present some numerical results
illustrating the effect of this resummation. Because its
implementation has not yet been completed for Yang-Mills
theory\cite{DusliGV1}, we have tried it on a much simpler
model\cite{DusliEGV1,EpelbG1,DusliEGV2} --a real scalar field with a
$\phi^4$ coupling--, that shares with QCD some important features:
\begin{itemize}
\item Scale invariance at the classical level in $3+1$ dimensions.
\item Instabilities in the classical equations of motion, due to
  parametric resonance\footnote{Parametric resonance has been studied
    extensively in other
    situations\cite{Son1,GreenKLS1,AllahBCM1,Frolo1,BergeS3}, especially
    in inflationary cosmology.}.
\end{itemize}
Many aspects of this resummation can be studied by considering a
system confined in a fixed volume, while other questions require to
consider a system that expands longitudinally, as in a high energy
collision.

\subsection{Evolution of a fixed volume system}
Consider first a system in a fixed volume. It is initialized
with a large background field that mimics the glasma field of heavy
ion collisions, to which we superimpose Gaussian fluctuations given by
eq.~(\ref{eq:resum1}). Then, each field configuration evolves
according to the classical equation of motion, and observables are
evaluated in its solution. A Monte-Carlo sampling is performed
in order to average over the Gaussian ensemble of fluctuations.

\begin{figure}[htbp]
\begin{center}
\resizebox*{6cm}{!}{\rotatebox{-90}{\includegraphics{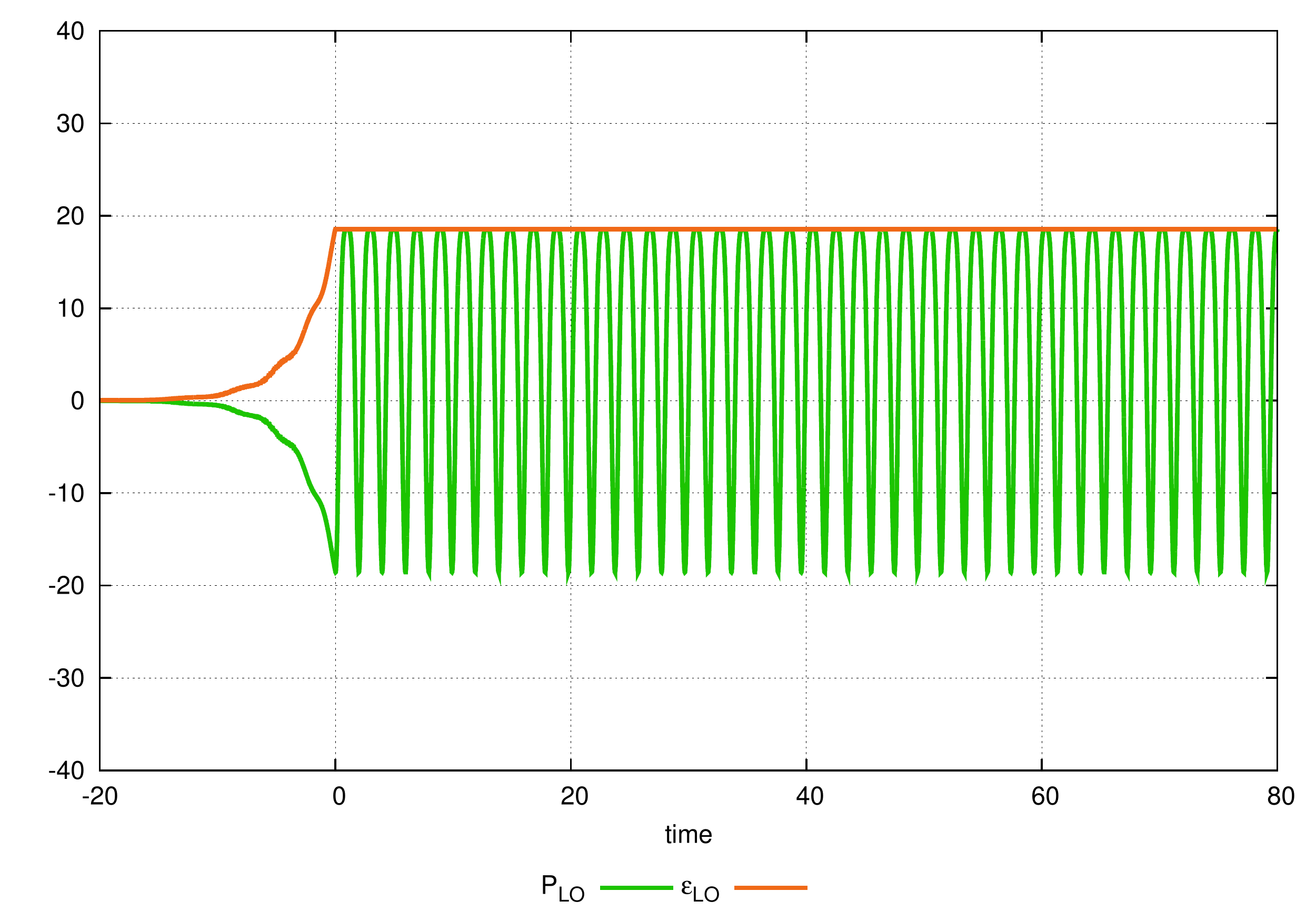}}}\hfil
\resizebox*{6cm}{!}{\rotatebox{-90}{\includegraphics{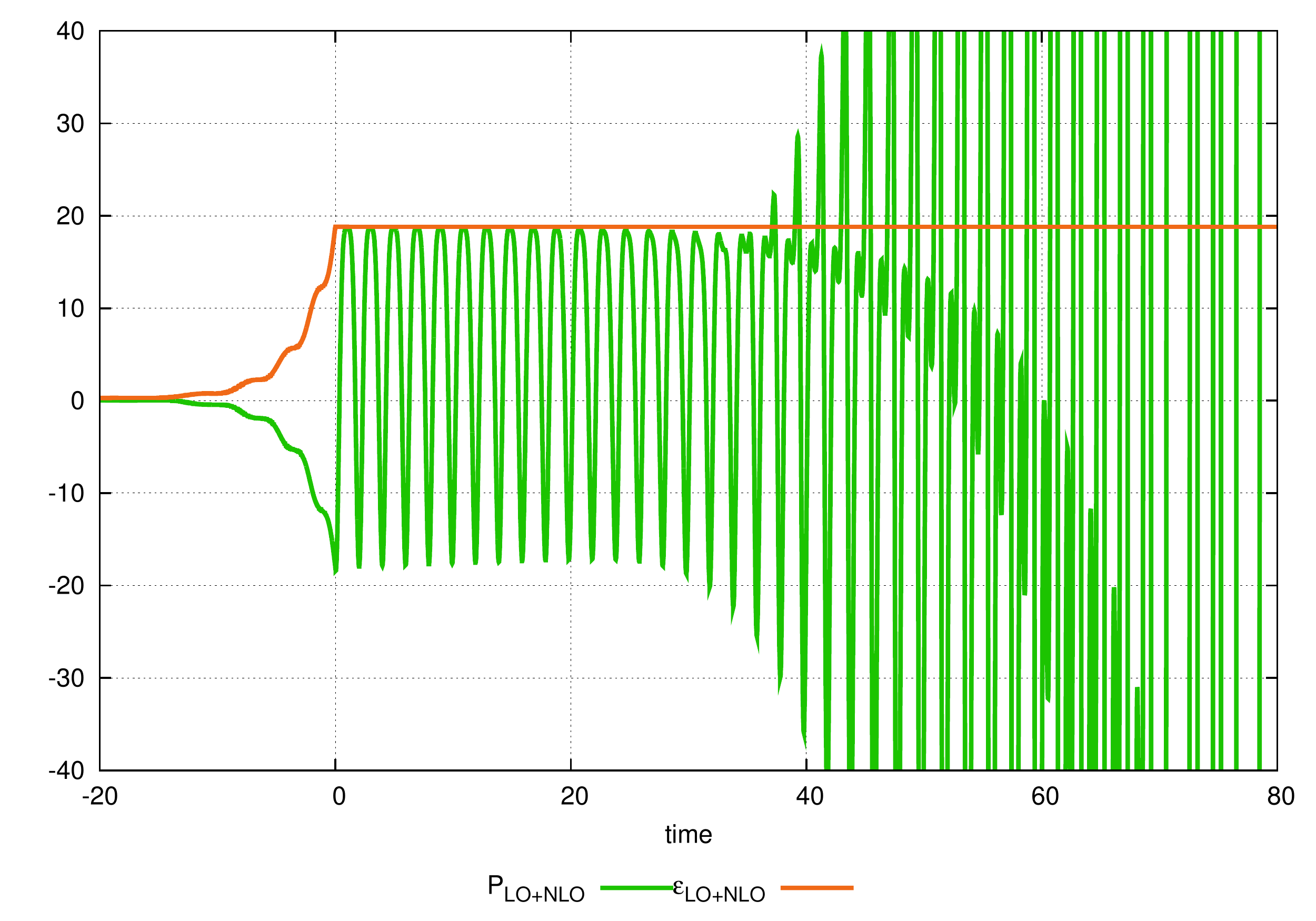}}}
\end{center}
\caption{\label{fig:phi4}Energy density and pressure in $\phi^4$
  scalar theory. Left: tree level result. Right: tree level + one-loop result.}
\end{figure}

Firstly, it is interesting to look at observables in a fixed loop
order expansion, in order to highlight the pathologies caused by the
presence of instabilities in the theory. This is illustrated in the
figure \ref{fig:phi4}, where we show the tree level (left) and
one-loop (right) results for the energy density and the pressure.
(The evolution really starts at $t=0$. At $t<0$, the system is
initialized by a source that drives the background field to a large
non-zero value.) One sees that the pressure oscillates in time -- this
means that there is no equation of state, since that would require a
one-to-one correspondence between the pressure and the energy
density. Even more problematic is the fact that at NLO, the
oscillations of the pressure grow exponentially in time, making the
predictions of the expansion in $g^2$ completely unreliable after a
finite time.

\begin{figure}[htbp]
\begin{center}
\resizebox*{7cm}{!}{\rotatebox{-90}{\includegraphics{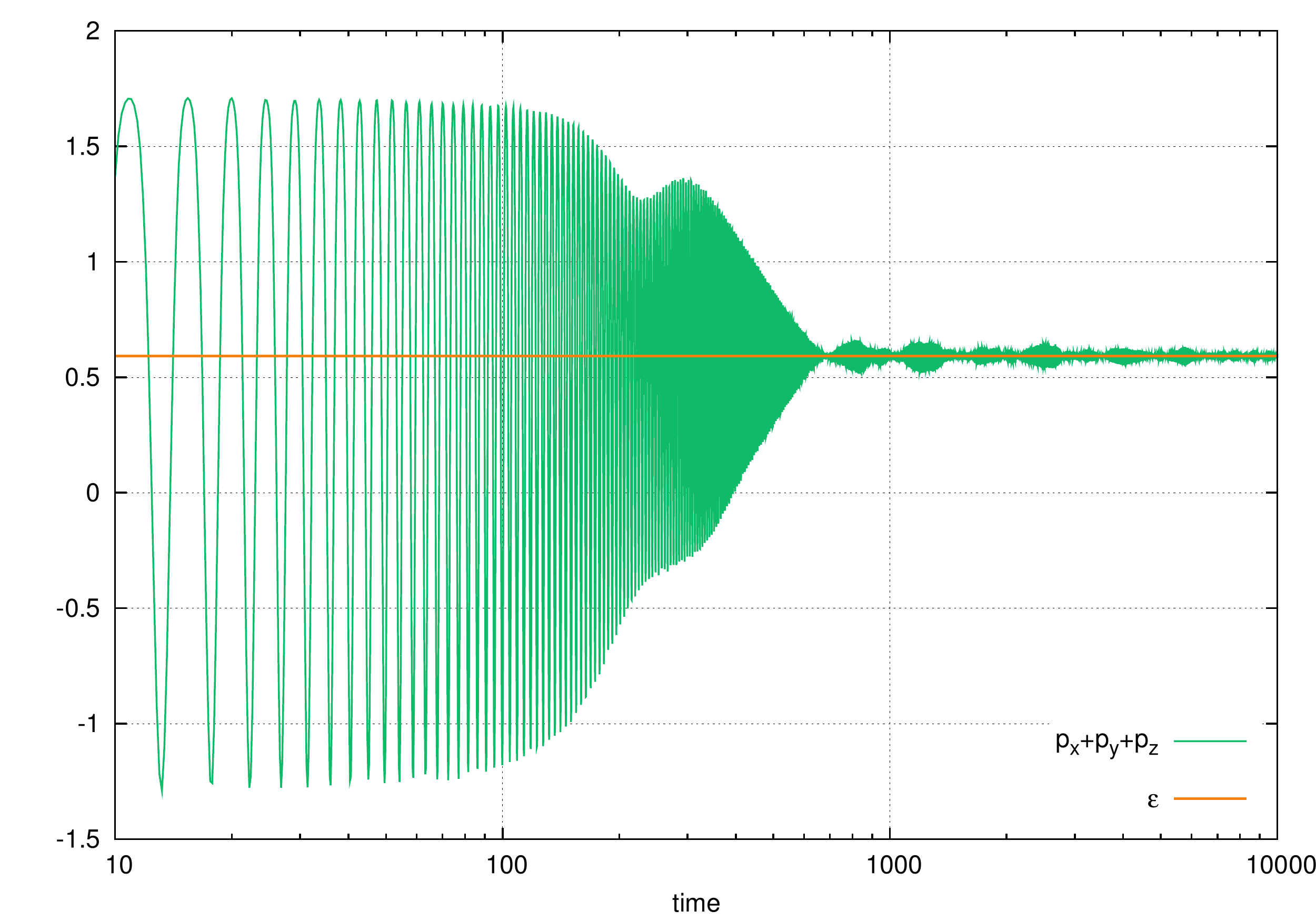}}}
\end{center}
\caption{\label{fig:relax}}
\end{figure}

\begin{figure}[htbp]
\begin{center}
\resizebox*{6cm}{!}{\rotatebox{-90}{\includegraphics{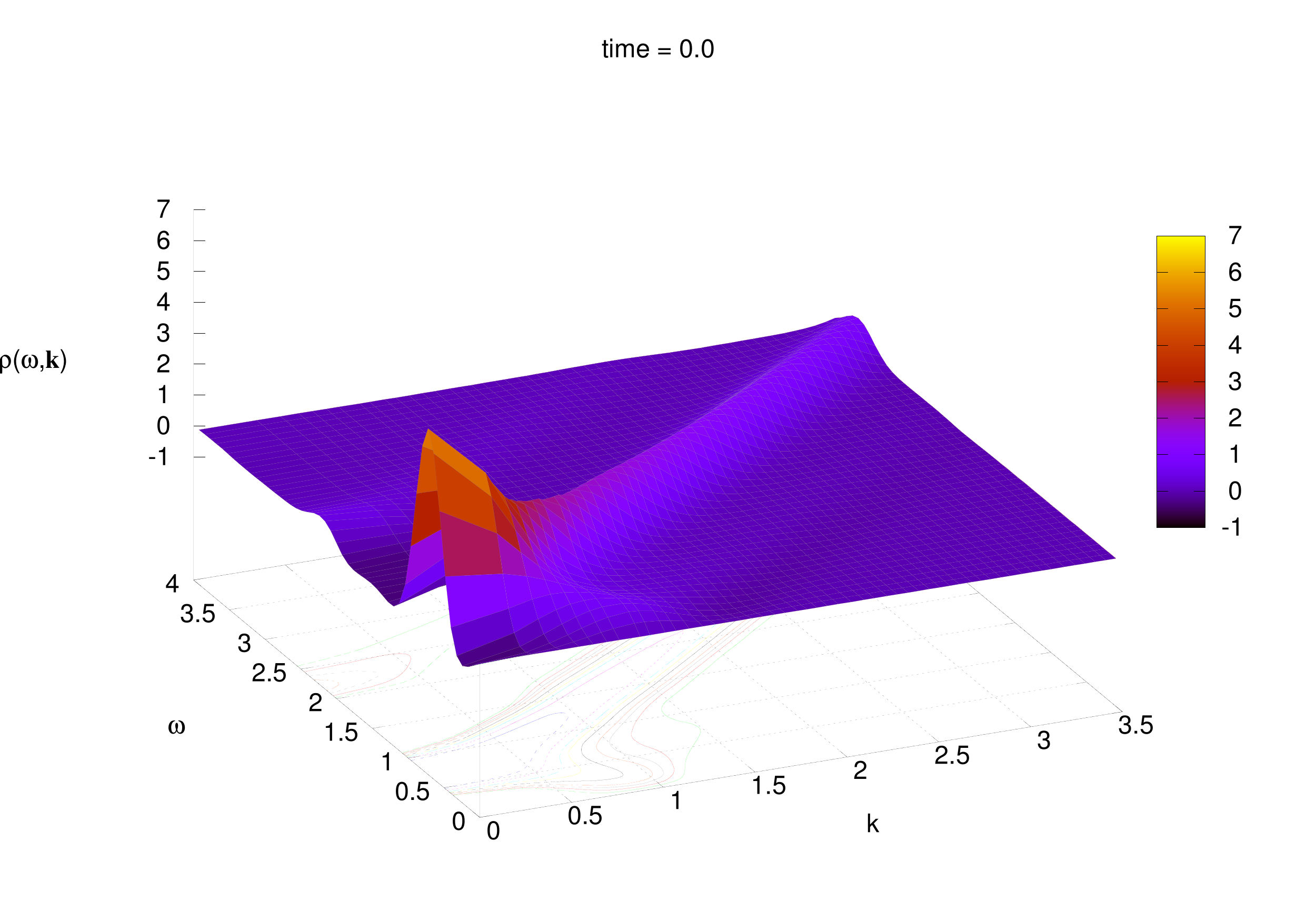}}}\hfil
\resizebox*{6cm}{!}{\rotatebox{-90}{\includegraphics{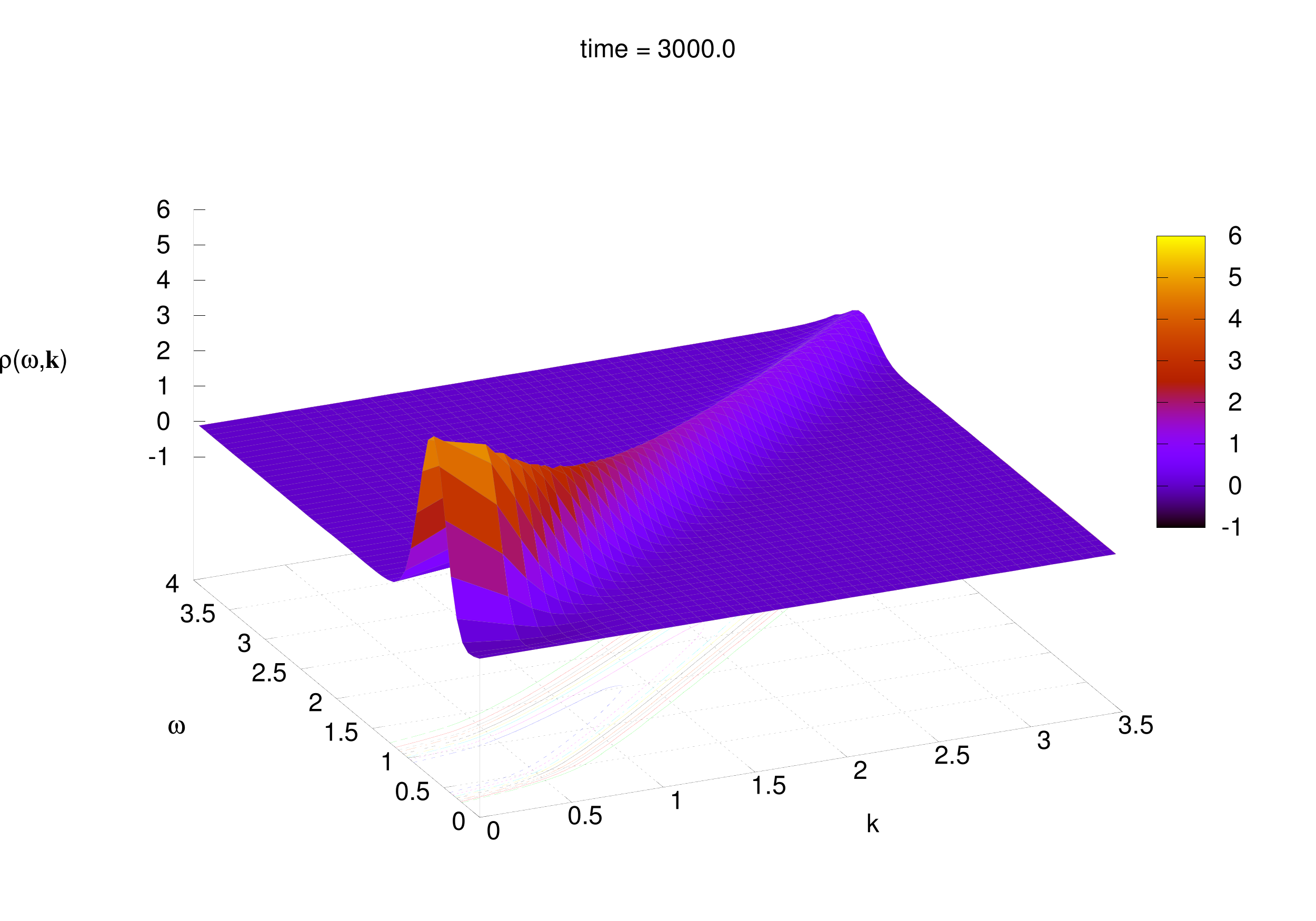}}}
\end{center}
\caption{\label{fig:spectral}Spectral function of the system at two
  different times. Left: t=0. Right: t=3000 (in lattice units).}
\end{figure}

In contrast, the resummed pressure behaves in a very different way, as
shown in the figure \ref{fig:relax}.
The oscillations are damped, so that the pressure reaches a fixed
value after a certain time. This asymptotic value of the pressure is
related to the energy density by a very simple {\sl equation of
  state},
\begin{equation}
\epsilon = 3P \; ,
\end{equation}
which is the expected relationship in a scale invariant theory in 3+1
dimensions. On similar timescales, the spectrum of excitations in the
system also becomes considerably simpler, as shown by the spectral
functions at two different times in the figure \ref{fig:spectral}.
At the initial time, the spectral function has a complicated structure
with more than one branch (including a branch that has a negative
slope). At later times (comparable to the relaxation time of the
pressure), only one branch remains in the spectral function,
suggesting that the system can be interpreted as a simple gas of
quasiparticles.

\begin{figure}[htbp]
\begin{center}
\resizebox*{6cm}{!}{\rotatebox{-90}{\includegraphics{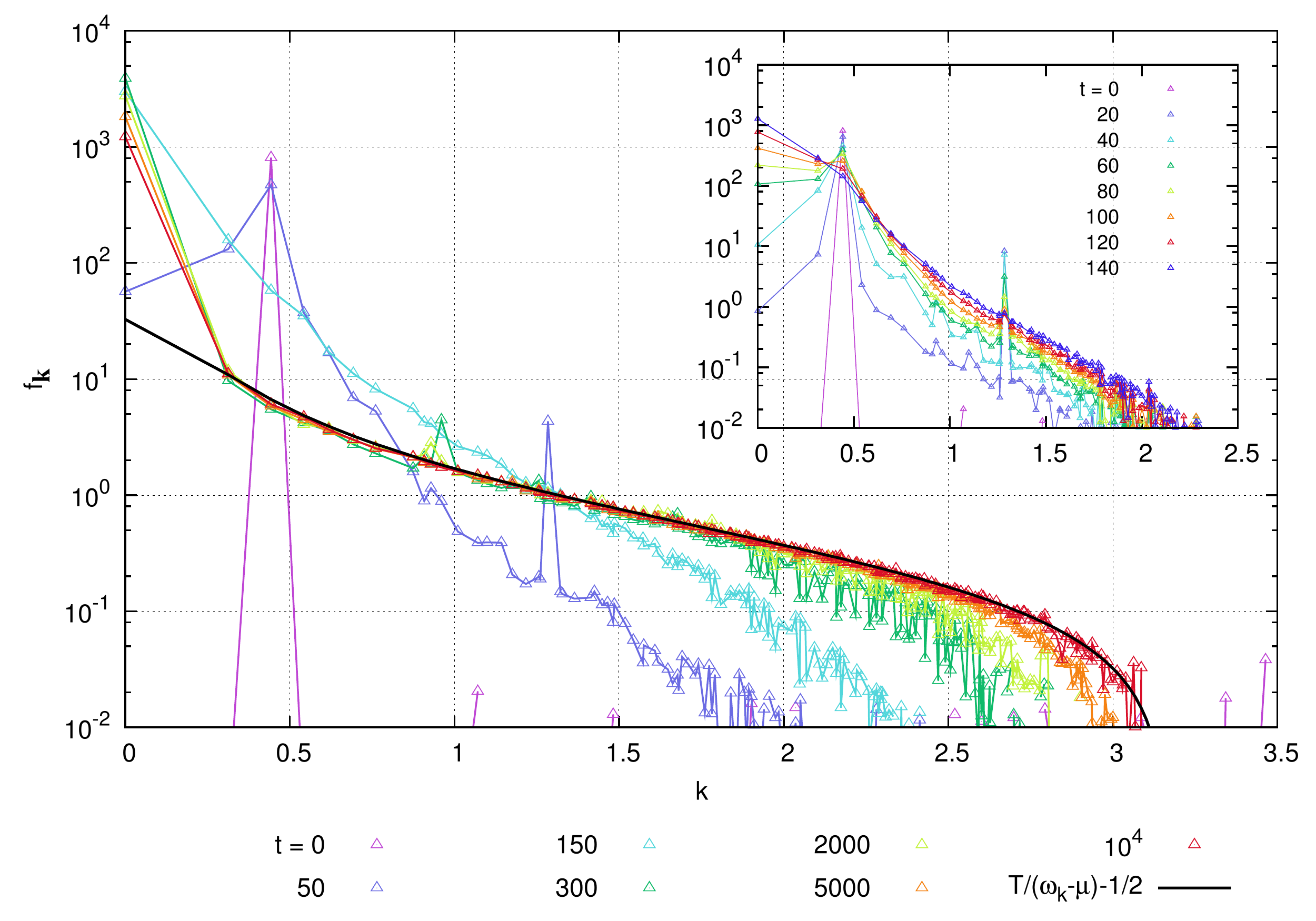}}}\hfil
\resizebox*{6cm}{!}{\rotatebox{-90}{\includegraphics{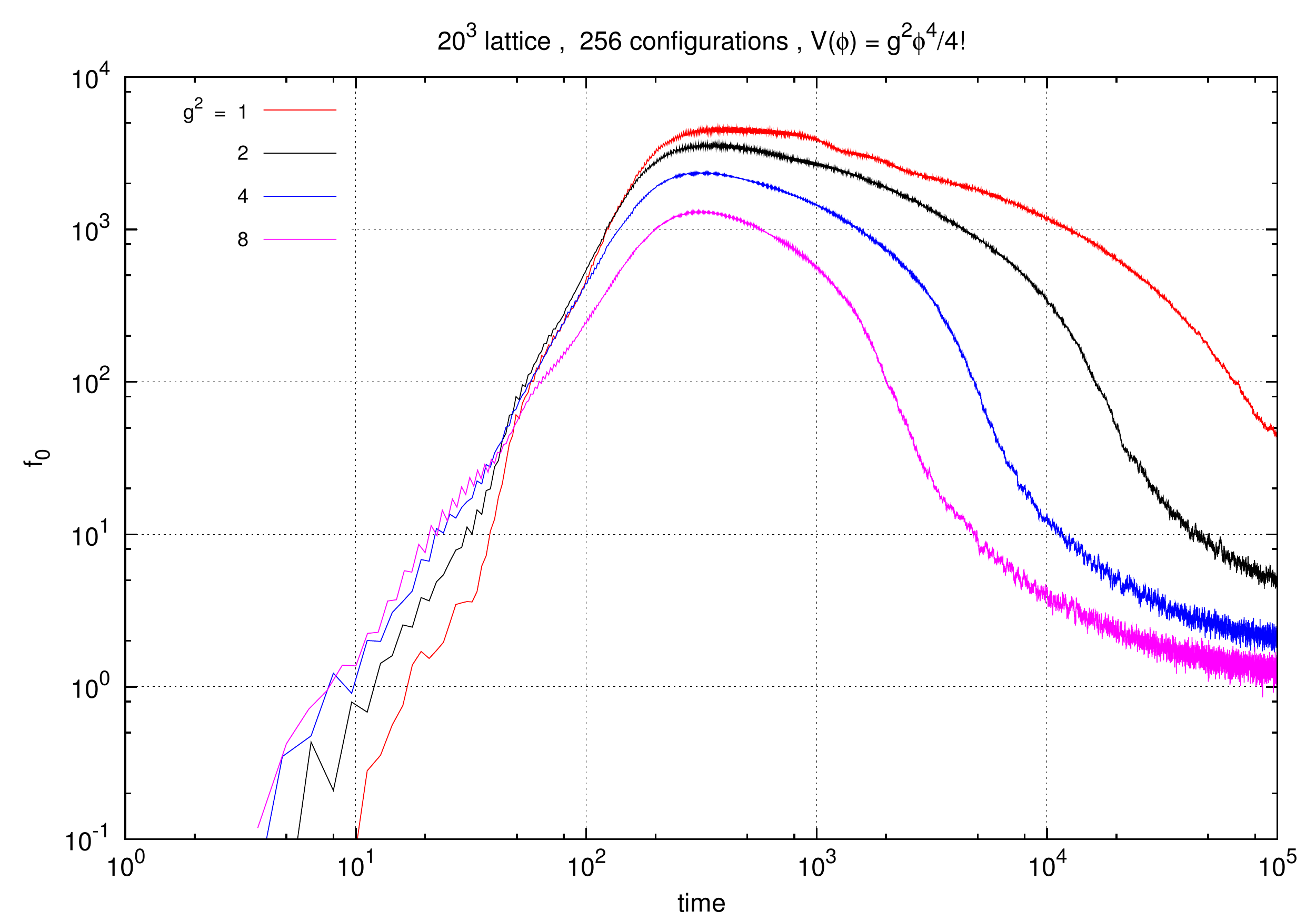}}}
\end{center}
\caption{\label{fig:fk}Left: occupation number as a function of
  momentum at various times. Right: time evolution of the occupation
  in the zero mode, for various values of the coupling constant.}
\end{figure}

However, the relaxation of the pressure does not imply that the system
has fully thermalized, as illustrated in the figure \ref{fig:fk}.
Here, the background field is chosen such that only a nonzero
momentum mode has a large occupation number at the initial time, while
all the other modes are empty. Very quickly, particles are produced
in all the other modes, including the zero mode, mostly by elastic
scatterings. At late times, one can fit the occupation number by a
function of the form,
\begin{equation}
F(\k) = \frac{T}{\omega_\k-\mu}-\frac{1}{2}\; ,
\end{equation}
where $\omega_\k^2=\k^2+m^2$ with $m$ the quasiparticle mass obtained
by studying the spectral function. $T$ and $\mu$ are fitted to the
numerical results. This functional form corresponds to the first two
terms of the expansion of a Bose-Einstein distribution in the soft
sector\footnote{The exponential tail of a Bose-Einstein distribution
  cannot be obtained in classical statistical field theory, because it
  is an approximation of the full theory which is valid only in the
  region where the occupation number is larger than
  unity.}. Interestingly, the best fit requires a nonzero chemical
potential. Since the theory under consideration has no conserved
particle number, this suggests that the system is not yet in chemical
equilibrium even at the largest times considered. This is expected for
the $\phi^4$ theory at weak coupling, because the inelastic
cross-section (that can equilibrate the particle number) is much
smaller than the elastic one.

Another interesting feature is the fact that the fitted value of the
chemical potential is very close to the mass of the
quasiparticles. This, combined with the fact that there seem to be an
excess of particles in the zero mode, suggests that the system
undergoes {\sl Bose-Einstein condensation}. We have confirmed this by
showing that the occupation number in the zero mode is proportional to
the volume of the system.  Of course, in a theory where the particle
number is not conserved, a Bose-Einstein condensate cannot be
stable. Eventually, the inelastic collisions will eliminate the excess
of particles that led to its formation, and dissolve the
condensate. This is visible in the right plot of the figure
\ref{fig:fk}, where the occupation number in the zero mode is shown as
a function of time. After a very rapid growth, $f(0)$ remains almost
constant for a rather long time, to eventually decrease on even longer
timescales (the timescales over which the inelastic processes become
effective). As expected, by increasing the coupling, the condensate
dissolves faster, because the inelastic collision rate is larger.

Whether a Bose-Einstein condensate of gluons can form in heavy ion
collisions is still an open question. The condition of overpopulation
is initially also satisfied by the glasma fields\cite{BlaizGLMV1},
but the inelastic rates are not parametrically suppressed compared to
the elastic ones. Therefore it may happen that the condensate
dissolves as fast as it forms, never really becoming a relevant
feature of the time evolution.

\subsection{Evolution of a longitudinally expanding system}
In nucleus-nucleus collisions at high energy, another important
question arises: is the longitudinal pressure comparable to the
transverse pressure? This is an important aspect in the applicability
of hydrodynamics: if the two pressures are two different, then the
viscous corrections are large. When discussing the glasma fields, we
have seen that at leading order, the longitudinal pressure is in fact
the opposite of the energy density at $\tau=0^+$. This is in fact a
generic feature in any system of longitudinally expanding
fields. Indeed, energy and momentum conservation lead to the following
equation (for a system homogeneous in the transverse plane)
\begin{equation}
\frac{\partial\epsilon}{\partial\tau}+\frac{\epsilon+P_{_L}}{\tau}=0\; .
\label{eq:hydro}
\end{equation}
In order to a get a finite $\epsilon$ at $\tau=0^+$, it is necessary
that $\epsilon+P_{_L}$ vanishes at $\tau=0^+$. As we have seen
earlier, this is true in the glasma. This is also the case in the toy
scalar model that we have considered in these numerical studies.

A central question is whether the Gaussian fluctuations that are
superimposed to the background field can cause the longitudinal
pressure to become equal to the transverse one. This question can be
addressed in the scalar $\phi^4$ model considered in this
section. We simply need to use $\tau,\eta$ coordinates, and to setup a
boost invariant background field, i.e. independent of $\eta$.
\begin{figure}[htbp]
\begin{center}
\resizebox*{9cm}{!}{\rotatebox{0}{\includegraphics{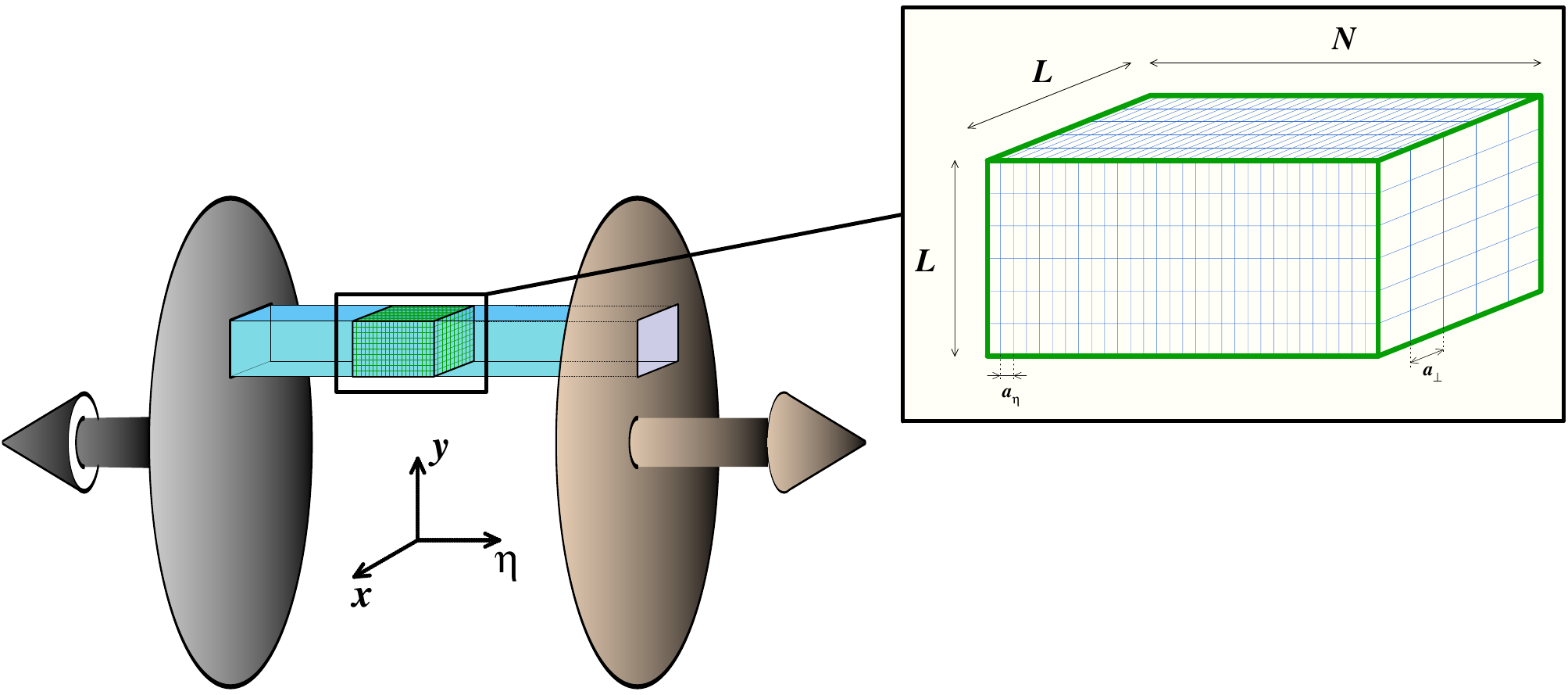}}}
\end{center}
\caption{\label{fig:simulexp}Setup for the lattice study of a system
  in longitudinal expansion, as in a high energy collision.}
\end{figure}
The connection between this kind of lattice simulation and an actual
collision is illustrated in the figure \ref{fig:simulexp}. It would be
too costly to simulate the entire volume of the matter produced in a
collision, therefore the lattice describes only a small patch of this
matter that has a fixed extent in $\eta$ and $\x_\perp$ (and
therefore stretches in $z$ as time increases).

\begin{figure}[htbp]
\begin{center}
\resizebox*{6cm}{!}{\includegraphics{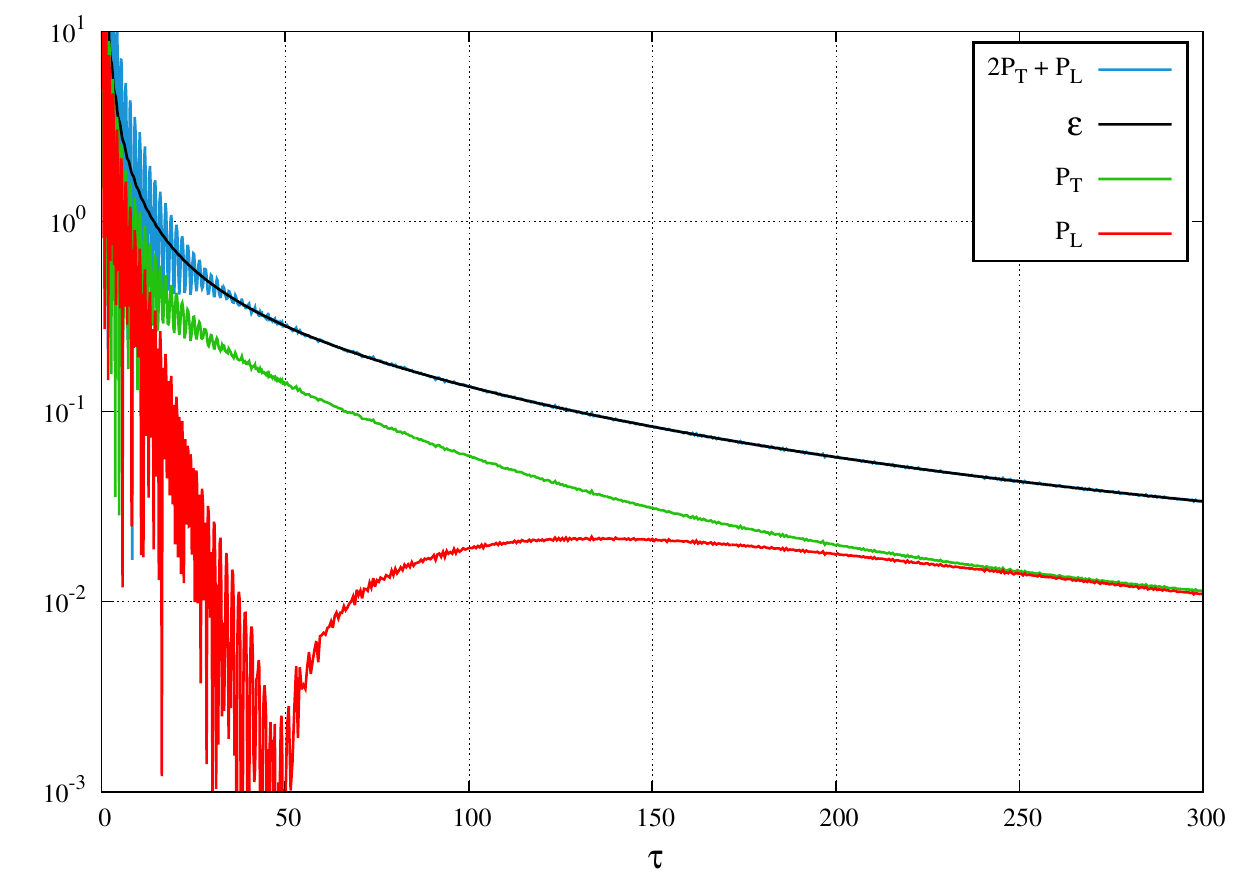}}\hfil
\resizebox*{6cm}{!}{\includegraphics{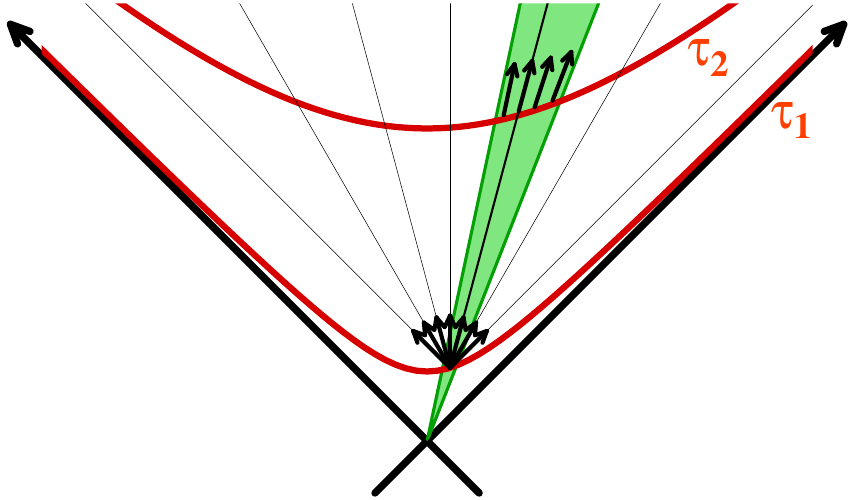}}
\end{center}
\caption{\label{fig:tmunu}Left: time evolution of the energy density,
  and of the transverse and longitudinal pressures, in a longitudinal
  expanding system of scalar fields. Right: illustration of the
  redshifting of the longitudinal momenta due to longitudinal
  expansion. The black arrows represent the momentum vectors of
  particles, and the thin lines their trajectories if they move
  freely.}
\end{figure}

The behavior of the energy density, and of the transverse and
longitudinal pressures, is shown in the plot on the left of the figure
\ref{fig:tmunu}. At very early times, the trace of the pressure tensor
($2P_{_T}+P_{_L}$) has oscillations. These oscillations are quickly
damped, and the system then obeys the equilibrium equation of state,
\begin{equation}
\epsilon = 2P_{_T}+P_{_L}\; .
\end{equation}
However, at this stage of the time evolution, the transverse and
longitudinal pressures still behave very differently. The longitudinal
pressure decreases much faster, and even becomes negligible compared
to the transverse pressure. This is explained naturally by the
redshifting of the longitudinal momenta in an expanding system. This
effect is illustrated in the right part of the figure \ref{fig:tmunu}:
if one starts with a broad distribution of longitudinal momenta in a
given slice of rapidity $\eta$, after some period of free streaming
only the particles that have a momentum rapidity $y$ equal to the
space-time rapidity $\eta$ remain in this slice. Thus, the
longitudinal pressure (defined in the local rest frame of the matter,
i.e. in a frame comoving along the rapidity slice) decreases rapidly.

However, at some point, the longitudinal pressure stops
decreasing. Instead, it increases exponentially and becomes very close
to the transverse pressure, leading to an almost perfectly isotropic
pressure tensor. This radical change of behavior of the longitudinal
pressure seems related to the instabilities present in the system: at
early times, the expansion rate (proportional to $\tau^{-1}$) of the
system is too large for the instabilities to be able to compete
efficiently against the redshifting. It is only at later times, when
the expansion rate has become low enough, that the instabilities
become the driving force in the system.

It is interesting to compare these results, obtained in classical
statistical field theory, with hydrodynamical evolution. Naturally,
since the transverse and longitudinal pressures are not equal, viscous
corrections must be included in the hydrodynamical description to
account for this difference. 
\begin{figure}[htbp]
\begin{center}
\resizebox*{6cm}{!}{\includegraphics{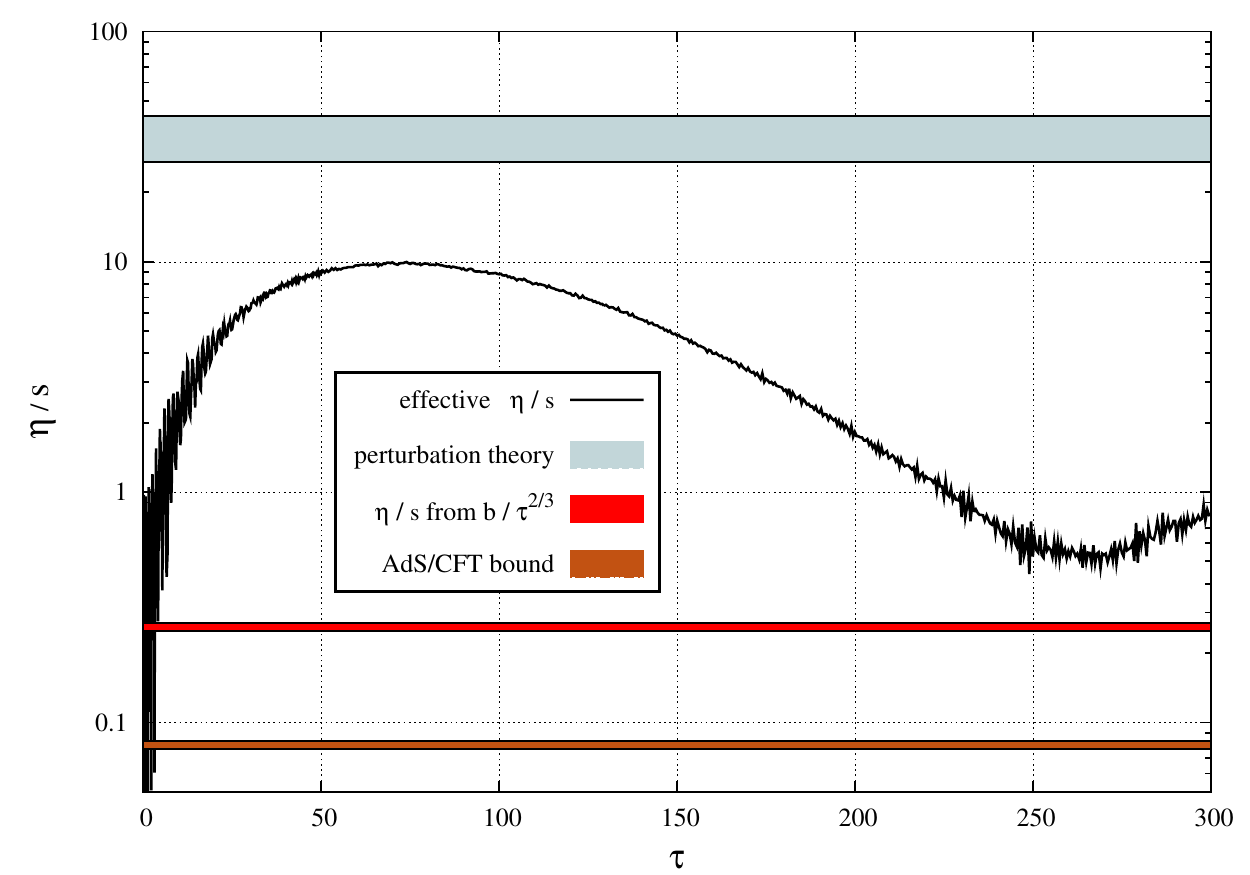}}
\hfill
\resizebox*{6cm}{!}{\includegraphics{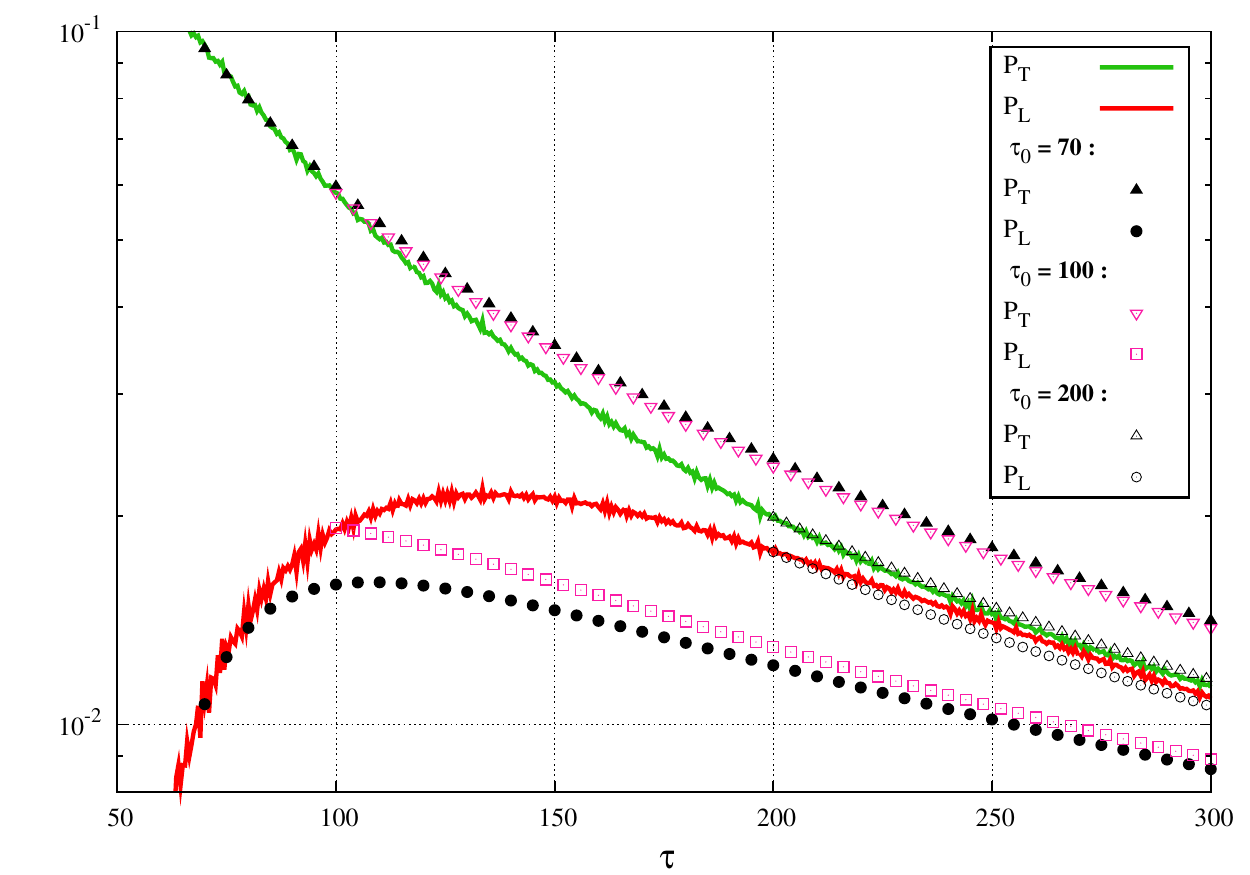}}
\end{center}
\caption{\label{fig:hydro}Left: estimate of the viscosity to entropy
  density ratio. Right: comparison between classical statistical field
  theory and first order hydrodynamics, for various hydro initial
  times.}
\end{figure}
In a system which is homogeneous in the transverse plane and boost
invariant, the simplest ansatz is to write
\begin{equation}
P_{_T}=\frac{\epsilon}{3}+\frac{2\eta}{3\tau}\quad,\quad
P_{_L}=\frac{\epsilon}{3}-\frac{4\eta}{3\tau}\; ,
\label{eq:1storder}
\end{equation}
where $\eta$ is the shear viscosity of the system (this corresponds to
{\sl first order hydrodynamics}). In an equilibrated scale invariant
fluid, $\eta$ is proportional to the entropy density $s$, and it is
customary to quantify how viscous a fluid is by the dimensionless
ratio $\eta/s$. If we assume Stefan-Boltzmann's law for the entropy
density, we have $s\approx \epsilon^{3/4}$, and since we know
$\epsilon, P_{_{T,L}}$ as a function of $\tau$, we can extract from
our results an effective value of $\eta/s$. The result of this
estimate is shown in the plot on the left of the figure
\ref{fig:hydro}. Clearly, the ratio $\eta/s$ is not constant. Its time
dependence could be attributed to several causes: first order
hydrodynamics (i.e. eq.~(\ref{eq:1storder})) may not be a valid
description of the system at these early times, and/or the system may
still be too far from local equilibrium.  Nevertheless, this estimate
is instructive because it leads to an $\eta/s$ ratio which is much
smaller than the perturbative estimate (blue band in the plot). This
may be a manifestation of the anomalously small viscosity that has
been conjectured\cite{AsakaBM1,AsakaBM2} for systems subject to
instabilities and turbulence. Note also that the ratio we have
extracted remains significantly above the value $1/4\pi$, obtained for
certain gauge theories in the limit of infinite coupling.

Another way to compare our results with first order hydrodynamics is
to solve eq.~(\ref{eq:hydro}) with the ansatz of
eq.~(\ref{eq:1storder}) for the longitudinal pressure, under the
assumption that the ratio $\eta/s$ is a constant. To close the
equation, we may assume again $s\approx \epsilon^{3/4}$. A starting
time $\tau_0$ must be chosen for this comparison, and the difference
between $P_{_T}$ and $P_{_L}$ at this initial time determines the
value of $\eta/s$. For $\tau>\tau_0$, the evolution of the system is
governed by eq.~(\ref{eq:hydro}). This comparison is shown in the plot
on the right of the figure \ref{fig:hydro}, for several initial times.
It appears that the relaxation $P_{_L}\to P_{_T}$ is much faster in
the glasma description than in hydrodynamics, presumably due to the
effect of instabilities. It is only if the hydrodynamical evolution
is started when the system is already well isotropized that it provides
a good description of the evolution of the system.  Naturally,
eq.~(\ref{eq:hydro}) is certainly valid because it is just a
consequence of energy and momentum conservation. Therefore, this
comparison questions the validity of the first order approximation to
describe the viscous corrections (and the non-constancy of the ratio
$\eta/s$ when extracted directly from $P_{_T}-P_{_L}$ points also in
the same direction). The other conclusion one can make from this
comparison, given the fact that hydrodynamics is the limit of kinetic
theory when the mean free path goes to zero, is that instabilities are
more efficient  than collisions to isotropize the system.

\section{Conclusions}
\label{sec:conclusions}
In this review, we have presented the Color Glass Condensate effective
theory, that describes the partonic content of a hadron or nucleus in
the saturated regime. Gluon saturation, that happens at high energy,
is characterized by large gluon occupation numbers, and by non-linear
effects not present in the dilute regime. In order to describe
collisions involving hadrons in this regime, one needs a framework
that provides a handle on multi-gluon Fock states in the hadron
wavefunction, and the tools to perform calculations with these states.

The Color Glass Condensate approximates the fast partons as stochastic
classical color sources, while the usual field description is retained
for the slow gluons. The requirement that observables be independent
on the cutoff that separates these two types of degrees of freedom
leads to a renormalization group equation (the JIMWLK equation) for the
distribution of the classical sources.

A crucial aspect of the physics of gluon saturation is that the
non-linear dynamics generates a semi-hard scale, the saturation
momentum. This scale, that increases with energy, plays a crucial role
because it sets the value of the strong coupling constant, allowing a
weak coupling treatment of gluon saturation. Nevertheless, even at
weak coupling, the study of hadronic or nuclear collisions in the
saturated regime is non-perturbative, because the smallness of the
coupling is compensated by the large occupation numbers for the gluons
below the saturation momentum.

The Color Glass Condensate framework proves especially useful in the
study of heavy ion collisions at high energy, because the bulk of
particle production in these collisions is controlled by the physics
of gluon saturation. Moreover, it has been shown that inclusive
observables can be factorized (at leading logarithmic accuracy so far)
as a convolution of universal distributions of color sources
representing the gluon content of the two colliding nuclei, and an
observable evaluated in the retarded classical field produced by these
sources.

The universality of these distributions of sources, and therefore of
the classical color field they produce, confers to these field
configurations --the {\sl glasma}-- a very central role in the CGC
description of heavy ion collisions. At very short times after the
collision, they are characterized by longitudinal chromo-electric and
chromo-magnetic fields that are nearly boost invariant, with a
transverse correlation length of the order of the inverse saturation
momentum. In particular, these fields that have long range
correlations in rapidity have been invoked in order to explain
the observed two-hadron correlations in heavy ion collisions.

The initial glasma fields form a system which is very from local
equilibrium: it is a coherent state rather than a mixed thermal state,
and it has a large negative longitudinal pressure. Moreover, these
fields are unstable against small perturbations of their initial
conditions. In fixed order CGC calculations, these unstable modes lead
to secular divergences in quantities such as the energy-momentum
tensor. This pathology can be cured by a resummation that amounts to
letting the initial glasma field fluctuate around its classical value,
with a Gaussian spectrum determined by a one-loop calculation. A
numerical implementation of this resummation in a much simpler scalar
field theory has recently shown that the instabilities lead to the
isotropization of the transverse and longitudinal pressures shortly
after the collision, and to full thermalization on longer
timescales. Whether this is also the case in QCD, and over which
timescale, is the subject of ongoing works.

\section*{Acknowledgements}
I would like to thank the organizers of the 22nd Jyv\"askyl\"a Summer
School, where these lectures have been delivered, and in particular
T. Lappi and T. Renk, as well as all the students for their questions
and remarks. This work is supported by the Agence Nationale de la
Recherche project \#~11-BS04-015-01.

%\bibliographystyle{unsrt}
%\bibliography{biblio}

\end{document}